\documentclass{article}
\usepackage[utf8]{inputenc}
\usepackage[paper=a4paper,left=25mm,right=25mm,top=25mm,bottom=25mm]{geometry}

\usepackage{xr-hyper}
\makeatletter

\usepackage{hyperref,xcolor}

\usepackage{siunitx}
\usepackage{pdflscape}
\usepackage{amsmath, amsfonts, amssymb, mathtools, amsthm}
\usepackage{cite}
\usepackage[super,sort&compress]{natbib}       
\usepackage{arydshln}
\usepackage{float}
\usepackage{dsfont}
\usepackage{authblk}
\usepackage{bbm} % \mathbb für Zeilen, für Indikatorfunktionen
\usepackage{multicol} % spaltenübergreifende Zellen in Tabellen
\usepackage{multirow} % reihenübergreifende Zellen in Tabellen
\usepackage{subcaption} % mehrere Plots in einer figure mit separaten captions
\usepackage{listings} %Code style
\usepackage{enumitem}

\DeclareMathOperator*{\argmax}{arg\,max}

\newtheorem{theorem}{Theorem}
\newtheorem{lemma}{Lemma}
\newtheorem{corollary}{Corollary}

\usepackage{rotating,graphicx} % Package for double graphics

\title{Adaptive weight selection for time-to-event data under non-proportional hazards}
\author[1]{Moritz Fabian Danzer}
\author[2]{Ina Dormuth}

\date{September 2024}

\affil[1]{Institute of Biostatistics and Clinical Research, University of Münster, 48149 Münster, Germany}
\affil[2]{Department of Statistics, TU Dortmund University, 44227 Dortmund, Germany}

\begin{document}

\maketitle

\begin{abstract}
When planning a clinical trial for a time-to-event endpoint, we require an estimated effect size and need to consider the type of effect. Usually, an effect of proportional hazards is assumed with the hazard ratio as the corresponding effect measure. Thus, the standard procedure for survival data is generally based on a single-stage log-rank test. Knowing that the assumption of proportional hazards is often violated and sufficient knowledge to derive reasonable effect sizes is usually unavailable, such an approach is relatively rigid. We introduce a more flexible procedure by combining two methods designed to be more robust in case we have little to no prior knowledge. First, we employ a more flexible adaptive multi-stage design instead of a single-stage design. %It offers the well-known advantages and flexibility of an adaptive design. 
Second, we apply combination-type tests in the first stage of our suggested procedure to benefit from their robustness under uncertainty about the deviation pattern. %They permit interim analyses at various time points, allowing for changes in study design, such as correcting the assumed effect size or resulting in possible early termination of the study.
%, which are more powerful under non-proportional hazards while maintaining a reasonable power under proportional hazards \cite{Dormuth:2023}. 
We can then use the data collected during this period to choose a more specific single-weighted log-rank test for the subsequent stages. In this step, we employ Royston-Parmar spline models to extrapolate the survival curves to make a reasonable decision. Based on a real-world data example, we show that our approach can save a trial that would otherwise end with an inconclusive result. Additionally, our simulation studies demonstrate a sufficient power performance while maintaining more flexibility.
	
\end{abstract}

\vskip .5in


\section{Introduction}
There are two common sources of uncertainty in the planning phase of a clinical trial with a survival endpoint. On the one hand, we need to identify the effect size on which case number planning is based. On the other hand, we must make assumptions about the type of effect. Traditionally, an effect of proportional hazards is considered, whereby the effect size is referred to as the hazard ratio. However, the assumption of proportional hazards must often be questioned. This is particularly the case if therapies with different mechanisms of action are to be compared.\\
Adaptive designs are widely accepted as a possible solution for the uncertainty regarding the effect size\cite{Bauer:2016, Wassmer:2016}. Such designs allow for interim analyses at multiple time points. In these interim analyses, the study can be terminated early, either with or without rejection of the null hypothesis, or the further course of the study can be adapted. This concerns, for example, the planning of future interim analyses or an adjustment of the sample size. For this purpose, the data collected to date on the endpoint to be investigated can be used as a basis for planning. This data can be used to calculate the conditional power of the design. The design is often adapted so that the conditional power reaches a specific target value. In particular, such an adaptive procedure allows us to revisit the assumptions regarding the effect size made at the beginning of the study and correct them if needed.

Stage-wise log-rank statistics are often employed in standard adaptive designs for survival data. Then, the initial sample size planning and the sample size recalculation are commonly based on the hazard ratio that can also be re-estimated during the trial. This is, e.g., demonstrated in Wassmer (2006)\cite{Wassmer:2006}. The technical foundation is given in Tsiatis (1982)\cite{Tsiatis:1982}.\\
As in the adaptive (multi-stage) approach described above, the standard log-rank test is commonly applied in single-stage designs, with one single analysis at the end of the study. This is because, in proportional hazard situations, the log-rank test is the optimal test in terms of power. When this assumption is violated, it might, however, lose power and thus lead to poor test decisions. One way to increase the power for non-proportional hazard alternatives is by implementing a weight function. Multiple weight functions have been introduced over time \cite{Harrington:1982, tarone1977distribution}. More recent approaches facilitate the weight-choosing procedure and increase robustness against various hazard patterns by introducing combination-type tests \cite{Brendel:2014, DitzhausFriedrich2018, linAlternativeAnalysisMethods2020}. One combination approach is the multi-directional log-rank (\textit{mdir}) test, originally proposed by Brendel et al. \cite{Brendel:2014} and revisited by Ditzhaus and Friedrich \cite{DitzhausFriedrich2018}. This Wald-type test statistic consists of multiple weighted log-rank tests covering several alternatives and their linear combinations. Dormuth et al. \cite{Dormuth:2023} illustrated the robust power behavior of the two-sided version of the \textit{mdir}test. Ditzhaus et al. \cite{Ditzhaus:2019} extended the test for a one-sided testing problem.\\ %using bootstrapping.\\
Employing more robust testing procedures is also crucial in multi-stage designs \cite{Jimenez:2022}. Group sequential designs for weighted log-rank tests\cite{Hasegawa:2016} and the \textit{max combo} combination test\cite{Ghosh:2022} have already been proposed. One straightforward extension of such procedures is to select testing procedures for the forthcoming stages based on the available information. Of course, this only makes sense if the test statistics in each stage test the same hypothesis. Selecting the test depending on a proportional hazards check has already been considered \cite{Campbell:2014}. However, it must be taken into account here that such two-stage procedures generally increase the significance level, and a corresponding adjustment must be made. An adaptation of the testing procedure for survival endpoints has only been considered in Lawrence (2002)\cite{Lawrence:2002}. We will go beyond this work in several respects:\\  
Using a slightly more formal approach, we obtain second-stage test statistics explicitly as increments of the corresponding stochastic processes. An additional calculation of covariances is therefore not necessary due to the use of prominent asymptotic results\cite{Tsiatis:1982, Scharfstein:1997}. Furthermore, this approach also allows combination tests in individual design stages. In particular, we would like to use these in the first stage to apply a robust test under uncertainty about the type of deviation. For the following stages, we use the information available to select the best weighted log-rank test regarding conditional power. Therefore, we present flexible calculations for this quantity. The full first-stage primary endpoint data can be used to guide this decision. In addition, the basic design still allows for adjustments to the sample size or changes to the analysis schedule.\\
The manuscript is organized as follows. In Section \ref{sec:methods}, we give a brief overview of the methods and techniques we will combine to construct our adaptive testing procedure, which will be presented in Section \ref{sec:admdir}. We illustrate its application on a reconstructed data set in Section \ref{sec:example} and study its characteristics via simulation in Section \ref{sec:sim_study}. Finally, in section \ref{sec:discussion}, we summarize our findings, discuss them, and give outlooks for further research. Further material on our application example, the simulation study, and technical foundations are provided in the Supplementary Material.

\section{Methods}\label{sec:methods}

This section will present various methods and techniques we require to formulate our proposed method. We will start by introducing some notation.\\
Any patient $i\in \{1,\dots,n\}$ enters the trial at the random time $R_i \geq 0$ in calendar time. That patient is assigned to treatment group $Z_i \in \{0,1\}$. The patient experiences the event of interest at the random time $T_i \geq 0$ after recruitment. However, the observation of this event may be censored. This may either happen because of a random dropout or administrative censoring. The former occurs at $C^{\star}_i$. The latter depends on the time of analysis. If this analysis is performed at calendar time $t$, the latter censoring is given by $(t - R_i)_+$. Overall, censoring at this calendar date is thus given by $C_i(t)\coloneqq C^{\star}_i \wedge (t - R_i)_+$ where $\wedge$ indicates the minimum of two real numbers. The observation that can hence be made at that time is the probably censored event date $X_i(t)\coloneqq T_i \wedge C_i(t)$ and the corresponding indicator function $\delta_i(t)\coloneqq \mathbbm{1}_{\{T_i \leq C_i(t)\}}$.\\
We assume that the tuples $(R_i, Z_i, C^{\star}_i, T_i)$ are independent and identically distributed for all $i \in \{1,\dots,n\}$. In particular, they shall be independent replicates of some tuple $(R, Z, C^{\star}, T)$. Furthermore, the censoring through $C(t)$ shall constitute an independent censoring mechanism and we assume that  $R$ and $C^{\star}$ are independent random variables with cumulative distribution functions $F_R$ and $F_{C^{\star}}$, respectively.\\
Based on this, we can now define counting processes and at-risk indicators for the event of interest. For any $i \in \{1,\dots,n\}$, the multivariate process $(N_i(t,s))_{t \geq 0, s \geq 0}$ defined by
\begin{equation*}
N_i(t,s) \coloneqq \mathbbm{1}_{\{ T_i \leq s \wedge C_i(t) \}}
\end{equation*}
indicates whether the event of interest was observed before calendar time $t$ and before patient $i$ spent $s$ time units in the study. These processes can be aggregated over the complete study sample to obtain the overall number of events $N(t,s) \coloneqq \sum_{i=1}^n N_i(t,s)$ observed before calendar time $t$ and trial time $s$. Additionally, we define those processes in the subgroups of patients that are assigned to the same treatment, i.e.
\begin{equation*}
    N^{Z=k}(t,s) \coloneqq \sum_{i=1}^n N_i(t,s) \cdot \mathbbm{1}_{\{Z_i=k\}}
\end{equation*}
for treatment group $k\in \{0,1\}$.\\
Similarly, the multivariate process $(Y_i(t,s))_{t \geq 0, s \geq 0}$ indicates whether it is known at calendar time $t$ that patient $i \in \{1,\dots,n\}$ remained event-free in the trial just before $s$ time units after its enrollment, i.e.
\begin{equation*}
    Y_i(t,s) \coloneqq \mathbbm{1}_{\{N_i(t,s-) = 0\}} \cdot \mathbbm{1}_{\{s \leq C_i(t)\}}
\end{equation*}
The term $N_i(t,s-)$ denotes the left-hand limit in the second argument, i.e.
\begin{equation*}
    N_i(t,s-)\coloneqq \lim_{u \nearrow s} N_i(t,u).
\end{equation*}
As above, we can aggregate these quantities over the complete study sample or the two treatment groups to obtain the processes $(Y(t,s))_{t \geq 0, s \geq 0}$ respectively $(Y^{Z=k}(t,s))_{t \geq 0, s \geq 0}$ for $k \in \{0,1\}$.\\
Given these quantities, we can estimate the pooled and group-specific survival functions $S, S_k\colon[0,\infty) \to [0,1]$ of $T$ which are defined by $S(s)\coloneqq \mathbb{P}[T \geq s]$ and $S_k(s)\coloneqq \mathbb{P}[T \geq s|Z=k]$, respectively, using the information collected up to calendar time $t$ by
\begin{equation*}
    \hat{S}(t,s) \coloneqq \prod_{i \colon X_i(t) \leq s} \left( 1 - \frac{\delta_i(t)}{Y(t,X_i(t))} \right)
\end{equation*}
resp.
\begin{equation*}
    \hat{S}_k(t,s) \coloneqq \prod_{i \colon X_i(t) \leq s, Z_i = k} \left( 1 - \frac{\delta_i(t)}{Y^{Z=k}(t,X_i(t))} \right)
\end{equation*}
for $k \in \{0,1\}$. Corresponding cumulative distribution functions $F, F_k$ and estimators $\hat{F},\hat{F}_k$ are given by the probabilities of the respective complementary events.\\
The corresponding cumulative hazard functions $A,A_k\colon [0,\infty)\to[0,\infty)$ defined by $A(t)\coloneqq -\log(S(t))$ and $A_k(t)\coloneqq -\log(S_k(t))$, respectively, can be estimated at calendar time $t$ by the Nelson-Aalen estimates
\begin{equation*}
    \hat{A}(t,s)\coloneqq \int_{[0,s]} \frac{\mathbbm{1}_{\{ Y(t,u) > 0 \}}}{Y(t,u)} dN(t,u) = \sum_{i \colon X_i(t) \leq s} \frac{\delta_i(t)}{Y(t,X_i(t))}
\end{equation*}
resp.
\begin{equation*}
    \hat{A}_k(t,s)\coloneqq \int_{[0,s]} \frac{\mathbbm{1}_{\{ Y^{Z=k}(t,u) > 0 \}}}{Y^{Z=k}(t,u)} dN^{Z=k}(t,u) = \sum_{i \colon X_i(t) \leq s, Z_i=k} \frac{\delta_i(t)}{Y^{Z=k}(t,X_i(t))}
\end{equation*}
for $k \in \{0,1\}$.\\
In what follows, we present testing procedures to investigate the null hypothesis of equal survival distributions
\begin{equation}\label{eq:null_hyp}
    H_0\colon\{A_0 \equiv A_1\}=\{S_0 \equiv S_1\}.
\end{equation}
It will be tested against the one-sided alternative of superiority
\begin{equation}\label{eq:alt_hyp}
    H_{\geq}\colon\{A_0 \geq A_1, A_0 \not\equiv A_1\}=\{S_0 \leq S_1, S_0 \not\equiv S_1\}.
\end{equation}
In particular, weighted log-rank tests and combination tests can be applied to address this issue. These will be presented in the following subsections.

\subsection{Weighted log-rank tests}\label{subsec:weighted_logrank}
Weighted log-rank tests allow giving more emphasis to different parts of the survival function depending on the selection of appropriate weights \cite{fleming2011counting}. The non-standardized test statistic at calendar time $t$ is defined as
	\begin{equation}\label{eq:wlrt}
	T_{\hat{Q}}(t)\coloneqq n^{-\frac{1}{2}} \sum_{i \colon X_i(t)\leq s, \delta_i(t)=1} \hat{Q}(t,X_i(t)) \underbrace{\left(Z_i -   \frac{Y^{Z=1}(t,X_i(t))}{Y(t,X_i(t))} \right),}_{\text{observed event}-\text{expected event}}
	\end{equation}
with some weight function $\hat{Q} \colon [0,\infty)^2 \to \mathbb{R}$ that fulfills the standard assumptions for repeated testing with log-rank type tests\cite{Tsiatis:1982, Harrington:1982} (see also Supplementary Material, Section \ref{supp-sec:technical}). In particular, its large sample limit $Q$ shall not depend on calendar time $t$ and hence be only a function of $s$. Additionally, we require the weight functions to be positive such that the test is valid to test the null hypothesis \eqref{eq:null_hyp} against the one-sided alternative from \eqref{eq:alt_hyp}. A set of weight functions that satisfies these conditions are referred to as the Fleming-Harrington weights. The weight functions are based on pooled Kaplan-Meier estimates and parameters $\rho,\gamma \geq 0$:
	\begin{equation}\label{eq:fh_weights}
	\hat{Q}(t,s)=w^{(\rho,\gamma)}(\hat{F}(t,s-))\coloneqq \hat{F}(t,s-)^\rho \cdot \hat{S}(t,s-)^\gamma.
	\end{equation}
The test statistic can be standardized using the corresponding variance. For big enough sample sizes, the standardized test statistic is approximately standard normally distributed under the null. The $p$-values are then obtained employing the associated tables.
 
We obtain the classical log-rank test when setting $\rho = \gamma = 0$. The resulting test is optimal regarding power under proportional hazards \cite{fleming2011counting}. Higher values of $\rho$ result in a weight function that gives higher weights to events that occur later in time relative to earlier events. On the other side, higher values of $\gamma$ emphasize events that occur earlier in time. When both values are high, events that occur in the middle are given more weight than early and late events. The challenge in utilizing weighted log-rank tests often lies in selecting meaningful weights.

\subsection{mdir}\label{subsec:mdir}
One way to simplify the weight selection procedure is combining multiple weighted log-rank tests. 
Such methods allow using several weighted log-rank tests in one testing approach \cite{Brendel:2014, DitzhausFriedrich2018, linAlternativeAnalysisMethods2020}. One way of combining multiple weighted tests is in a Wald-type test statistic \cite{Brendel:2014, DitzhausFriedrich2018}. In the following, we define the multi-directional log-rank test (\textit{mdir}) in terms of the multivariate process
	\begin{equation*}
	\mathbf{T}_{\hat{\mathcal{Q}}} (t)\coloneqq (T_{\hat{Q}_1}(t),\dots,T_{\hat{Q}_m}(t))
	\end{equation*}
	where
	\begin{equation*}
	T_{\hat{Q}_\ell}(t)\coloneqq n^{-\frac{1}{2}}\sum_{i=1}^n \int_{[0,t]} \hat{Q}_\ell(t,s) \left(Z_i -   \frac{Y^{Z=1}(t,s)}{Y(t,s)} \right) dN_i(t,s),
	\end{equation*}
for a set of weights $\hat{\mathcal{Q}}\coloneqq \{\hat{Q}_1,\dots,\hat{Q}_m\}$. This process is asymptotically equivalent to a multivariate martingale and has asymptotically independent and jointly normally distributed increments (expanding on Tsiatis, 1982\cite{Tsiatis:1982}). These properties are essential for the adaptive approach introduced in Subsection \ref{sec:admdir}.\\ 
%The test statistic can then be derived for $m$ weighted log-rank tests as the quadratic form of the test statistics $((T_{\hat{Q}_1}(t),\dots,T_{\hat{Q}_m}(t)))$ and the empirical covariance matrix:  
%\[W_n = (T_{\hat{Q}_1}(t),\dots,T_{\hat{Q}_m}(t))~\hat{\boldsymbol{\Sigma}}(t)^-~ ((T_{\hat{Q}_1}(t),\dots,T_{\hat{Q}_m}(t)))^T.\]
%This quadratic form follows a $\chi_m^2$-distribution under the null. \\
In adaptive designs, we are only interested in one-sided testing. This allows us to avoid situations where we obtain a final significant result but with effects pointing in different directions in subsequent stages. Thus, in the following, we focus on the one-sided test statistic of the \textit{mdir} test.\\
Ditzhaus and Pauly (2019) \cite{Ditzhaus:2019} derived the one-sided test statistic from the initial set of weights. The main idea is to restrict the space in which the test statistic is spanned to only positive values. This results in the test statistic
	\begin{align*}
		W(t) \coloneqq \max \{&0, \mathbf{T}_{\hat{\mathcal{L}}}(t)^T \hat{\boldsymbol{\Sigma}}_{\hat{\mathcal{L}}}^{-}(t) \mathbf{T}_{\hat{\mathcal{L}}}(t)%\colon\\
		\colon \emptyset \neq \hat{\mathcal{L}} \subseteq \hat{\mathcal{Q}}; \hat{\boldsymbol{\Sigma}}_{\hat{\mathcal{L}}}^{-}(t) \mathbf{T}_{\hat{\mathcal{L}}}(t) \geq 0\},
	\end{align*}
 with $\mathbf{T}_{\hat{\mathcal{L}}}(t) = (T_{\hat{Q}}(t) )_{\hat{Q} \in \hat{\mathcal{L}}}$ for all $t\geq 0$. The covariance matrix can be consistently estimated by $\hat{\boldsymbol{\Sigma}}(t)$. Its entries are given by
 \begin{equation*}
     (\hat{\boldsymbol{\Sigma}}(t))_{\ell \ell^{\star}} = n^{-1} \sum_{i=1}^n \int_{[0,t]} \hat{Q}_\ell(t,X_i(t)) \hat{Q}_{\ell^{\star}}(t,X_i(t)) \frac{Y^{Z=1}(t,s)}{Y(t,s)}\left(1 -   \frac{Y^{Z=1}(t,s)}{Y(t,s)} \right)  dN_i(t,s).
 \end{equation*}
 Furthermore, $\hat{\boldsymbol{\Sigma}}(t)^-$ denotes the Moore-Penrose inverse of $\hat{\boldsymbol{\Sigma}}(t)$.\\
 The authors state that an asymptotic limit distribution was not derived and propose a wild bootstrap approach with Rademacher weights instead. For more details, see Ditzhaus and Pauly (2019) \cite{Ditzhaus:2019}. The test is implemented in the R package \texttt{mdir.logrank}. \cite{mdir.logrank}

\subsection{Royston-Parmar splines}\label{subsec:rp_splines}
%Nicht-parametrische vs parametrische Schätzung
As described at the beginning of this section, the pooled and group-specific survival functions $S, S_k$ can be estimated non-parametrically by Kaplan-Meier estimators. Such a non-parametric estimation can potentially be inefficient compared to a parametric estimation approach if the distribution of $T$ resp. $T|Z=k$ lies within the parametric family assumed for the estimation process. Additionally, a parametric approach allows extrapolation of the survival curve beyond the time horizon that is present in the data. This is because the estimated parameters directly specify a distribution on $[0,\infty)$.\\%Two prominent parametric families in the context of survival data are exponential and Weibull distributions. The Weibull distribution depends on a two-dimensional parameter $(\lambda,\theta) \in (0,\infty)^2$ that determines the cumulative hazard function by
%\begin{equation*}
%    A(t)=\lambda t^{\theta}.
%\end{equation*}
%The family of exponential distributions can be regarded as a subfamily of Weibull distributions with $\theta=1$.\\
Despite the potential efficiency, such parametric approaches can be criticized for being too restrictive in terms of the shape of the distribution. Hence, important characteristics of the survival mechanism could potentially not be captured \cite{Latimer:2022}. Consequently, it is difficult to quantify differences among treatment groups. For example, estimation within the family of exponential distributions in several treatment groups directly leads to the assumption of proportional hazards.\\
As a more flexible alternative, Royston \& Parmar introduced an estimation procedure based on natural cubic splines \cite{Royston:2002}. A transformation of the survival function $S$, given by a link function $g:[0,1] \to \mathbb{R}$, is modeled by
\begin{equation*}
    g(S(t;\mathbf{z})) = g(S_{\text{baseline}}(t)) + \boldsymbol{\beta}^T \mathbf{z} = s(x;\phi) + \boldsymbol{\beta}^T \mathbf{z}
\end{equation*}
where $S(t;\mathbf{z})$ denotes the survival distribution given covariates $\mathbf{z}$ and $x=\log(t)$. The function $s\colon \mathbb{R} \times \mathbb{R}^{p+2} \to \mathbb{R}$ is a natural cubic spline which is parameterized by $\phi \in \mathbb{R}^{p+2}$. The number of internal knots $p$ determines the number of polynomials employed in the cubic spline.\\ %For the boundary knots $-\infty < \kappa_{\text{min}} < \kappa_{\text{max}} < \infty$ and $d$ internal knots $\kappa_{\text{min}} < \kappa_1 < \dots < \kappa_d < \kappa_{\text{max}}$ the natural cubic spline function is given by 
%\begin{equation*}
%    s(x;\phi)= \phi_0 + \phi_1x + \sum_{i=1}^d \phi_{i+1} v_i(x)
%\end{equation*}
%where $v_i$ is the $i$-th basis function that is defined by
%\begin{equation*}
%    v_i(x)=(x-\kappa_i)_+^3 - \lambda_i(x-\kappa_{\text{min}})_+^3 - (1-\lambda_i)(x-\kappa_{\text{max}})_+^3
%\end{equation*}
%with $\lambda_i \coloneqq (\kappa_{\text{max}} - \kappa_i)/(\kappa_{\text{max}} - \kappa_{\text{min}})$ and $u_+\coloneqq \max(0,u)$ for any $u \in \mathbb{R}$. 
When fitting such a model, it is suggested to place the boundary knots at the smallest and the largest uncensored logarithmized survival time and the internal knots evenly at the centiles of the uncensored logarithmized survival times \cite{Royston:2002}. For example, for $p=3$, one would place the internal knots at the two quartiles and the median of the uncensored logarithmized survival times.\\
For the scale on which the function should be modeled, three prominent suggestions have been proposed in the literature \cite{Royston:2002, Royston:2011} that are also implemented in software \cite{Jackson:2016}. They are shown in Table \ref{table:rp_scales}.\\
\begin{table}[h]
    \centering
    \begin{tabular}{l|c}
         scale & link function $g(u)$ \\
         \hline
         hazard& $\log(-\log(u))$ \\
         odds& $\log(1/u - 1)$\\
         normal& $-\Phi^{-1}(u)$
    \end{tabular}
    \caption{Popular link functions for Royston-Parmar splines; naming according to the function \texttt{flexsurvspline} in the R package \texttt{flexsurv} \cite{Jackson:2016}}
    \label{table:rp_scales}
\end{table}
The choice of $p$ and the scale can be guided by criteria such as the AIC (Akaike information criterion) or BIC (Bayesian information criterion). However, it is warned against using these criteria mechanically, and an informal choice based on the appearance of the fitted survival functions is also suggested\cite{Royston:2002, Royston:2011}. Simulation studies have shown that the correct choice of knots is not mandatory because the splines are flexible enough if a sufficient number of knots is used\cite{Rutherford:2015}.\\
Like simple parametric models, Royston-Parmar splines admit an extrapolation of the survival curve beyond the available time horizon. It should be mentioned here that the transformed survival function beyond the upper boundary knot is linear. 

\subsection{Adaptive designs}\label{subsec:ad}
We briefly outline some cornerstones for constructing group-selective adaptive trial designs with unblinded interim analyses. For the sake of simplicity, we restrict ourselves to two-stage designs with one interim and one final analysis. In these two stages, we define a test statistic and a corresponding $p$-value to test a null hypothesis $H_0$. These $p$-values will be denoted by $p_1$ and $p_2$, respectively. At best, these are independent and uniformly distributed on $[0,1]$. However, it is possible to relax this assumption to what is known as the \textit{p-clud} property \cite{Brannath:2012}. This property is fulfilled if
\begin{equation*}
    \mathbb{P}_{H_0}[p_1\leq u]\leq u \quad \text{and} \quad \mathbb{P}_{H_0}[p_2\leq u|p_1=v]\leq u \qquad \forall 0\leq u,v \leq 1.
\end{equation*}
If this is warranted, an adaptive design can be defined by a combination function $C\colon[0,1]^2\to[0,1]$ that is non-decreasing in $p_1$ and $p_2$ and continuous in $p_2$, bounds $\alpha_0$ and $\alpha_1$ for stopping of the trial in the interim analysis and a critical value $c$ for the combined $p$-value in the final analysis. In such a design, the trial will stop with the rejection of $H_0$ in the interim analysis if $p_1 \leq \alpha_1$, and it stops for futility, i.e., with acceptance of $H_0$ if $p_1 \geq \alpha_0$. The null hypothesis can be rejected at the final analysis if $C(p_1, p_2)\leq c$ and otherwise it stops without rejection of $H_0$. It adheres to the nominal type I error level $\alpha$ if
\begin{equation}\label{eq:ad_t1er}
    \alpha = \alpha_1 + \int_{\alpha_1}^{\alpha_0} \int_0^1 \mathbbm{1}_{\{C(p_1, p_2) \leq c\}} dp_1\,dp_2.
\end{equation}
Popular choices for $C$ are the combination function arising from Fisher's product test $C(p_1, p_2)=p_1 p_2$ \cite{Fisher:1970, Bauer:1989} and the inverse normal combination function \cite{Lehmacher:1999}
\begin{equation}\label{eq:inverse_normal}
    C(p_1, p_2)=1 - \Phi\left(w_1 \cdot \Phi^{-1}(1-p_1) + w_2 \cdot \Phi^{-1}(1-p_2) \right)
\end{equation}
where $\Phi$ and $\Phi^{-1}$ denote the cumulative distribution function and the quantile function of the standard normal distribution, respectively, and the weights $w_1, w_2 \geq 0$ underly the constraint $w_1^2 + w_2^2 = 1$.\\
The choice of the sequential decision bounds will determine the values of $\alpha_1$ and $c$. One can use standard bounds according to the sequential plans of Pocock \cite{Pocock:1977} or O'Brien and Fleming \cite{OBrien:1979} or some $\alpha$-spending approach \cite{Lan:1983}.\\
All the methods mentioned here are implemented in the comprehensive R package \texttt{rpact} \cite{rpact}. More details on group-sequential adaptive designs going far beyond what we sketched here can be found in Wassmer and Brannath (2016) \cite{Wassmer:2016}.\\
The feature of adaptive designs that needs to be emphasized again for our purposes is the capability to redesign the second stage using the data collected on the primary endpoint up to the interim analysis.

\section{Adaptive testing procedure}\label{sec:admdir}
In this section, we present our adaptive testing procedure. While focusing on technical aspects here, it will also be presented in an example based on real data in Section \ref{sec:example}.\\
To put it briefly, we conduct a multi-stage adaptive design that addresses the uncertainty about the type of effect by application of a combination testing procedure in the first stage and uses the information from this early stage to choose a well-suited weighted log-rank test for later stages. For the sake of simplicity, we will restrict ourselves to considering two-stage designs. However, an extension to more than two stages follows straightforward.\\
At the beginning of the trial, we fix two sets of weight functions $\hat{\mathcal{Q}}_{\text{mdir}}\coloneqq \{\hat{Q}_{\text{mdir}, 1},\dots,\hat{Q}_{\text{mdir}, m_1}\}$ and $\hat{\mathcal{Q}}_{\text{cand}}\coloneqq \{\hat{Q}_{\text{cand}, 1},\dots,\hat{Q}_{\text{cand}, m_2}\}$ and its union $\hat{\mathcal{Q}}_{\text{all}} \coloneqq \hat{\mathcal{Q}}_{\text{mdir}} \cup \hat{\mathcal{Q}}_{\text{cand}}$. The only things that need to be ensured are that all weights meet standard conditions required for repeated significance testing with log-rank type tests \cite{Harrington:1982} (see also Supplementary Material, Section \ref{supp-sec:technical}) and linear independence of the weights in $\hat{\mathcal{Q}}_{\text{mdir}}$ \cite{Ditzhaus:2019}.\\
Additionally, we set an interim analysis date $t_1$. Based on the results shown in Section \ref{supp-sec:technical} of the Supplementary Material, the calendar time process
\begin{equation*}
    (\mathbf{T}_{\hat{\mathcal{Q}}_{\text{all}}}(t))_{t \geq 0},\;\text{with}\; \mathbf{T}_{\hat{\mathcal{Q}}_{\text{all}}}(t)=(T_{\hat{Q}}(t))_{\hat{Q} \in \hat{\mathcal{Q}}_{\text{all}}} \, \forall t\geq 0.
\end{equation*}
is asymptotically equivalent to a multi-dimensional Gaussian process. In particular, its asymptotically independent increments asymptotically follow a joint normal distribution.\\
Additionally, we fix decision bounds for the first stage $\alpha_0$ (futility) and $\alpha_1$ (efficacy), a combination function $C$ and a decision bound $c$ for the combined $p$-value. These quantities are chosen such that \eqref{eq:ad_t1er} is maintained for the prefixed type 1 error rate $\alpha$. The analyses shall take place at calendar times $t_1$ (interim analysis) and $t_2$ (final analysis).\\
At the time of the interim analysis, the weighted testing procedure for the next stage is determined. Therefore, let $D$ be a random variable that assumes values in $\{1,\dots,m_2\}$ and is measurable w.r.t. the information about the primary endpoint collected up to the interim analysis.\\
In the first stage, we obtain the test statistic and $p$-value 
\begin{align*}
    S_1 &\coloneqq \max \{0, \mathbf{T}_{\hat{\mathcal{L}}}(t_1)^T \hat{\boldsymbol{\Sigma}}_{\hat{\mathcal{L}}}^{-}(t_1) \mathbf{T}_{\hat{\mathcal{L}}}(t_1)\colon\emptyset \neq \hat{\mathcal{L}} \subseteq \hat{\mathcal{Q}}_{\text{mdir}}; \hat{\boldsymbol{\Sigma}}_{\hat{\mathcal{L}}}^{-}(t_1) \mathbf{T}_{\hat{\mathcal{L}}}(t_1) \geq 0\}\quad\text{ resp. }\\
    p_1 & \coloneqq \inf\{p \in [0,1]\,\colon\, q_{1-p}^G \geq S_1\}
\end{align*}
where $q_{1-p}^G$ is the quantile of the distribution resulting from the wild bootstrap procedure. Together with the second stage test statistic and $p$-value
\begin{align}
    S_{2,\hat{Q}_{\text{cand}, D}} & \coloneqq (T_{\hat{Q}_{\text{cand}, D}}(t_2) - T_{\hat{Q}_{\text{cand}, D}}(t_1))/\sqrt{\hat{\Sigma}_{\hat{Q}_{\text{cand}, D}}(t_2) - \hat{\Sigma}_{\hat{Q}_{\text{cand}, D}}(t_1)}\label{eq:std_increment}\quad\text{ resp. }\\
    p_2 & \coloneqq 1-\Phi(S_2)\nonumber,
\end{align}
we obtain an adaptive testing approach that asymptotically keeps the type I error level $\alpha$ when applied as described in \ref{subsec:ad}. I.e., we stop the trial for futility at the interim analysis if $p_1 > \alpha_0$, we reject $H_0$ at the interim analysis if $p_1 \leq \alpha_1$ and we can reject the trial at the final analysis if neither of those is the case and $C(p_1, p_2)\leq c$.\\
Although the design would also permit sample size recalculation (e.g., by extending the recruitment period) and concomitantly postpone the final analysis date, we only consider adapting the weight in the testing procedure. 

\subsection{Determination of the second stage test statistic}\label{subsec:second_stage}
The critical question here concerns the choice of test statistics in the second stage. As mentioned above, we can use all the information about the primary endpoint collected up to the interim analysis. Of course, this consideration is only necessary if we proceed to a second stage, i.e., if $\alpha_1 < p_1 \leq \alpha_0$. Then, we suggest the following procedure:\\
Based on the $p$-value $p_1$ of the first stage, we can compute the conditional error probability
\begin{equation*}
    \tilde{\alpha}_2\coloneqq\mathbb{P}_{H_0}[C(p_1,p_2) \leq c|p_1] = \int_0^1 \mathbbm{1}_{\{C(p_1,u) \leq c\}} du = \sup\{u \in [0,1]\colon C(p_1,u) \leq c\}.
\end{equation*}
If $p_2 < \tilde{\alpha}_2$, we will reject \eqref{eq:null_hyp} in favour of \eqref{eq:alt_hyp} at the final analysis. Our aim is, therefore, to estimate the probability of this event. The associated considerations, which we now present, are based on the calculations shown in Yung and Liu (2020)\cite{Yung:2020}.\\
We fit a Royston-Parmar spline model for each group as presented in Section \ref{subsec:rp_splines}. Please note that we do not fit one joint model for the two groups. This would mean that the difference between the groups would follow a fixed pattern, e.g., a pattern of proportional hazards if the hazard scale was chosen. The hyperparameters of the spline model (number of knots, scale) are chosen based on information criteria or based on visual inspection (see Section \ref{sec:example} for exemplary applications). To highlight quantities based on estimates or assumptions made at the interim analysis, we will add a tilde to all of them. Hence, we denote the resulting pooled and group-specific survival, density and hazard functions by $\tilde{S}$ and $\tilde{S}_k$, $\tilde{f}$ and $\tilde{f}_k$, and $\tilde{\lambda}$ and $\tilde{\lambda}_k$, respectively. We assume we can identify the large sample limit $Q$ of $\hat{Q}$ under knowledge of the true distribution of $T$ and $T|Z=k$. This is, e.g., the case for the Fleming-Harrington weights presented in \eqref{eq:fh_weights}. We do not know the true distribution, so we insert $\tilde{S}_k$ instead and denote the resulting weight functions by $\tilde{Q}$. \\
The same could be done to assess the distribution of the recruitment date $R$ and random dropout $C^{\star}$. We denote our planning assumptions about the distribution functions of these two random variables at the time of the interim analysis by $\tilde{F}_R$ and $\tilde{F}_{C^{\star}}$, respectively. Based on those assumptions, the probability that some individual is allocated to treatment group $k$ and has spent at least $s$ time units at risk (i.e., without any censoring and without experiencing the event of interest) at calendar time $t$ is given by
\begin{equation*}
    \tilde{\pi}_k(t,s)\coloneqq \mathbb{P}[Z=k]\tilde{F}_R((t-s)_+)(1 - \tilde{F}_{C^{\star}}(s))\tilde{S}_k(s).
\end{equation*}
Now, we estimate the drift of the weighted log-rank test \eqref{eq:wlrt} by
\begin{equation*}
    \tilde\xi_{\tilde{Q}}(t)\coloneqq \int_0^t \tilde{Q}(s) \cdot \frac{\tilde{\pi}_0(t,s) \tilde{\pi}_1(t,s)}{\tilde{\pi}_0(t,s) + \tilde{\pi}_1(t,s)} \cdot (\tilde{\lambda}_0(s) - \tilde{\lambda}_1(s)) \,ds.
\end{equation*}
Its asymptotic variance is estimated by
\begin{equation*}
    \tilde{\sigma}_{\tilde{Q}}^2(t)\coloneqq \int_0^t \tilde{Q}(s)^2 \cdot \left(\frac{\tilde{\pi}_0(t,s) \tilde{\pi}_1(t,s)}{\tilde{\pi}_0(t,s) + \tilde{\pi}_1(t,s)}\right)^2 \cdot \tilde{F}_R((t-s)_+)(1 - \tilde{F}_{C^{\star}}(s)) \cdot ((1-r)\tilde{f}_0(s) + r\tilde{f}_1(s))\, ds.
\end{equation*}
where $r$ denotes the proportion of patients allocated to treatment group $k=1$. Based on the property of asymptotically independent and normally distributed increments of the process $(T_{\tilde{Q}}(t))_{t\geq 0}$, the standardized increment \eqref{eq:std_increment} would then follow the distribution
\begin{equation*}
    \tilde{S}_{2,\tilde{Q}} \sim \mathcal{N}\left(\frac{\tilde{\xi}_{\tilde{Q}}(t_2) - \tilde{\xi}_{\tilde{Q}}(t_1)}{\sqrt{\tilde{\sigma}_{\tilde{Q}}^2(t_2) - \tilde{\sigma}_{\tilde{Q}}^2(t_1)}},1\right)
\end{equation*}
To maximize the power of our procedure, one would choose the weight for which the largest conditional power is assumed based on our planning assumptions, i.e.
\begin{equation*}
    D\coloneqq \argmax_{d \in \{1,\dots,m_2\}} 1 - \Phi\left(\Phi^{-1}(1 - \tilde{\alpha}_2) - \frac{\tilde{\xi}_{\tilde{Q}_{\text{cand}, d}}(t_2) - \tilde{\xi}_{\tilde{Q}_{\text{cand}, d}}(t_1)}{\sqrt{\tilde{\sigma}_{\tilde{Q}_{\text{cand}, d}}^2(t_2) - \tilde{\sigma}_{\tilde{Q}_{\text{cand}, d}}^2(t_1)}} \right).
\end{equation*}
This selection process is demonstrated in Section \ref{sec:example}.
%\subsection{Exhausting the implicit binding futility}

%Muss das sein? Ggf. streichen!

%Small sample size -> small number of knots, prevent overfitting
%Seperate fit for both groups -> No assumption abouttype of deviation (standard model with treatment as covariate leads to proportional hazards or odds when using the corresponding link function)

\section{Real data example}\label{sec:example}
We illustrate the proposed procedure based on reconstructed data from the FAKTION trial \cite{Jones:2020} (NCT01992952). In this multicentre, randomized, placebo-controlled, phase 2 trial, the addition of capivasertib to fulvestrant was investigated in patients with advanced breast cancer. It was found that the experimental therapy extended progression-free survival, which was the primary endpoint. However, no effect could be proven for overall survival, possibly the primary endpoint in a consecutive phase III trial. Therefore, we reanalyze overall survival data using the method presented above. For further details, we refer to the published results of this trial \cite{Jones:2020}.\\
We obtained the individual patient data using the algorithm by Guyot et al. (2012)\cite{guyotEnhancedSecondaryAnalysis2012}. The obtained data includes the observed survival time and a group and censoring indicator. However, the recruitment dates could not be reconstructed. Hence, we processed the data to receive the required data structure, including enrollment. The observed censoring suggests that recruitment took place evenly over the entire duration of the study (see Supplementary Material, Section \ref{supp-sec:example}). Hence, we assume that new patients were recruited over the entire duration of the trial.\\ 
For patients with censored event time $C_i$, the recruitment date was set to $t_2 - C_i$ where $t_2=35.98$ is the calendar date of the final analysis. For patients with uncensored event time $T_i$ a recruitment date that is uniformly distributed on the interval $[0, t_2 - T_i]$ was simulated.
%Neues Design mit Zwischenanalyse ein Jahr vor Ende, inverse normal combination mit gleichen Gewichten, O'Brien-Fleming-Ablehnregion
We considered an adaptive design with one interim analysis after 24 months, i.e., approximately one year before the final analysis. Stage-wise $p$-values are combined with the inverse normal combination function from \eqref{eq:inverse_normal} with equal weights $w_1 = w_2 = 1/\sqrt{2}$. Decision bounds were calculated according to the design of O'Brien and Fleming \cite{OBrien:1979} without any futility bound. Accordingly, the null hypothesis \eqref{eq:null_hyp} will be rejected in favor of the alternative \eqref{eq:alt_hyp} if $p_1\leq 0.002583$ or $C(p_1, p_2) \leq 0.023996$.\\
When applying the adaptive testing procedure proposed in Section \ref{sec:admdir}, we consider the set of candidate weights $\hat{\mathcal{Q}}_{\text{cand}}=\{w^{(0,0)} \circ \hat{F}, w^{(1,0)} \circ \hat{F}, w^{(0,1)} \circ \hat{F}, w^{(1,1)} \circ \hat{F}\}$ from the Fleming-Harrington family from \eqref{eq:fh_weights}. Here, $\circ$ denotes function composition. For the first stage, we consider different choices of $\hat{\mathcal{Q}}_{\text{mdir}}$. To this end, we look at all elements of the power set of $\hat{\mathcal{Q}}_{\text{cand}}$ that contain the standard log-rank weight $w^{(0,0)} \circ \hat{F}$. Resulting $p$-values can be found in the second column of Table \ref{table:example_p}.\\
For the interim analysis data, we fit Royston-Parmar splines to the interim data as presented in Section \ref{subsec:rp_splines}. This is done for the number of internal knots $p \in \{0,1,2\}$ and all three scales presented in Table \ref{table:rp_scales}. The final model used for conditional power calculation is chosen based on the AIC. The model with the lowest AIC has 0 internal knots and is computed on the normal scale. It is displayed in Figure \ref{fig:rp_interim}. Fitted curves for other parameter configurations can be found in Supplementary Figure \ref{supp-fig:interim_fits}. The extrapolation performance can be judged based on Supplementary Figure \ref{supp-fig:interim_fits_ep}. AIC values for all models are displayed in Supplementary Table \ref{supp-table:rp_aic}. Of course, we are convinced that a medical expert should also be consulted at this stage to assess the plausibility of the extrapolated curves. In this particular case, the extrapolation of the model selected on the basis of the AIC proves to be plausible.\\
For this model, conditional power calculations as presented in Section \ref{subsec:second_stage} were executed. The conditional power computations favor the $(1,1)$-weighted Fleming-Harrington test statistic for the second stage. However, the conditional power values differ greatly between the models. For the model that is chosen based on the AIC (see Figure \ref{fig:rp_interim}), the conditional power ranges between 73.45\% and 83.34\%, depending on the chosen test statistic for the first stage. The results of the conditional power calculations for all modeling parameter choices and first-stage test statistics choices can be found in \ref{supp-table:cond_power}.\\
\begin{figure}[h]
\centering
\includegraphics[width=0.7\textwidth]{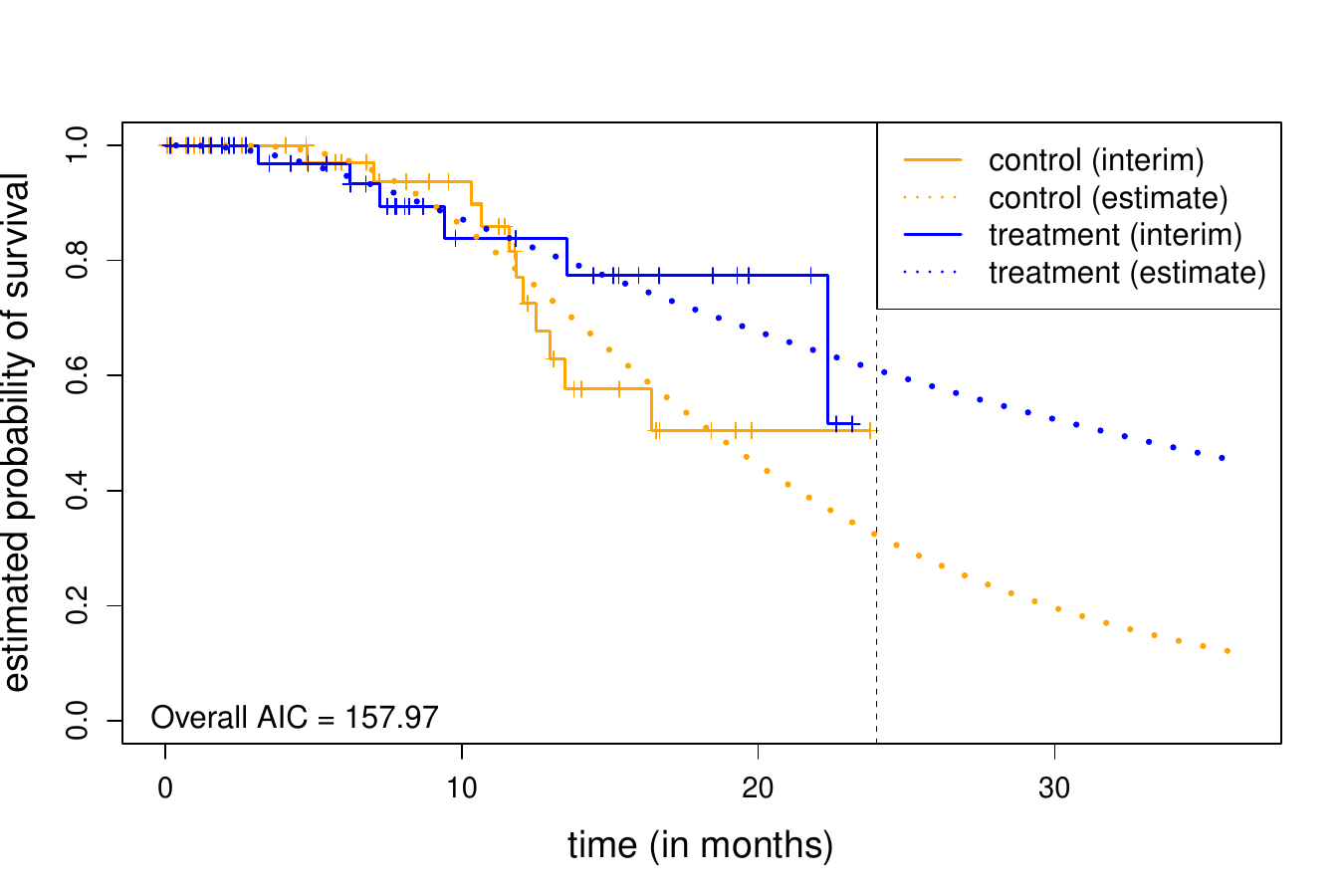}
\caption{Interim data with fitted Royston-Parmar spline models with 0 interior knots on the normal scale. This model has the lowest AIC among all models considered. Vertical line at 24 months indicates the calendar date of the interim analysis.}
\label{fig:rp_interim}
\end{figure}
\begin{table}[h]
    \centering
    \begin{tabular}{l|c|c|c|c|c}
         \multirow{2}{*}{$\hat{\mathcal{Q}}_{\text{mdir}}$} & \multirow{2}{*}{$p_1$} & \multicolumn{4}{c}{$C(p_1,p_2)$}\\
         \cline{3-6}
         &&$(0,0)$&$(0,1)$&$(1,0)$&$(1,1)$\\
         \hline
         $(0,0),(1,0),(0,1),(1,1)$& $0.133$ & $0.021$ & $0.034$ & $0.021$ & $0.009$ \\
         $(0,0),(1,0),(0,1)$& $0.168$ & $0.027$ & $0.043$ & $0.027$ & $0.012$\\
         $(0,0),(1,0),(1,1)$& $0.116$ & $0.018$ & $0.030$ & $0.018$ & $0.008$\\
         $(0,0),(0,1),(1,1)$& $0.125$ & $0.020$ & $0.032$ & $0.019$ & $0.009$\\
         $(0,0),(1,0)$& $0.152$ & $0.024$ & $0.039$ & $0.024$ & $0.011$\\
         $(0,0),(0,1)$& $0.188$ & $0.031$ & $0.048$ & $0.030$ & $0.014$\\
         $(0,0),(1,1)$& $0.110$ & $0.017$ & $0.028$ & $0.017$ & $0.008$\\
         $(0,0)$& $0.172$& $0.028$ & $0.044$ & $0.027$ & $0.013$
    \end{tabular}
    \caption{$p$-values from first stage data for different choices of $\hat{\mathcal{Q}}_{\text{mdir}}$ and combined $p$-value for different choices of the single weighted log-rank test for the second stage test}
    \label{table:example_p}
\end{table}
Based on the decision rule lined out above, the combined $p$-values in Table \ref{table:example_p} indicate that the null hypothesis can be rejected whenever the $(1,1)$-weighted Fleming-Harrington test is chosen as the test statistic for the second stage. Interestingly, this is the test statistic suggested by our approach, as lined out above. Whenever, the $(1,1)$-weighted Fleming-Harrington test is included in the combination test of the first stage, a rejection can also be achieved if the standard log-rank test or the $(1,0)$-weighted Fleming-Harrington test is chosen for the second stage. Additionally, rejection also occurs if the weights $(0,0)$ and $(1,0)$ are combined in the first stage and $(1,0)$ is chosen for the second stage. The trial would always end with the acceptance of the null hypothesis if the $(0,1)$-weighted Fleming-Harrington test is selected for the second stage.\\
In summary, we could show that an adaptation of the weight could have rejected the hypothesis of equal distributions of overall survival in the two treatment groups. In particular, applying our proposed procedure would have yielded such a result.\\
We emphasize that these results rely on simulated recruitment dates for uncensored patients. Nevertheless, this is necessary for demonstration purposes. In Section \ref{supp-subsec:sim_dependence} of the Supplementary Material, the influence of this simulation is assessed.

\section{Simulations}\label{sec:sim_study}

In this simulation study, we want to examine compliance with our adaptive selection procedure's nominal type I error rate. Furthermore, we compare power curves for various testing procedures to illustrate the merits of an adaptive selection procedure.\\
In both parts, the calendar time schedule of our fictional clinical trial will be the same. We consider a design with one interim analysis after $t_1=5$ and a final analysis after $t_2=8$ years. We will not consider any sample size adaptation with an accompanying shift of $t_2$ as we only want to focus on the advantages and disadvantages of the selection procedure. Nevertheless, this possibility exists and represents an undisputed advantage of our design. The participating individuals enter the trial uniformly up until $a=6$ years have passed. We only consider administrative censoring and no additional loss to follow-up.\\
The time-to-event variable in the control group is exponentially distributed with parameter $-\log(1 - 0.3)$, i.e., an annual event rate of $30\%$. For the flexible estimation and extrapolation of the survival curves, the same nine Royston-Parmar spline models as in Section \ref{sec:example} will be considered. For each run, the best among these models will be determined based on the AICs of the models. The test for the second stage is then chosen based on conditional power calculations based on the extrapolated curves from this particular model.\\
For the two-stage designs examined here, combinations of stagewise $p$-values and sequential decision bounds are chosen as in the previous example from Section \ref{sec:example}.

\subsection{Empirical type I error rates}

For the first stage, we consider $7$ different combination tests and $8$ differently weighted log-rank tests based on Fleming-Harrington weights for the second stage. In particular, we set $\hat{\mathcal{Q}}_{\text{cand}}=\{w^{(0,0)} \circ \hat{F}, w^{(1,0)} \circ \hat{F}, w^{(2,0)} \circ \hat{F}, w^{(3,0)} \circ \hat{F}, w^{(1,1)} \circ \hat{F}, w^{(0,1)} \circ \hat{F}, w^{(0,2)} \circ \hat{F}, w^{(0,3)} \circ \hat{F}\}$. The sample sizes for each group vary in the set $\{50, 100, 200, 500\}$. We only considered balanced group sizes. 
The results are based on 10,000 simulation runs. Hence, for a true underlying rate of $0.025$, the empirical rate lies within the interval $[0.0219, 0.0281]$ with a probability of $95\%$.\\
The empirical rejection rates for any pre-fixed combination of test statistics in the two stages (i.e., not determined by a selection procedure at the interim analysis but already predefined at the start of the trial) can be found in Supplementary Tables \ref{supp-table:t1e_all_n50} - \ref{supp-table:t1e_all_n500}.

\begin{table}[h]
    \centering
    \begin{tabular}{l|c|c|c|c}
         \multirow{2}{*}{$\hat{\mathcal{Q}}_{\text{mdir}}$} &  \multicolumn{4}{c}{$n$}\\
         \cline{2-5}
         &$100$&$200$&$400$&$1000$\\
         \hline
         $(0,0),(1,0),(0,1),(1,1)$& $ 0.0238 $&$ 0.0256 $&$ 0.0248 $&$ 0.0260 $\\
         $(0,0),(1,0),(0,1)$& $ 0.0241 $&$ 0.0269 $&$ 0.0237 $&$ 0.0260 $\\
         $(0,0),(1,0),(1,1)$& $ 0.0238 $&$ 0.0259 $&$ 0.0240 $&$ 0.0264 $\\
         $(0,0),(0,1),(1,1)$& $ 0.0242 $&$ 0.0277 $&$ 0.0244 $&$ 0.0259 $\\
         $(0,0),(1,0)$& $ 0.0233 $&$ 0.0259 $&$ 0.0235 $&$ 0.0279 $\\
         $(0,0),(0,1)$& $ 0.0251 $&$ 0.0273 $&$ 0.0230 $&$ 0.0269 $\\
         $(0,0),(1,1)$& $ 0.0258 $&$ 0.0267 $&$ 0.0231 $&$ 0.0265 $\\
         $(0,0)$& $ 0.0269 $&$ 0.0285 $&$ 0.0235 $&$ 0.0265 $
    \end{tabular}
    \caption{Empirical type I error rates}
    \label{table:type_1_error_rates}
\end{table}

All rates shown here lie in the confidence interval mentioned above. We can observe a slight inflation of the empirical type I error level for small sample sizes. We can assume that this is due to the lack of agreement between the actual distribution of the weighted log-rank test statistics and its asymptotical approximation by a normal distribution for small sample sizes. This is supported by the fact that we can see similar inflations for the fixed combinations displayed in the Supplementary Tables \ref{supp-table:t1e_all_n50} -  \ref{supp-table:t1e_all_n500}. However, this fact is well-known. For small sample sizes or high censoring percentages, this problem could easily be solved by applying a permutation version of the log-rank test\cite{Neuhaus:1993}. Slight inflations of the type I error levels in Table \ref{table:type_1_error_rates} do not exceed the reported rates. As those rates refer to group sequential designs without any adaptation, we can claim that the adaptive weight choice does not introduce any further complications regarding type I error levels.

\subsection{Power comparisons}\label{subsec:power}

\begin{figure}
	\centering
	\begin{subfigure}[t]{0.328\textwidth}
		{\Large \textsf{\textbf{(A)}}}
	\end{subfigure}
	\begin{subfigure}[t]{0.328\textwidth}
		{\Large \textsf{\textbf{(B)}}}
	\end{subfigure}
    \begin{subfigure}[t]{0.328\textwidth}
		{\Large \textsf{\textbf{(C)}}}
	\end{subfigure}
	\begin{subfigure}[tb]{0.328\textwidth}
		\includegraphics[width=\linewidth]{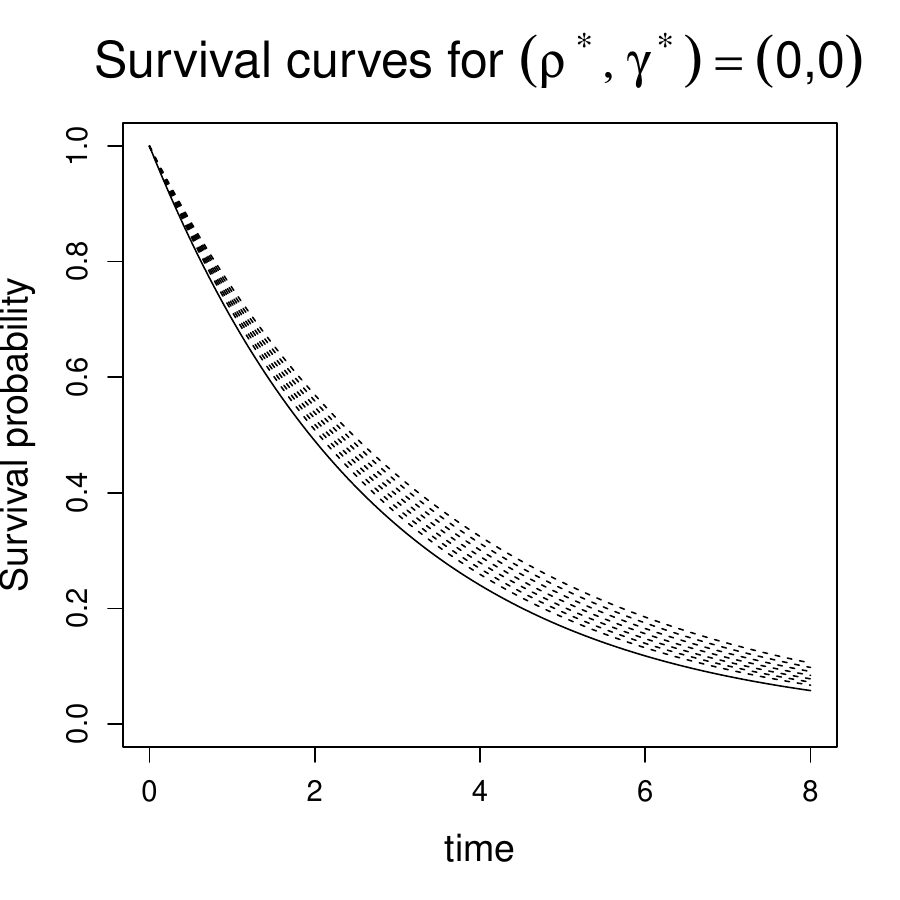}
	\end{subfigure}
	\begin{subfigure}[tb]{0.328\textwidth}
		\includegraphics[width=\linewidth]{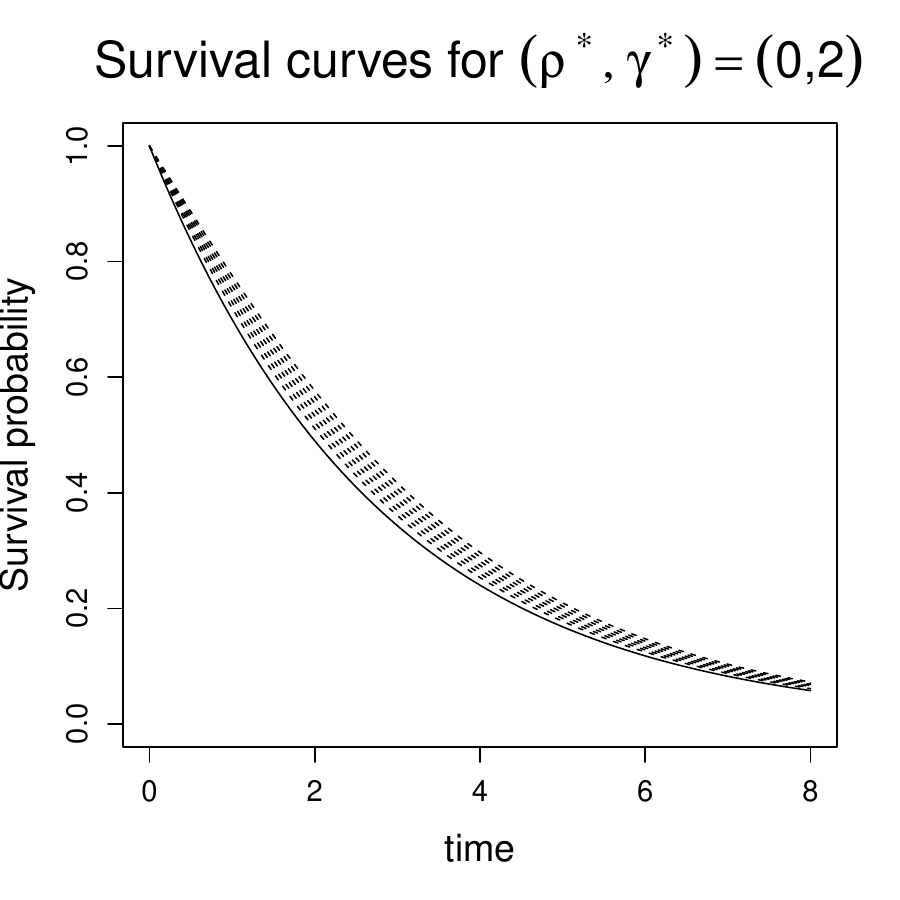}
	\end{subfigure}
    \begin{subfigure}[tb]{0.328\textwidth}
		\includegraphics[width=\linewidth]{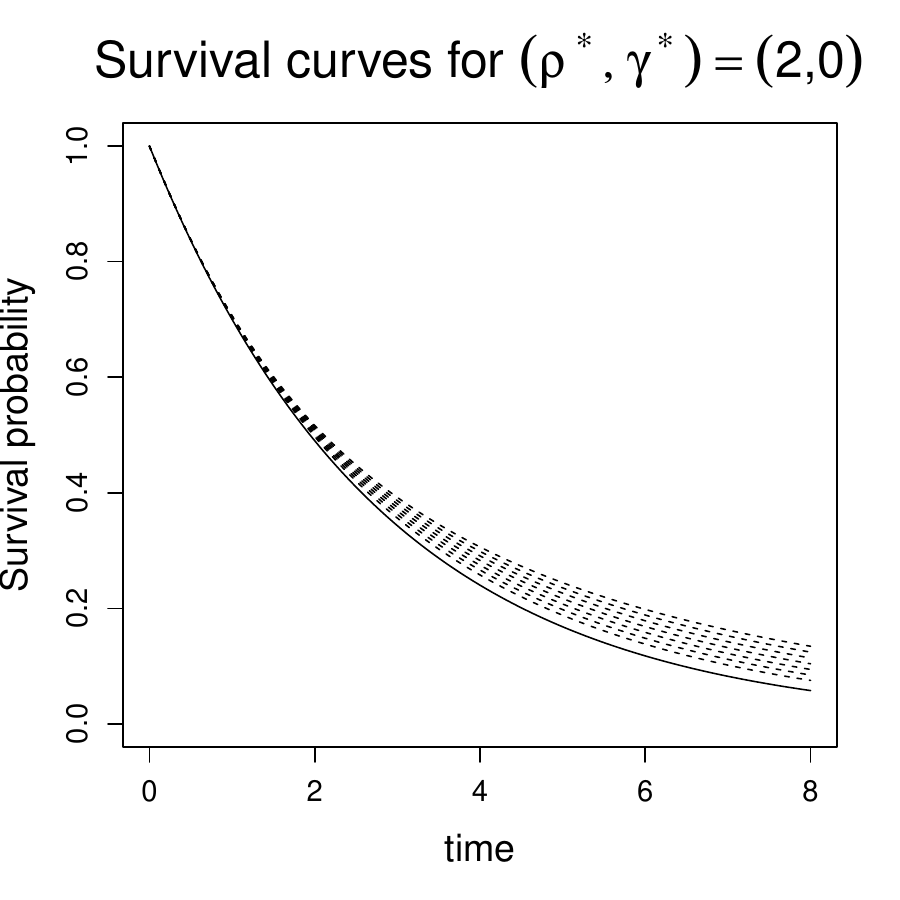}
	\end{subfigure}
	\caption{Survival curves for three types of deviation of the distribution in the experimental group from the distribution in the control group. The solid line gives the survival curve in the control group. Dashed lines are survival curves in the experimental group for the seven effect sizes $\{0.4\cdot\theta_0, 0.6\cdot\theta_0, 0.8\cdot\theta_0, \theta_0, 1.2\cdot\theta_0, 1.4\cdot\theta_0, 1.6\cdot\theta_0\}$. \textbf{(A)} Survival curves in the proportional hazards case $((\rho^{\star},\gamma^{\star}) = (0,0))$ \textbf{(B)} Survival curves in the early effect case $((\rho^{\star},\gamma^{\star}) = (0,2))$ \textbf{(C)} Survival curves in the late separation case $((\rho^{\star},\gamma^{\star}) = (2,0))$}
	\label{fig:survival curves}
\end{figure}

\begin{figure}
	\centering
	\begin{subfigure}[t]{0.5\textwidth}
		{\Large \textsf{\textbf{(A)}}}
	\end{subfigure}
	\begin{subfigure}[t]{0.49\textwidth}
		{\Large \textsf{\textbf{(B)}}}
	\end{subfigure}
	\begin{subfigure}[tb]{0.5\textwidth}
		\includegraphics[width=\linewidth]{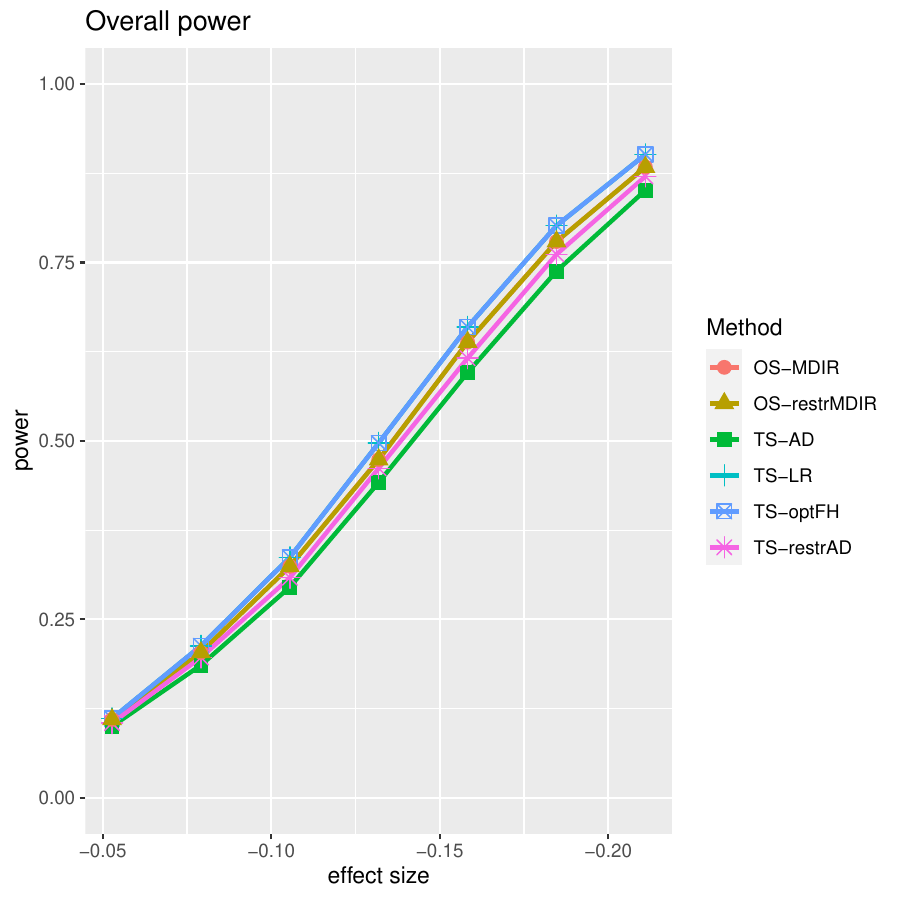}
	\end{subfigure}
	\begin{subfigure}[tb]{0.49\textwidth}
		\includegraphics[width=\linewidth]{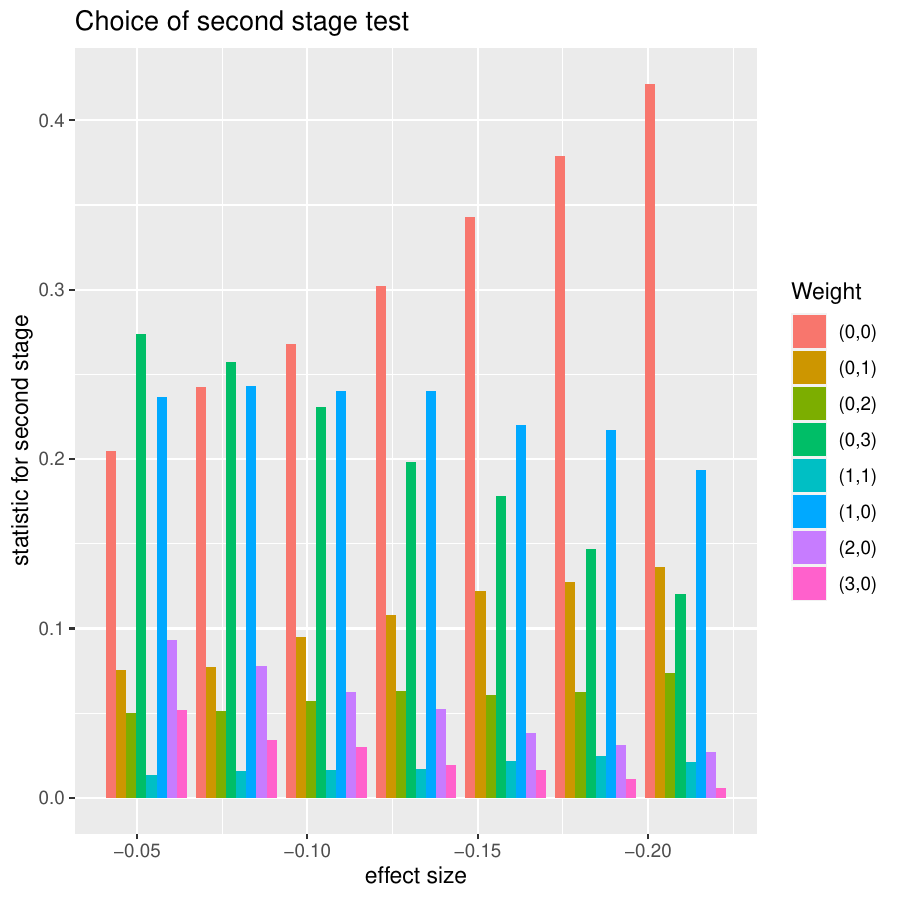}
	\end{subfigure}
	\caption{\textbf{(A)} Power curve for six testing procedures in case of proportional hazards $((\rho^{\star}, \gamma^{\star})=(0,0))$. Please note that the two procedures, TS-AD and TS-optFH, and the two procedures, OS-MDIR and OS-restrMDIR, coincide in this case. \textbf{(B)} Relative frequencies of the choice of single-weighted tests for the second stage of the TS-AD testing procedure.}
	\label{fig:power_ph}
\end{figure}

\begin{figure}
	\centering
	\begin{subfigure}[t]{0.5\textwidth}
		{\Large \textsf{\textbf{(A)}}}
	\end{subfigure}
	\begin{subfigure}[t]{0.49\textwidth}
		{\Large \textsf{\textbf{(B)}}}
	\end{subfigure}
	\begin{subfigure}[tb]{0.5\textwidth}
		\includegraphics[width=\linewidth]{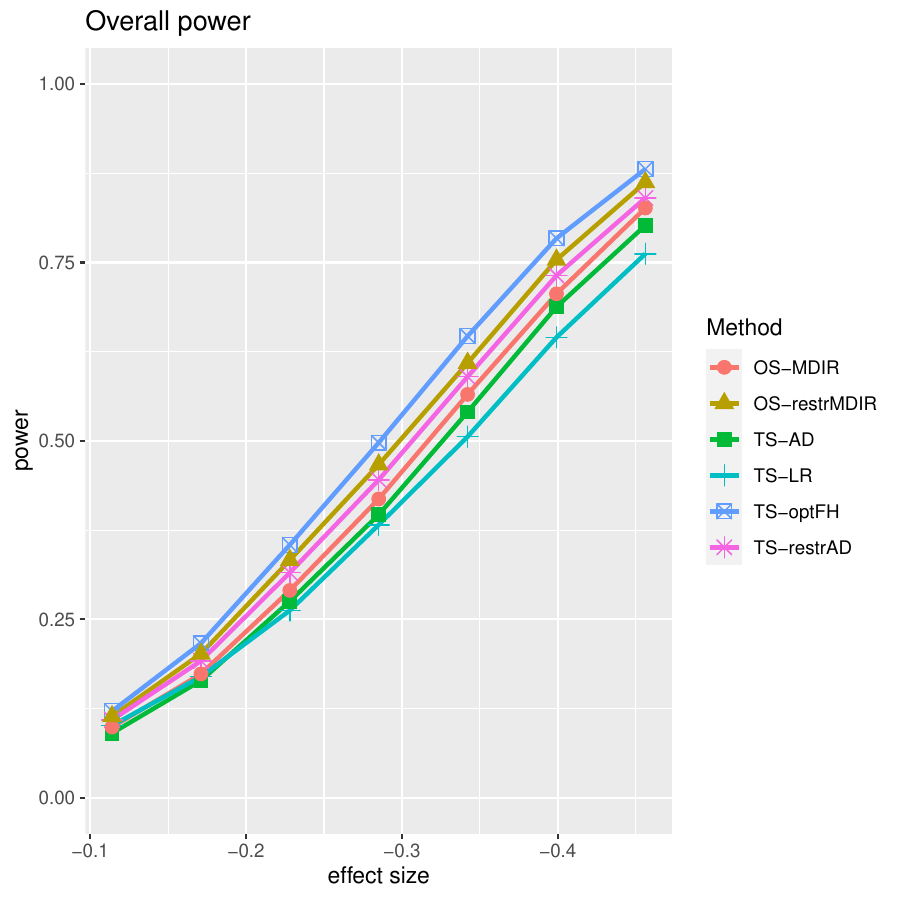}
	\end{subfigure}
	\begin{subfigure}[tb]{0.49\textwidth}
		\includegraphics[width=\linewidth]{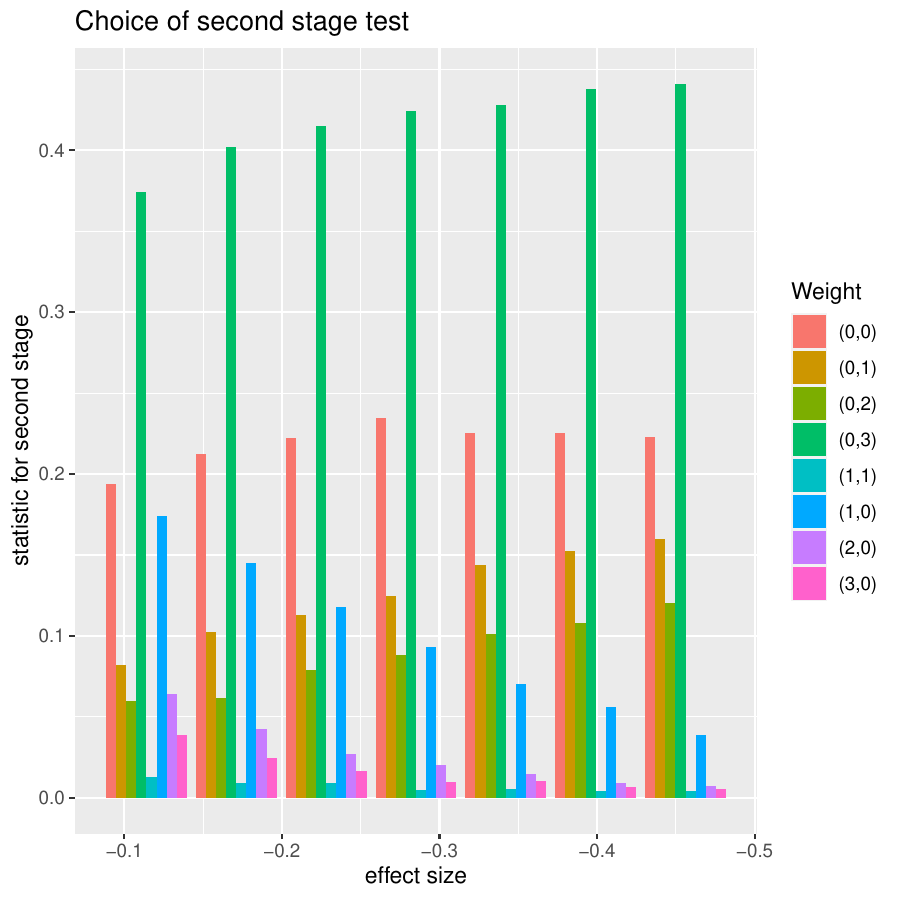}
	\end{subfigure}
	\caption{\textbf{(A)} Power curve for six testing procedures in case of an early effect $((\rho^{\star}, \gamma^{\star})=(0,2))$. \textbf{(B)} Relative frequencies of the choice of single-weighted tests for the second stage of the TS-AD testing procedure.}
	\label{fig:power_early}
\end{figure}

\begin{figure}
	\centering
	\begin{subfigure}[t]{0.5\textwidth}
		{\Large \textsf{\textbf{(A)}}}
	\end{subfigure}
	\begin{subfigure}[t]{0.49\textwidth}
		{\Large \textsf{\textbf{(B)}}}
	\end{subfigure}
	\begin{subfigure}[tb]{0.5\textwidth}
		\includegraphics[width=\linewidth]{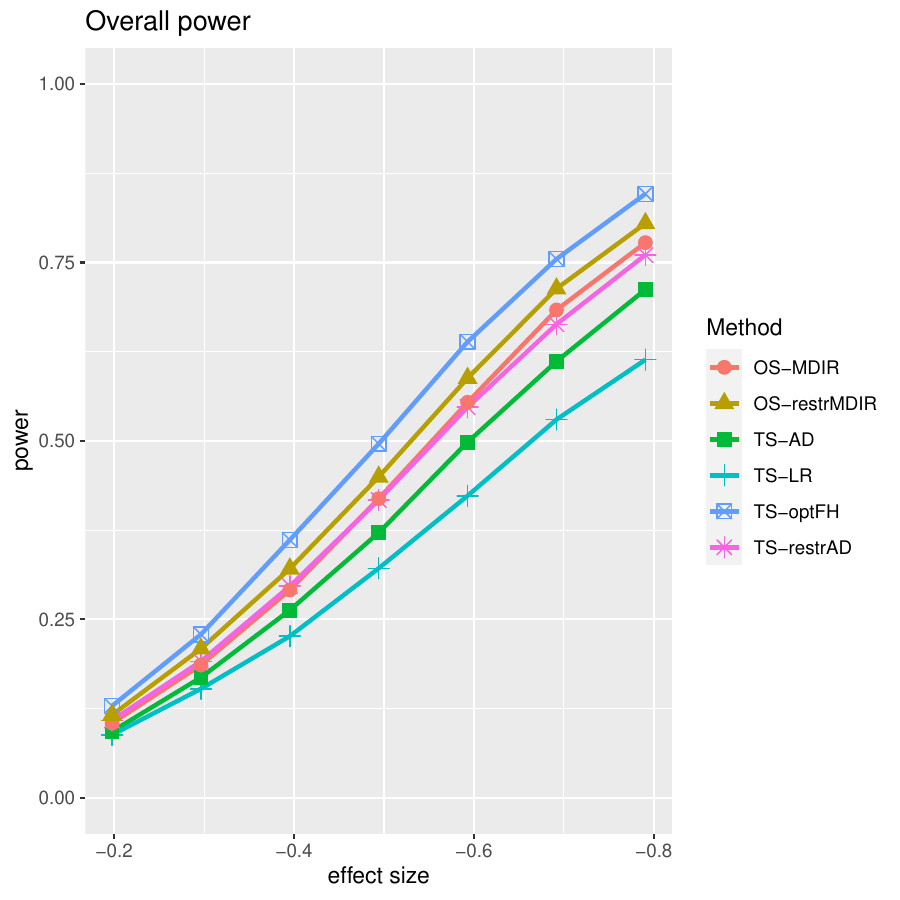}
	\end{subfigure}
	\begin{subfigure}[tb]{0.49\textwidth}
		\includegraphics[width=\linewidth]{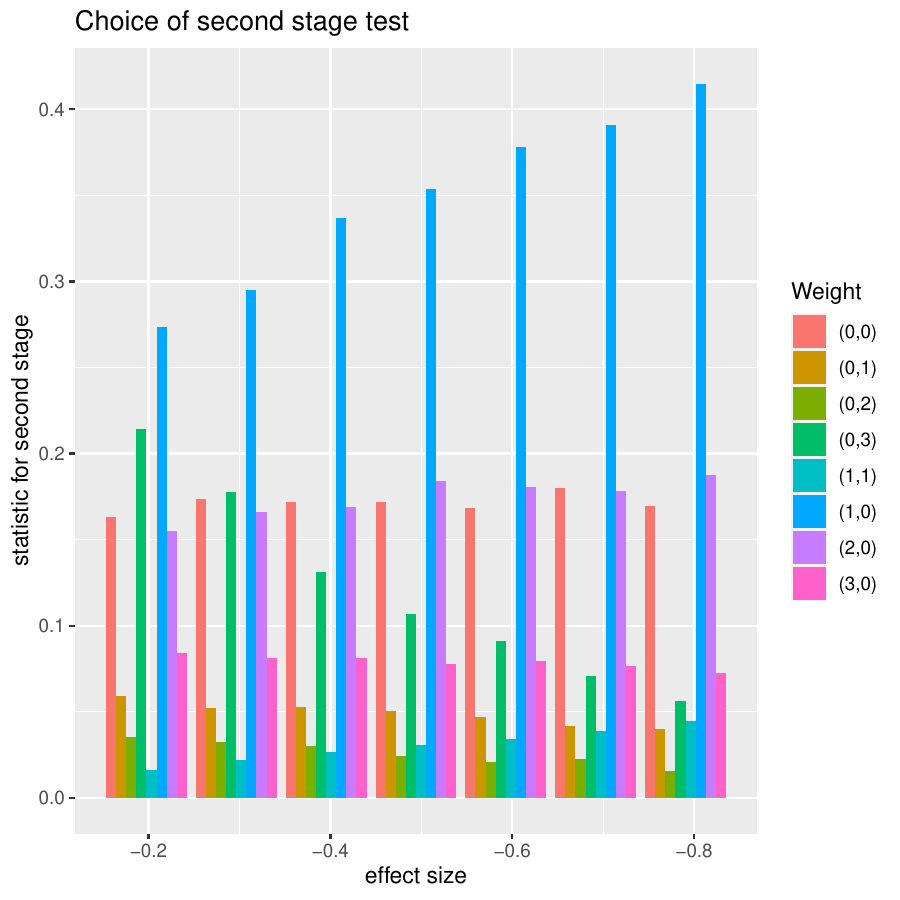}
	\end{subfigure}
	\caption{\textbf{(A)} Power curve for six testing procedures in case of a late separation $((\rho^{\star}, \gamma^{\star})=(2,0))$. \textbf{(B)} Relative frequencies of the choice of single-weighted tests for the second stage of the TS-AD testing procedure.}
	\label{fig:power_late}
\end{figure}

We examine $7$ different types of deviations of the distribution in the experimental group from the distribution in the control group. For each type, the strength of the deviation will be given by a parameter $\theta$. These types of deviations are chosen in such a way that one particular test based on Fleming-Harrington weight $w^{(\rho^{\star}, \gamma^{\star})} \circ \hat{F}$ will be optimal. We consider combinations $(\rho^{\star}, \gamma^{\star}) \in \{(0,0), (1,0), (2,0), (3,0), (0,1),\allowbreak (0,2), (0,3)\}$. For $\rho^{\star} = \gamma^{\star} = 0$, the type of deviation is just given by proportional hazards, and the standard log-rank test is optimal. For $\rho^{\star} > \gamma^{\star} = 0$ and $\gamma^{\star} > \rho^{\star} = 0$, the construction of the corresponding mechanisms is described in Garès et al. (2017)\cite{Gares:2017} and Chapter 7.4 of Fleming and Harrington (2011)\cite{fleming2011counting}, respectively. Thus, $\theta = 0$ yields no difference between the survival curves in the two groups, and the advantage of the experimental group increases as $\theta$ decreases.\\
All simulation runs' sample size is 500 patients per group. We consider $7$ different values of $\theta$ for each type of deviation. At first, we determine some value $\theta_0$ s.t. the two-stage test with the weighted log-rank test, that would be optimal in this case (i.e., $w^{(\rho^{\star}, \gamma^{\star})} \circ \hat{F}$) achieves an overall power of 50\%. Some analytical calculations can accomplish this. In the simulations, we then consider values of $\theta$ in the set $\{0.4\cdot\theta_0, 0.6\cdot\theta_0, 0.8\cdot\theta_0, \theta_0, 1.2\cdot\theta_0, 1.4\cdot\theta_0, 1.6\cdot\theta_0\}$ in order to cover a broad power range.\\
For the sake of brevity, we only show the results for deviations with $(\rho^{\star}, \gamma^{\star}) \in \{(0,0), (2,0), (0,2)\}$ in the main manuscript, see Figures \ref{fig:power_ph}-\ref{fig:power_late}. Survival curves and results for the other types can be found in Section \ref{supp-subsubsec:add_types} of the Supplementary Material. The selection comprises scenarios with proportional hazards and late and early effects. For these three types, the corresponding survival curves can be found in Figure \ref{fig:survival curves}. \\
For each type, we show two graphs. Subfigure \textbf{(A)} shows the power curves of six different testing procedures, which are abbreviated as follows:

\begin{table}[H]
    \centering
    \begin{tabular}{lp{10cm}}
       Abbreviation  &  Description\\ 
       \hline 
     \vspace{0.3cm} \textbf{OS-MDIR}    & One-stage testing procedure with the \textit{mdir} combination test based on the weights $w^{(0,0)} \circ \hat{F}$, $w^{(1,0)} \circ \hat{F}$ and $w^{(0,1)} \circ \hat{F}$ \\ \vspace{0.3cm}
      \textbf{OS-restrMDIR} & One-stage testing procedure with an \textit{mdir} combination test with a restricted set of weights in some cases ($w^{(0,0)} \circ \hat{F}$, $w^{(1,0)} \circ \hat{F}$ if $\rho^{\star} > \gamma^{\star} = 0$ and $w^{(0,0)} \circ \hat{F}$, $w^{(0,1)} \circ \hat{F}$ if $\gamma^{\star} > \rho^{\star} = 0$) \\ \vspace{0.3cm}
    \textbf{TS-AD} & Two-stage adaptive design with \textit{mdir} combination test as for OS-MDIR in the first stage and a selection of the test for the second stage among the weights in $\hat{\mathcal{Q}}_{\text{cand}}$ as defined above\\ \vspace{0.3cm}
    \textbf{TS-LR} & Two-stage standard log-rank test \\ \vspace{0.3cm}
    \textbf{TS-optFH} & Two-stage weighted log-rank test with the optimal weighting by $w^{(\rho^{\star}, \gamma^{\star})} \circ \hat{F}$ \\ \vspace{0.3cm}
    \textbf{TS-restrAD} & Two-stage adaptive design with \textit{mdir} combination test as for the design OS-restrMDIR in the first stage and a selection of the test for the second stage among a restricted subset of $\hat{\mathcal{Q}}_{\text{cand}}$ that only includes weights with $\rho \gtrless \gamma$ and the standard weight if $\rho^{\star} \gtrless \gamma^{\star}$ and weights with $|\rho-\gamma|\leq 1$ if $\rho^{\star} = \gamma^{\star} = 0$.
    \end{tabular}
    \caption{Overview of the different testing procedures.}
    \label{tab:Mdes}
\end{table}

% \begin{description}
%     \item[OS-MDIR] One-stage testing procedure with the \textit{mdir} combination test based on the weights $w^{(0,0)} \circ \hat{F}$, $w^{(1,0)} \circ \hat{F}$ and $w^{(0,1)} \circ \hat{F}$
%     \item[OS-restrMDIR] One-stage testing procedure with an \textit{mdir} combination test with a restricted set of weights in some cases ($w^{(0,0)} \circ \hat{F}$, $w^{(1,0)} \circ \hat{F}$ if $\rho^{\star} > \gamma^{\star} = 0$ and $w^{(0,0)} \circ \hat{F}$, $w^{(0,1)} \circ \hat{F}$ if $\gamma^{\star} > \rho^{\star} = 0$)
%     \item[TS-AD] Two-stage adaptive design with \textit{mdir} combination test as for OS-MDIR in the first stage and a selection of the test for the second stage among the weights in $\hat{\mathcal{Q}}_{\text{cand}}$ as defined above
%     \item[TS-LR] Two-stage standard log-rank test
%     \item[TS-optFH] Two-stage weighted log-rank test with the optimal weighting by $w^{(\rho^{\star}, \gamma^{\star})} \circ \hat{F}$
%     \item[TS-restrAD] Two-stage adaptive design with \textit{mdir} combination test as for the design OS-restrMDIR in the first stage and a selection of the test for the second stage among a restricted subset of $\hat{\mathcal{Q}}_{\text{cand}}$ that only includes weights with $\rho \gtrless \gamma$ and the standard weight if $\rho^{\star} \gtrless \gamma^{\star}$ and weights with $|\rho-\gamma|\leq 1$ if $\rho^{\star} = \gamma^{\star} = 0$.
% \end{description}
TS-optFH serves as a benchmark because the optimal test is used here. In comparison, applying TS-LR shall illustrate the disadvantages of not applying techniques that can guard against deviations from proportional hazards. To quantify the advantages of using a robust \textit{mdir} combination test, we consider a one-stage version of the \textit{mdir} test (OS-MDIR). Suppose it can be anticipated that a particular deviation from proportional hazards is more likely. In that case, choosing the weights for the \textit{mdir} combination test according to this prior knowledge can be beneficial (OS-restrMDIR). Analogously, for the two-stage design, one can use the robust \textit{mdir} and choose the second stage test out of a wide variety of single-weighted tests (TS-AD) or make pre-selections according to the anticipated possible effects (TS-restrAD).\\ 
Please note that the overall power curves do not consider any additional advantages of the adaptive designs, e.g., early rejection or sample size recalculations. For comparisons between one-stage \textit{mdir} combination tests and one-stage weighted log-rank tests we refer to Dormuth et al. (2023)\cite{Dormuth:2023}.\\
The second graph \textbf{(B)} always illustrates the choice of the test statistic for the second stage of the testing procedure TS-AD. It is shown how often each of the second stage tests from $\hat{\mathcal{Q}}_{\text{cand}}$ is chosen (using the selection procedure described above) in those simulated trials which proceeded to the second stage.\\
There is not much difference between the six approaches concerning proportional hazard scenarios. As expected, the two-stage standard log-rank test performs best. The two-stage designs with a combination test in the first and a weight selection in the second stage perform very similarly to the one-stage \textit{mdir} combination tests. Of course, some efficiency of the adaptive selection designs is lost because of a non-optimal test selection for the second stage. However, the optimal choice is made quite often. Additionally, the procedures TS-AD and TS-restrAD exhibit early rejection rates of about 9\% for the effect size $\theta_0$ and even more than $30\%$ for the effect size $1.6 \cdot \theta_0$.\\
For the scenarios of early separation, TS-LR now performs worst among the six competitors. The correctly weighted log-rank test performs best. The two procedures OS-MDIR and TS-AD perform very similarly. The weight selection graph shows that suitable tests are often chosen for the design TS-AD. However, the test based on the weights $w^{(0,3)} \circ \hat{F}$ is chosen more often than the optimal weighting scheme $w^{(0,2)} \circ \hat{F}$. We assume this behavior is rooted in the spline models, but this speculation has to be investigated further in future research. The two power curves for the restricted procedures are also almost equal. As the pre-selection of weights introduces useful information for the procedure, these perform slightly better (about five percentage points) than the unrestricted procedures OS-MDIR and TS-AD. Similarly to the case of proportional hazards, the procedures TS-AD and TS-restrAD exhibit early rejection rates of about $10\%$ for the effect size $\theta_0$ and even more than $35\%$ for the effect size $1.6 \cdot \theta_0$.\\

In the late separation scenario, TS-LR performs worst, and TS-optFH performs best again. The adaptive designs TS-AD and TS-restrAD have less overall power than the one-stage counterparts OS-MDIR and OS-restrMDIR. This may be because it is more difficult to identify a well-fitting test here since, at the time of the interim analysis, less information is available about the period in which the hazards in the two groups differ greatly. The selection procedure seems to prefer the test based on the weight $w^{(1,0)} \circ \hat{F}$ to the optimal test. Again, this behavior might result from the spline extrapolation. Additionally, we could observe that the model selection based on the AIC often prefers models with 0 interior knots (see Supplementary Material, Section \ref{supp-subsubsec:model_choice}). Unfortunately, these models often fail to detect late separation of the survival curves (see Supplementary Material, Section \ref{supp-subsubsec:modelwise_test_choice}). However, all tests perform remarkably better than TS-LR. The early rejection rates are smaller than in the previous scenarios, with about $3\%$ for the effect size $\theta_0$ and even more than $7\%$ for the effect size $1.6 \cdot \theta_0$. This is again due to the sparse information about late events at the date of the interim analysis.\\
In all scenarios, we could see that the adaptive selection procedure can close the power gap between one- and two-stage designs that is caused by the known inefficiencies of adaptive designs \cite{Tsiatis:2003}. However, as seen in Figures \ref{fig:power_early} and \ref{fig:power_late}, our selection procedure often selects well-suited but not optimal tests. This shows there is room for improvement in the selection procedure.

\section{Discussion}\label{sec:discussion}
In the previous sections, an adaptive design for a survival time endpoint was presented and examined, allowing for adjustment of test statistics at the time of interim analyses and the use of combination tests. Such a combination test is particularly beneficial in the first stage. In contrast, in the second stage, the information already collected can be used to select a suitable test for the further course. Our application example demonstrated that adapting the weight can save a trial that would otherwise end with an inconclusive result. Our simulation studies demonstrated that our two-stage procedures are superior to the two-stage log-rank test in non-proportional hazard settings. At the same time, they do not lose much power in a proportional hazard setting. Compared to one-stage combination tests, they provide much more flexibility as other adaptations are still allowed\cite{Wassmer:2016}.\\
However, we could observe that pre-selection of the involved weights based on some initial assumptions can markedly improve the performance. In this work, we only consider Fleming-Harrington weights in the combination tests and our selection procedure due to their popularity. The Fleming-Harrington weights were subject to increased criticism when applied to prove superiority due to their inconsistency regarding the scoring in late effect situations \cite{Magirr:2023}. However, our procedure can also be used for differently weighted log-rank tests that take this issue into account\cite{Magirr:2019} as shown in Section \ref{supp-subsec:modest_weights} of the Supplementary Material. This concerns the combination tests as well as the weight selection procedure. \\
The extrapolation of the survival curve of the variable under investigation beyond the time horizon observed so far is crucial for the decision to be made during the interim analysis. We applied the model of Royston and Parmar \cite{Royston:2002} as it is flexible, provides extrapolation, and incorporates some standard parametric distributions. We did not tailor the spline approach to our specific data set up to obtain a fair comparison between the two-stage and the one-stage procedures. However, we assume that this might lead to the selection of a weighted log-rank test that is not the optimal test by design. To tackle this issue, further investigation of the proposed selection procedure of Royston-Parmar splines or alternative extrapolation methods is necessary. In principle, our proposed procedure allows for any other extrapolation approach. For example, the Kaplan-Meier estimator could be combined with a parametric tail for extrapolation\cite{Gelber:1993}, or a penalized version of the chosen approach\cite{Liu:2018} can be employed. Of course, it would also be desirable to incorporate expert knowledge or prior knowledge from other data sets using Bayesian methods\cite{Jackson:2023}. We assume that such methods can further improve the extrapolation and, thus, the entire procedure. However, we need to make sure that information beyond the primary endpoint should not be incorporated into the decision process about the adaptation as it might compromise the type I error rate\cite{Bauer:2004} if nor precautions are taken against this\cite{Danzer:2023}.\\
The analysis schedule also plays a major role here. Suppose the survival curves of the two groups separate so late that no corresponding observations can be made until the interim analysis. In that case, no reasonably informed decision can be made about how to continue the study.\\
While we have limited ourselves to calculating conditional power as an instrument for determining adaptation, other tools are also possible. It is well-known that the conditional power tends to assume quite extreme values \cite{Bauer:2006}. Similar concepts, such as the predictive power\cite{Spiegelhalter:1986}, are less prone to this problem. Such an approach could also be excellently combined with a Bayesian approach to the survival extrapolation as mentioned above\cite{Jackson:2023}.\\
Finally, we would like to point out that the main aim of the proposed procedure is to be as flexible and robust as possible. This inevitably leads to a loss of efficiency. First, this applies to adaptive designs in general since the fixed weights selected for the combination function do not necessarily correspond to the amount of information that comes in from the individual stages\cite{Tsiatis:2003}. This challenge is even more accentuated here, as the information processes of differently weighted log-rank tests are not proportional to each other and generally do not correspond to the number of events observed\cite{Kundu:2021}. We would also like to mention again that combination tests have a good sensitivity to a wide range of alternatives\cite{Dormuth:2023} but are never optimal.\\
We regard our contribution as methodological phase I/II in the sense of Heinze et al. (2022) \cite{heinze2022phases}. From our point of view, the adaptation of test statistics has been woefully neglected, although discussions about the choice of an appropriate test in situations of non-proportional hazards continue. We believe that combination tests and adaptive designs complement each other here naturally. Hence, we would like to present this fundamental possibility here. Nevertheless, we are aware that further research is still required. This includes the incorporation of different weight functions \cite{Magirr:2023}, consideration of other extrapolation approaches \cite{Gelber:1993, Liu:2018, Jackson:2023} as well as fine-tuning the selection of the final extrapolation model and the investigation of alternative quantities for decision making regarding design adaptations \cite{Spiegelhalter:1986}. Furthermore, we want to investigate different approaches to combine the stagewise p-values and derive recommendations on executing the sample size reestimation within the framework.

\section*{Acknowledgments}
The work of Moritz Fabian Danzer was funded by the Deutsche Forschungsgemeinschaft (DFG, German Research
Foundation) - Project number 413730122.\\
Parts of the calculations for this publication were performed on the HPC cluster PALMA II of the University of Münster, subsidised by the DFG (INST 211/667-1).\\
In memory of our esteemed colleague, Marc Ditzhaus, who passed away in September 2024. His invaluable
input will always be remembered, and he will be deeply missed.

\section*{Conflict of Interest}
The authors have declared no conflict of interest. 

\bibliography{Lib}
\bibliographystyle{ama}

\newpage
\clearpage

\begin{center}
\huge
Adaptive weight selection for time-to-event data under non-proportional hazards - Supplementary Material
\end{center}

\renewcommand{\tablename}{Supplementary Table} 
\renewcommand{\figurename}{Supplementary Figure}
\renewcommand{\thetable}{S\arabic{table}}
\renewcommand{\thefigure}{S\arabic{figure}}
\renewcommand{\thesection}{\Alph{section}}
\setcounter{section}{0}

\section{Technical appendix}\label{supp-sec:technical}
In this section of the Supplementary Material, we state some technical results that justify the validity of our adaptive weight selection procedure. We adopt the notation from the main manuscript.

\begin{lemma}\label{lemma:mv_conv_in_prob}
	Let $(\mathbf{X}^{(n)})_{n \geq 0}$ be a sequence of $\mathbb{R}^d$-valued random vectors s.t. $X_c^{(n)} \overset{\mathbb{P}}{\to} X_c$ for each $c \in \{1,\dots, d\}$ as $n\to \infty$. Then it also holds
	\begin{equation*}
		\mathbf{X}^{(n)} \overset{\mathbb{P}}{\to} \mathbf{X} \eqqcolon (X_1,\dots X_d)
	\end{equation*}
\end{lemma}
as $n \to \infty$ in $\mathbb{R}^d$.
\begin{proof}
	As all norms are equivalent on $\mathbf{R}^d$, it is enough to show it for the 1-norm, i.e.
	\begin{equation*}
		\mathbb{P}\left[\sum_{c=1}^d |X_c^{(n)} - X_c| > \varepsilon \right] \to 0
	\end{equation*}
	for any $\varepsilon > 0$. Because $\sum_{c=1}^d |X_c^{(n)} - X_c| > \varepsilon$ implies that there is at least one $c$ s.t. $|X_c^{(n)} - X_c| > \varepsilon/d$, we get
	\begin{align*}
		&\mathbb{P}\left[\sum_{c=1}^d |X_c^{(n)} - X_c| > \varepsilon \right]\\
		\leq & \mathbb{P}\left[\cup_{c=1}^d |X_c^{(n)} - X_c| > \varepsilon/d \right]\\
		\leq & \sum_{c=1}^d \mathbb{P}[|X_c^{(n)} - X_c| > \varepsilon/d]
	\end{align*}
	As all of the summands in the last sum converge to $0$, the sum becomes arbitrary small for increasing $n$.
\end{proof}

Extending the notation from the manuscript, we define the bivariate $\mathbb{R}$-valued stochastic process $(T_{\hat{Q}}(t,s))_{t,s \geq 0}$ by
\begin{equation*}
    T_{\hat{Q}}(t,s)\coloneqq n^{-\frac{1}{2}}\sum_{i=1}^n \int_{[0,s]} \hat{Q}(t,u) \left(Z_i -   \frac{Y^{Z=1}(t,u)}{Y(t,u)} \right) dN_i(t,u),
\end{equation*}
that sums up the information available at calendar time $t$ about the time-to-event endpoint until trial time $s$. This process has to be adapted to the bivariate filtration $(\mathcal{F}(t,s))_{t,s \geq 0}$ with
\begin{equation*}
    \mathcal{F}(t,s) = \sigma\left(\cup_{i=1}^n \mathcal{F}_i(t,s) \right),
\end{equation*}
i.e. these $\sigma$-algebras are generated by patient-specific $\sigma$-algebras. These $\mathcal{F}_i(t,s)$ are in turn generated by the random variables
\begin{equation*}
    \begin{split}
        \mathbbm{1}_{\{R_i \leq t\}}, R_i \cdot \mathbbm{1}_{\{R_i \leq t\}}, \mathbbm{1}_{C_i^{\star} \leq s \wedge (t-R_i)_+}, C_i^{\star}\cdot \mathbbm{1}_{C_i^{\star} \leq s \wedge (t-R_i)_+},\\
        \mathbbm{1}_{T_i \leq s \wedge C_i(t)}, T_i^\cdot \mathbbm{1}_{T_i \leq s \wedge C_i(t)}.
    \end{split}
\end{equation*}
The following results are valid under the null hypothesis of equal distributions of the time-to-event variable $T$
\begin{theorem}\label{supp-thm:asymptotic_equivalence}
    If for all $t \geq 0$, the assumptions
    \begin{enumerate}[label = A\arabic*]
        \item For any $\tau < t$ \begin{equation*}
            \sup_{0\leq s \leq \tau} |\hat{Q}(t,s) - Q(t,s)| \overset{\mathbb{P}}{\to} 0
        \end{equation*}
        \item In its second argument, $\hat{Q}(t,s)$ is bounded over $[0,t]$, is left-continuous and has right hand limits
        \item For any $\tau_1 > 0$ and $\tau_2 < t$
        \begin{equation*}
            \sup_{\tau_1 \leq s \leq \tau_2} \left| \frac{Y^{Z=1}(t,s)}{Y(t,s)} - \frac{y^{Z=1}(t,s)}{y(t,s)} \right| \overset{\to}{\mathbb{P}}
        \end{equation*}
        where $y(t,s)\coloneqq \mathbb{E}[Y(t,s)]$ and $y^{Z=1}(t,s)\coloneqq \mathbb{E}[Y^{Z=1}(t,s)]$
    \end{enumerate}
    are fulfilled, then the process $(T_{\hat{Q}}(t))_{t\geq 0}$ is asymptotically equivalent to the process $(\tilde{T}_{Q}(t))_{t\geq 0}$, i.e.
    \begin{equation*}
        (T_{\hat{Q}}(t) - \tilde{T}_{Q}(t))\overset{\mathbb{P}}{\to} 0 \qquad{\forall t\geq 0}
    \end{equation*}
    This process is defined by
    \begin{equation*}
        \tilde{T}_{Q}(t,s)\coloneqq n^{-\frac{1}{2}}\sum_{i=1}^n \int_{[0,s]} Q(t,s) \left(Z_i - \frac{Y^{Z=1}(t,s)}{Y(t,s)} \right) dM_i(t,s),
    \end{equation*}
    where $(M_i(t,s))_{t,s \geq 0}$ is the counting process martingale based on the counting process $(N_i(t,s))_{t,s \geq 0}$.
\end{theorem}
For a proof, we refer to Theorem 1 of the Supplementary Material of \cite{Danzer:2023}. In particular, it should be noted that the random quantities $\hat{Q}$, $Y$ and $Y^{Z=1}$ have been replaced by deterministic quantities. Applying Lemma \ref{lemma:mv_conv_in_prob}, we obtain the following Corollary.

\begin{corollary}
    The multivariate processes $(\mathbf{T}_{\hat{\mathcal{Q}}}(t))_{t \geq 0}$ as in Section \ref{subsec:mdir} and the process $(\tilde{\mathbf{T}}_{\mathcal{Q}}(t))_{t \geq 0}$, where $\mathcal{Q}$ denotes the set of deterministic limit functions of those functions in $\hat{\mathcal{Q}}$ and $\tilde{\mathbf{T}}_{\mathcal{Q}}(t) \coloneqq (\tilde{T}_Q)_{Q \in \mathcal{Q}}$, are asymptotically equivalent.
\end{corollary}

\begin{lemma}
If all functions $Q \in \mathcal{Q}$ and $y^{Z=1}(t,s)/y(t,s)$ are independent of their first arguments, the multivariate processes $(\tilde{\mathbf{T}}_{\mathcal{Q}}(t))_{t \geq 0}$ is a martingale w.r.t. the filtration $(\mathcal{F}(t))_{t \geq 0}$ that comprises all available information in calendar time, i.e. $\mathcal{F}(t)\coloneqq \mathcal{F}(t,t)$.
\end{lemma}
The proof follows analogously to the proof of Lemma 2 in Danzer et al. (2023)\cite{Danzer:2023}.\\
Please note that the assumptions made in this Lemma are naturally fulfilled in our applications. For Fleming-Harrington weights, the limit functions are given by $F(t)^\rho\cdot S(t)^\rho$ and under the null hypothesis $y^{Z=1}(t,s)/y(t,s)$ reduces to the (constant) probability that an individual is assigned to the treatment group.

\begin{theorem}
As $n \to \infty$, $(\tilde{\mathbf{T}}_{\mathcal{Q}}(t))_{t \geq 0}$ converges in distribution to a Gaussian mean-zero vector martingale on some interval $[0,t_{\text{mas}}]$ with the $|\mathcal{Q}|\times|\mathcal{Q}|$-matrix-valued covariance function $\boldsymbol{\Sigma}_{\mathcal{Q}}\colon [0, t_{\text{max}}] \to \mathbb{R}^{|\mathcal{Q}|\times|\mathcal{Q}|}$ given by
\begin{equation*}
    (\boldsymbol{\Sigma(t)})_{kl} \coloneqq \int_{[0,t]} Q_k(s)Q_l(s) \cdot \mathbb{P}[s \leq C(t) \wedge T] \cdot \mathbb{P}[Z=1](1-\mathbb{P}[Z=1]) dA(s).
\end{equation*}
\end{theorem}
The proof follows analogously to the proof of Theorem 2 in Danzer et al. (2023)\cite{Danzer:2023}. The covariance function can be consistently estimated as stated in Section \ref{subsec:mdir}. 

\begin{corollary}
For a sequence of analysis dates $0 \eqqcolon t_0 < t_1 < \dots < t_m$ in calendar time, the test multivariate test statistics $\mathbf{T}_{\hat{\mathcal{Q}}}$ are asymptotically jointly normally distributed with asymptotically independent increments, i.e.
\begin{align*}
    (\mathbf{T}_{\hat{\mathcal{Q}}}(t_1), \dots, \mathbf{T}_{\hat{\mathcal{Q}}}(t_m))& \overset{\mathcal{D}}{\to} \mathcal{N}(0, \boldsymbol{\Sigma}_{\mathcal{Q}, \text{acc}})\\
    (\mathbf{T}_{\hat{\mathcal{Q}}}(t_1) - \mathbf{T}_{\hat{\mathcal{Q}}}(t_0), \dots, \mathbf{T}_{\hat{\mathcal{Q}}}(t_m) - \mathbf{T}_{\hat{\mathcal{Q}}}(t_{m-1}))& \overset{\mathcal{D}}{\to} \mathcal{N}(0, \boldsymbol{\Sigma}_{\mathcal{Q}, \text{inc}})
\end{align*}
where both $\boldsymbol{\Sigma}_{\mathcal{Q}, \text{acc}}$ and $\boldsymbol{\Sigma}_{\mathcal{Q}, \text{inc}}$ are $m|\mathcal{Q}| \times m|\mathcal{Q}|$ matrices consisting of $m^2$ blocks of size $|\mathcal{Q}|\times |\mathcal{Q}|$. The block in row $r_1$ and column $r_2$ of $\boldsymbol{\Sigma}_{\mathcal{Q}, \text{acc}}$ is given by $\boldsymbol{\Sigma}_{\mathcal{Q}}(t_{r_1} \wedge t_{r_2})$ and $\boldsymbol{\Sigma}_{\mathcal{Q}, \text{inc}}$ is a block diagonal matrix with $\boldsymbol{\Sigma}_{\mathcal{Q}, \text{inc}}=\text{diag}(\boldsymbol{\Sigma}_{\mathcal{Q}}(t_1) - \boldsymbol{\Sigma}_{\mathcal{Q}}(t_0), \dots, \boldsymbol{\Sigma}_{\mathcal{Q}}(t_m) - \boldsymbol{\Sigma}_{\mathcal{Q}}(t_{m-1}))$\\
Accordingly,
\begin{equation*}
    (\Psi_1(\mathbf{T}_{\hat{\mathcal{Q}}}(t_1) - \mathbf{T}_{\hat{\mathcal{Q}}}(t_0)), \dots, \Psi_m(\mathbf{T}_{\hat{\mathcal{Q}}}(t_m) - \mathbf{T}_{\hat{\mathcal{Q}}}(t_{m-1})))
\end{equation*}
forms a set of asymptotically independent random variables for any set of Borel-measurable functions $\{\Psi_1,\dots,\Psi_m\}$ with $\Psi_j\colon \mathbb{R}^{|\mathcal{Q}|} \to \mathbb{R}^{k_j}$ for any $k_j \in \mathbb{N}$ for all $j \in \{1,\dots,m\}$.
\end{corollary}
As in Corollary 3 of Danzer et al. (2023), this is a consequence of the previous results.\\
The last statement allows us to apply the results and techniques of Brendel et al. (2014) and Ditzhaus \& Friedrich (2019)\cite{Brendel:2014, Ditzhaus:2019} also to the increments of the multivariate test statistics.

%Asymptotische Unabhängigkeit von Zeit-Stufen-Teststatistik von Erst-Stufen-Information durch Anwendung von Dynkin's \pi-\sigma-Theorem

\newpage

\section{Additional details to the real data example in Section \ref{sec:example} of the main manuscript}\label{supp-sec:example}
In Section \ref{sec:example} of the main manuscript, we reanalysed reconstructed data from the FAKTION trial \cite{Jones:2020} (NCT number NCT01992952). Further details, that are also mentioned in the main manuscript, will be given here.

\subsection{Recruitment and censoring mechanism}
\begin{figure}[ht]
    \centering
    \includegraphics[width = .9\textwidth]{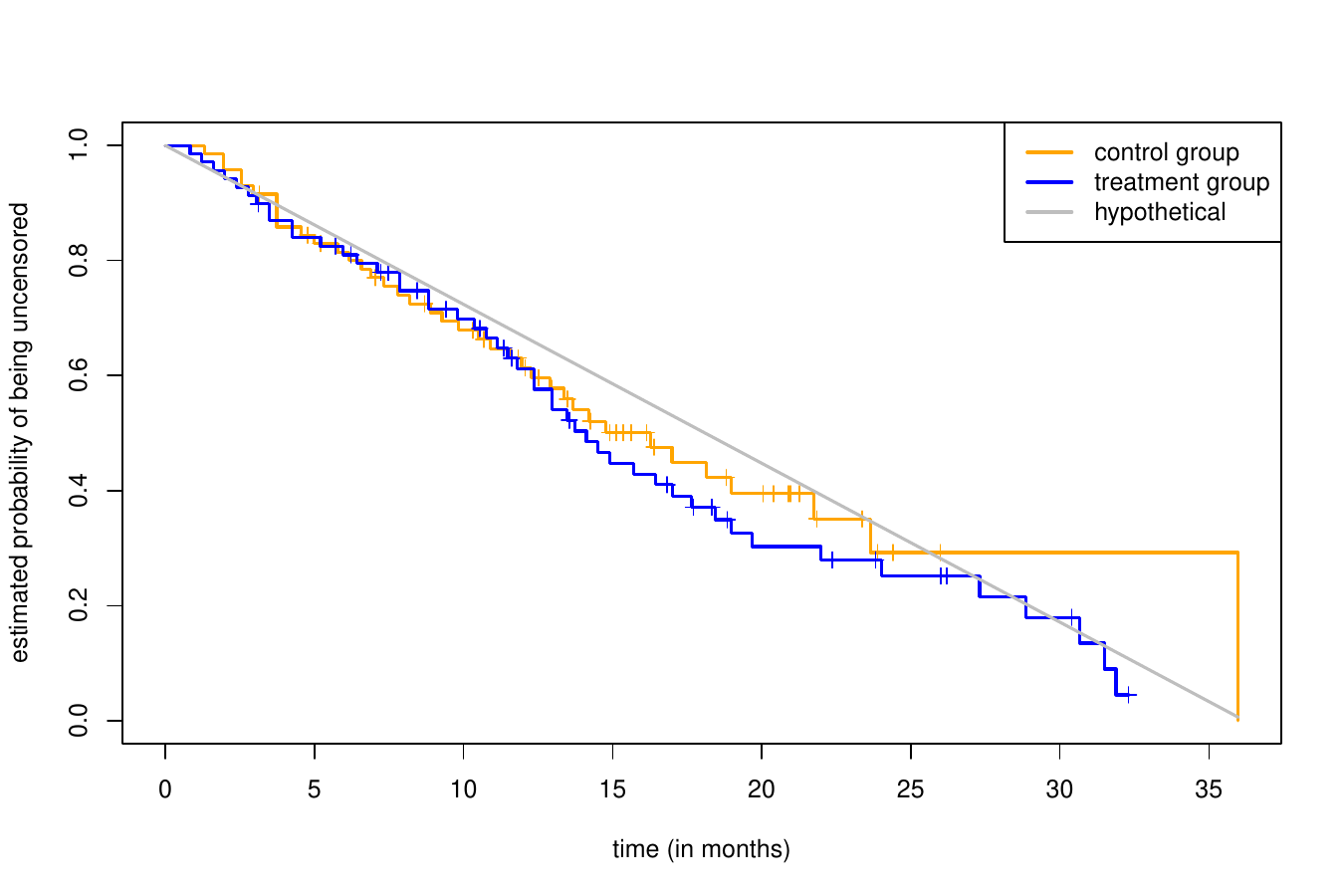}
    \caption{Kaplan-Meier estimates of the survival function of the group-wise censoring distribution from reconstructed data. Grey line indicates hypothetical function if recruitment occurs uniformly over the whole trial duration and no additional loss to follow-up occurs.}
    \label{supp-fig:km_cens}
\end{figure}

Here, we give some additional details on how and on what basis we reconstructed/simulated recruitment dates. Recruitment dates are required to apply administrative censoring at the interim analysis date to obtain hypothetical interim data.\\
According to the reconstructed data, the last observation has been censored at $t_2\coloneqq35.98$ months, which is approximately 36 months. This corresponds to the total recruitment duration specified in the published manuscript \cite{Jones:2020}. Additionally, the hypothetical censoring distribution at the final analysis that emerges from a uniform recruitment over the whole trial duration is very similar to the estimated distribution from reconstructed data (see Figure \ref{supp-fig:km_cens}). Hence, we assume that $C^{\star}=\infty$ with probability 1 and $R \sim \text{Unif}[0,t_2]$ according to the notation introduced in Section \ref{sec:methods}.\\
Therefore, we set $R_i = t_2 - X_i(t_2)$ if $\delta_i(t_2)=0$ and simulate $R_i \sim [0, t_2 - X_i(t_2)]$ if $\delta_i(t_2)=1$. This simulation is in accordance with our preceding comments as $R|R \leq r \sim \text{Unif}[0,r]$ for any uniformly distributed random variable $R$ on $[0, r_{\text{max}}]$ with $r \leq r_{\text{max}}$.\\
Obviously, the final results will depend on the simulated recruitment dates for uncensored observations. This dependence will be investigated further in Section \ref{supp-subsec:sim_dependence}.

\subsection{Royston-Parmar spline fits for interim data}
\begin{figure}[ht]
    \centering
    \includegraphics[width = .9\textwidth]{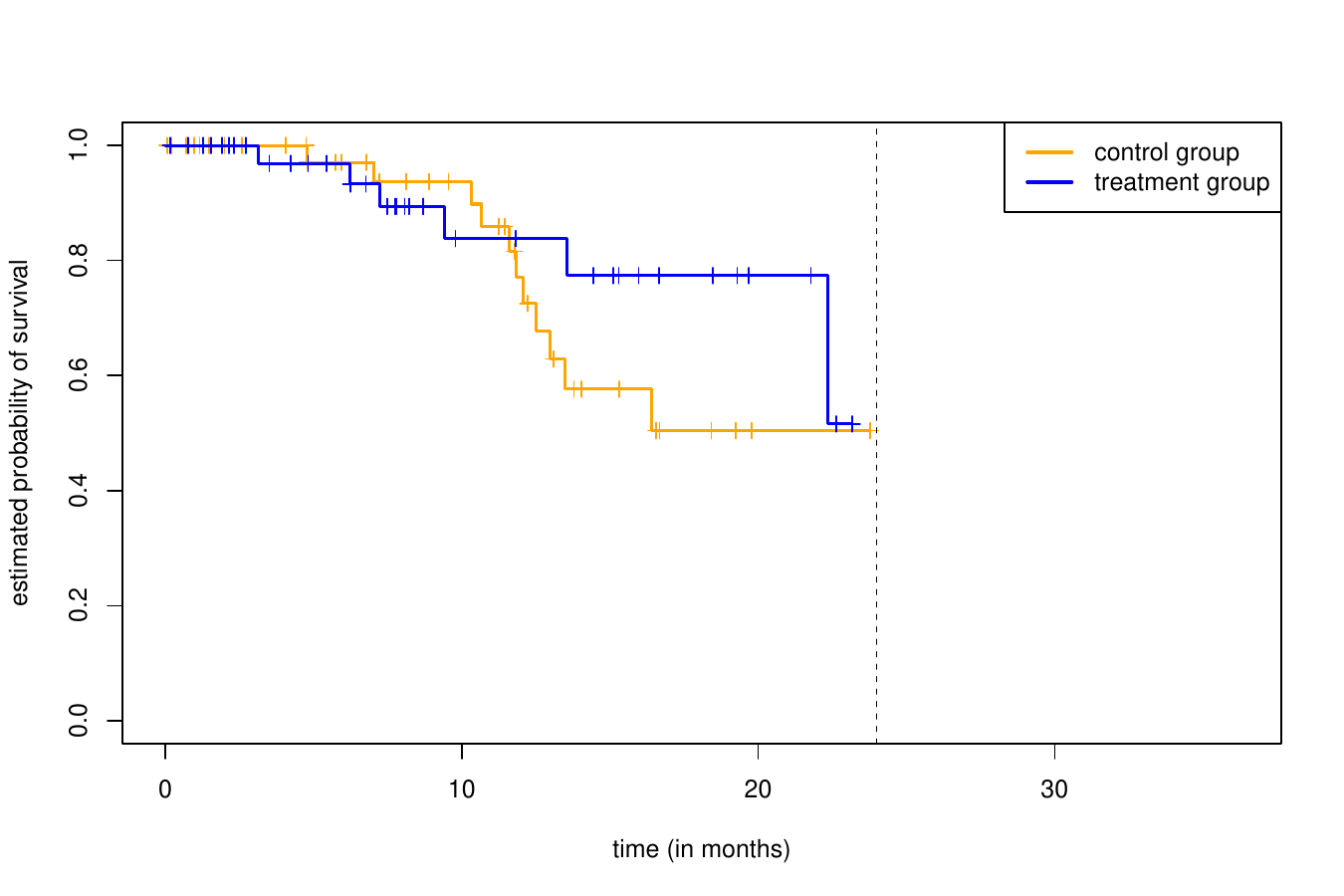}
    \caption{Group-wise Kaplan-Meier estimates from interim data.}
    \label{supp-fig:km_interim}
\end{figure}

\begin{landscape}

\begin{figure}

%First row (p=0)
\centering
\begin{subfigure}{0.45\textwidth}
    \includegraphics[width=\textwidth]{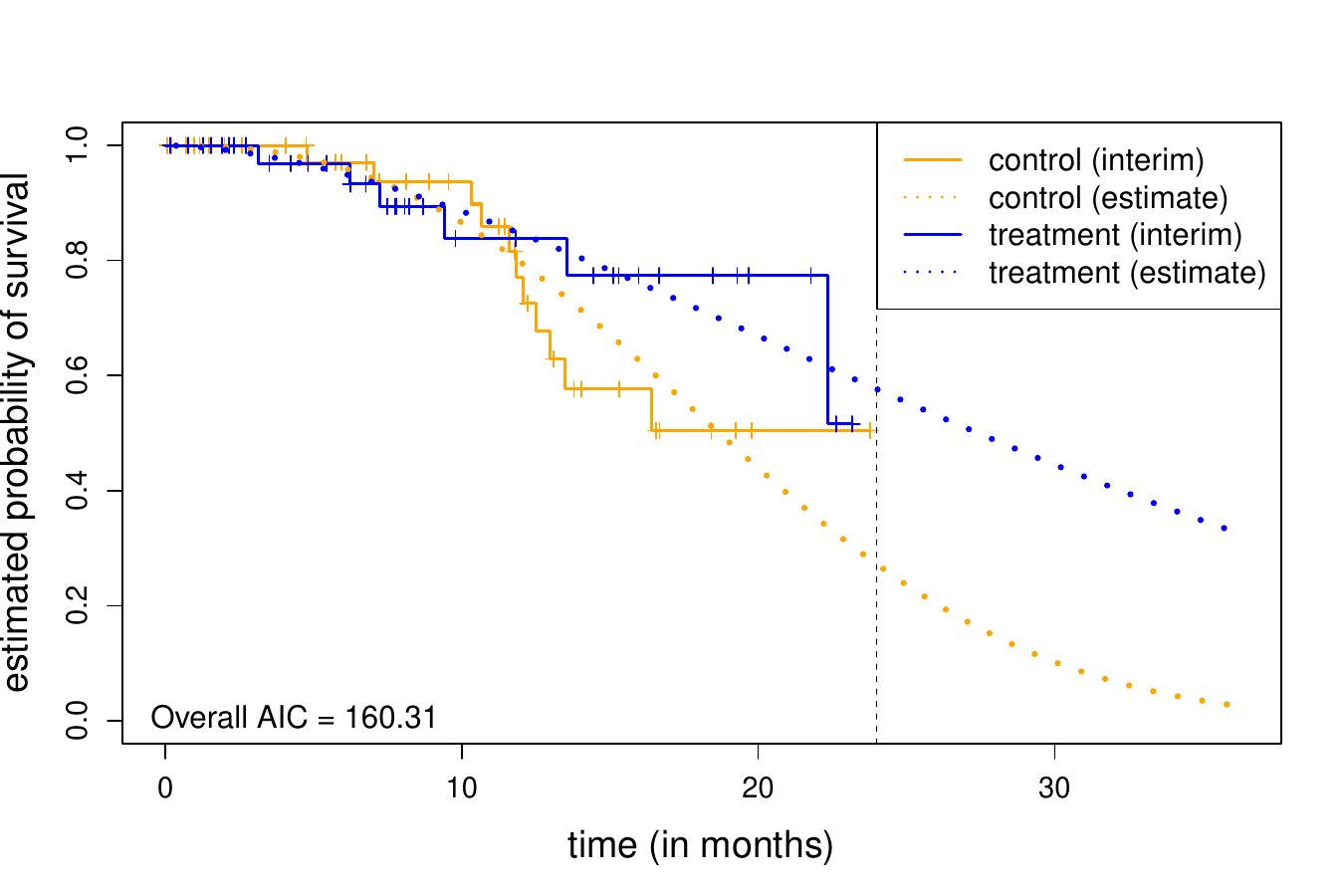}
    \caption{$p=0$, hazard scale}
\end{subfigure}
\hfill
\begin{subfigure}{0.45\textwidth}
    \includegraphics[width=\textwidth]{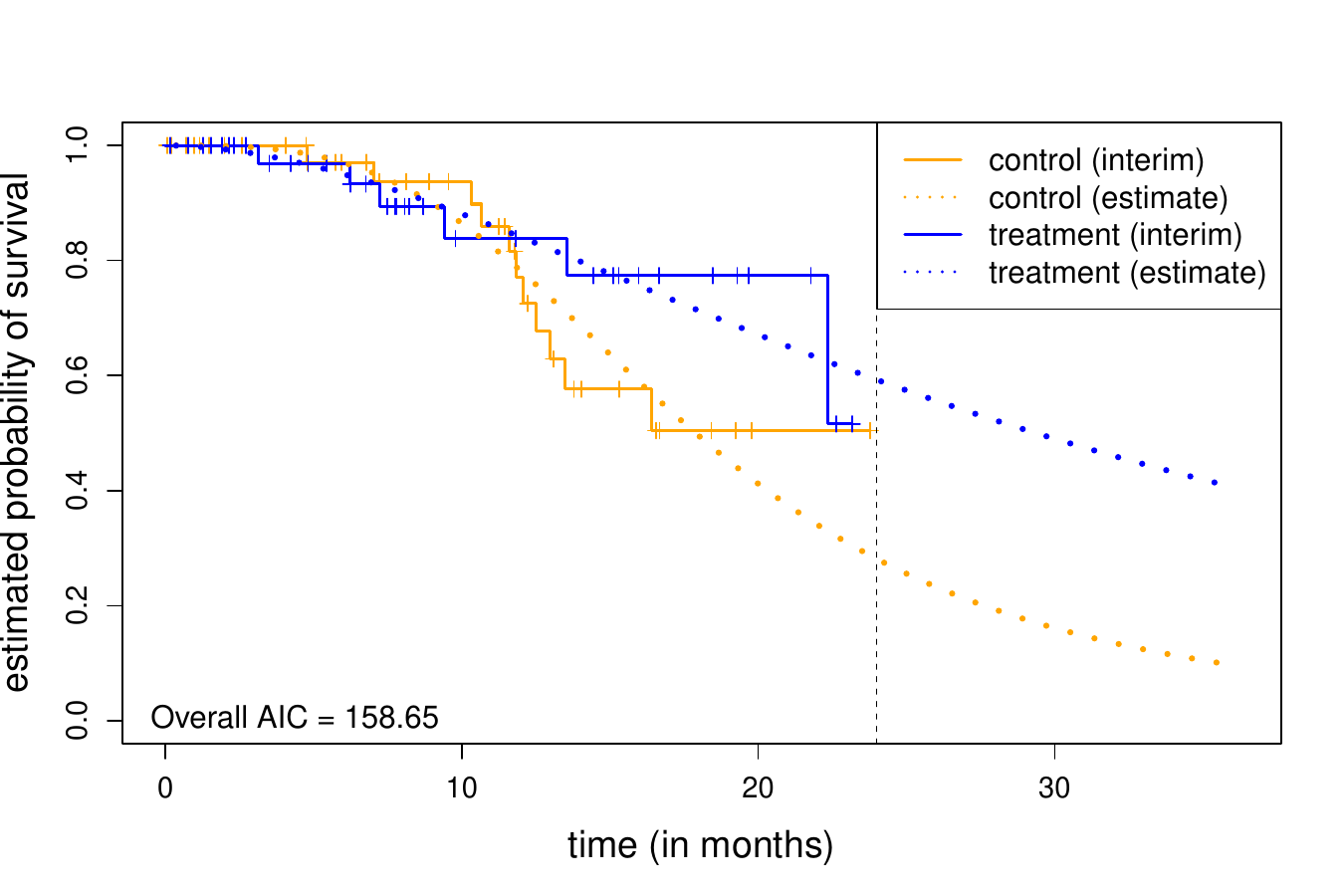}
    \caption{$p=0$, odds scale}
\end{subfigure}
\hfill
\begin{subfigure}{0.45\textwidth}
    \includegraphics[width=\textwidth]{rp_interim/os_interim_rp_k0_normal.pdf}
    \caption{$p=0$, normal scale}
\end{subfigure}

%Second row (p=1)
\begin{subfigure}{0.45\textwidth}
    \includegraphics[width=\textwidth]{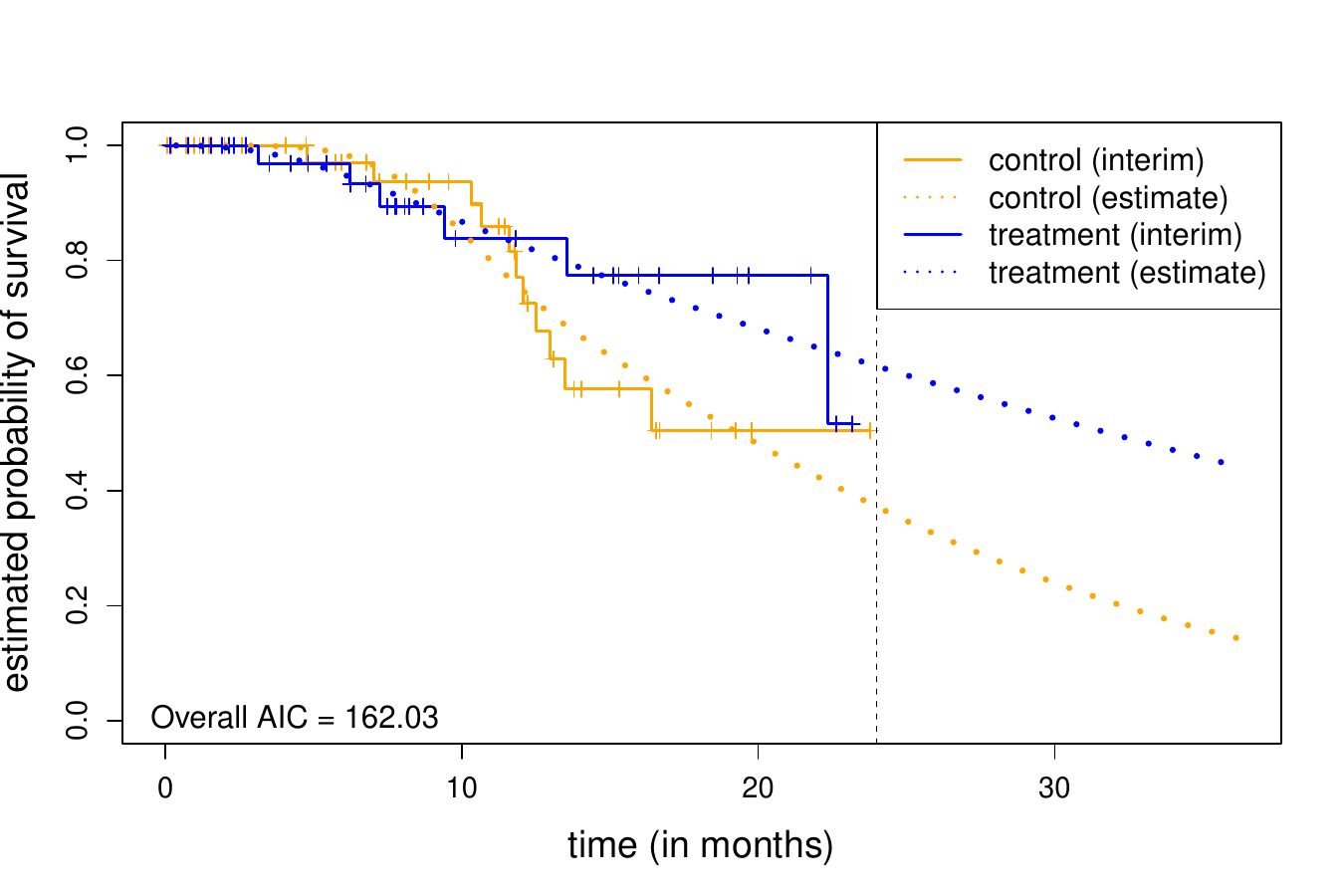}
    \caption{$p=1$, hazard scale}
\end{subfigure}
\hfill
\begin{subfigure}{0.45\textwidth}
    \includegraphics[width=\textwidth]{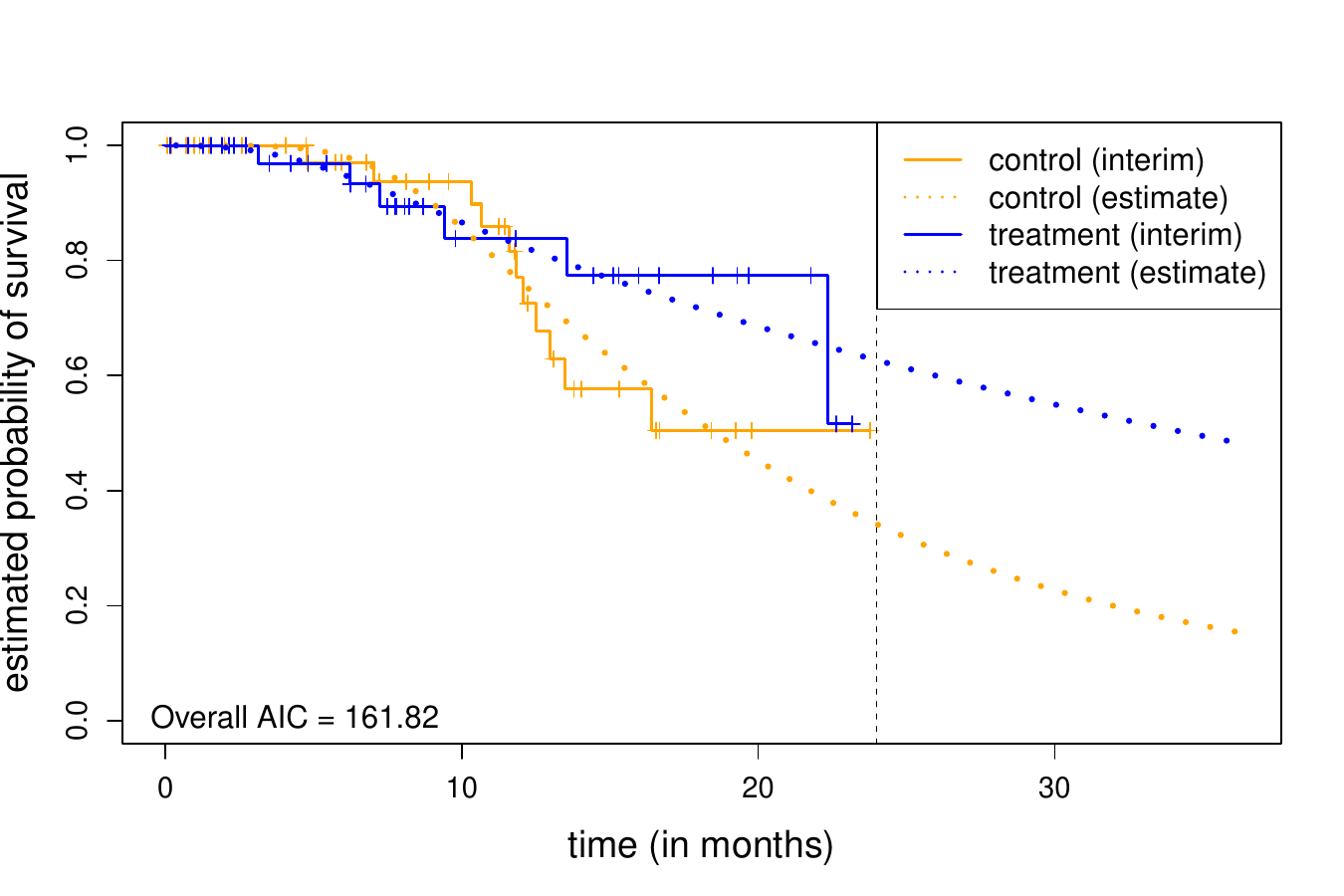}
    \caption{$p=1$, odds scale}
\end{subfigure}
\hfill
\begin{subfigure}{0.45\textwidth}
    \includegraphics[width=\textwidth]{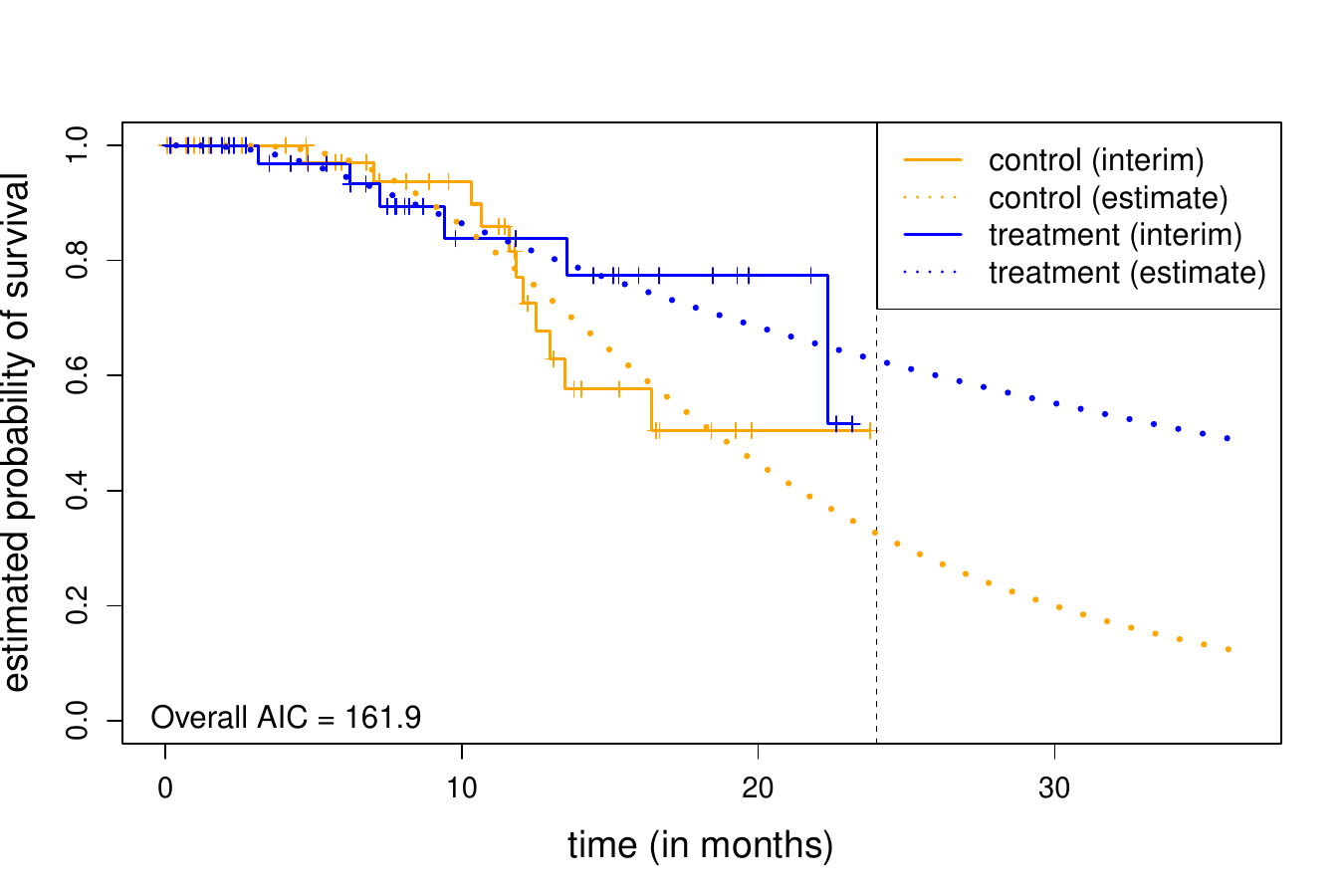}
    \caption{$p=1$, normal scale}
\end{subfigure}

%Third row (p=2)
\begin{subfigure}{0.45\textwidth}
    \includegraphics[width=\textwidth]{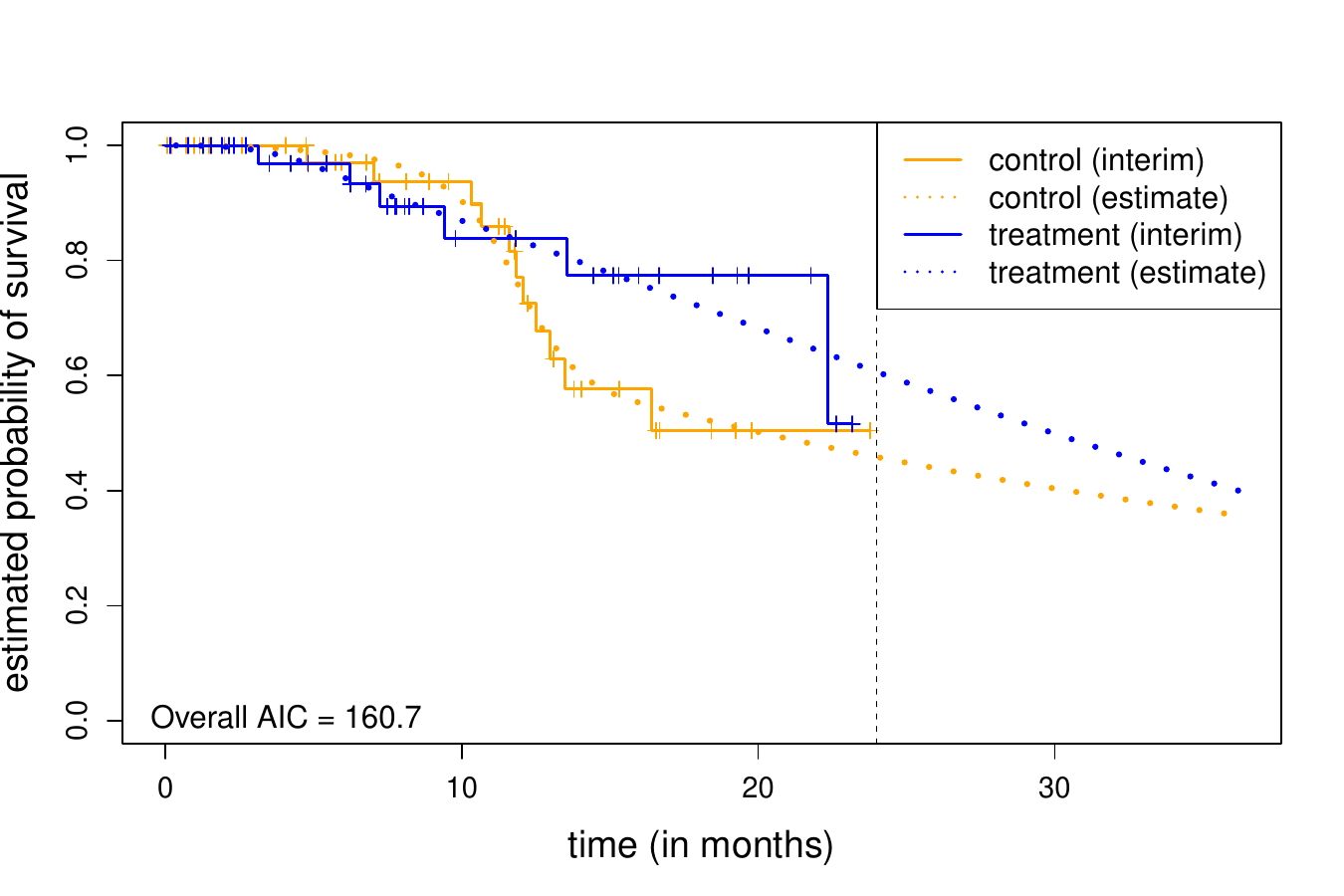}
    \caption{$p=2$, hazard scale}
\end{subfigure}
\hfill
\begin{subfigure}{0.45\textwidth}
    \includegraphics[width=\textwidth]{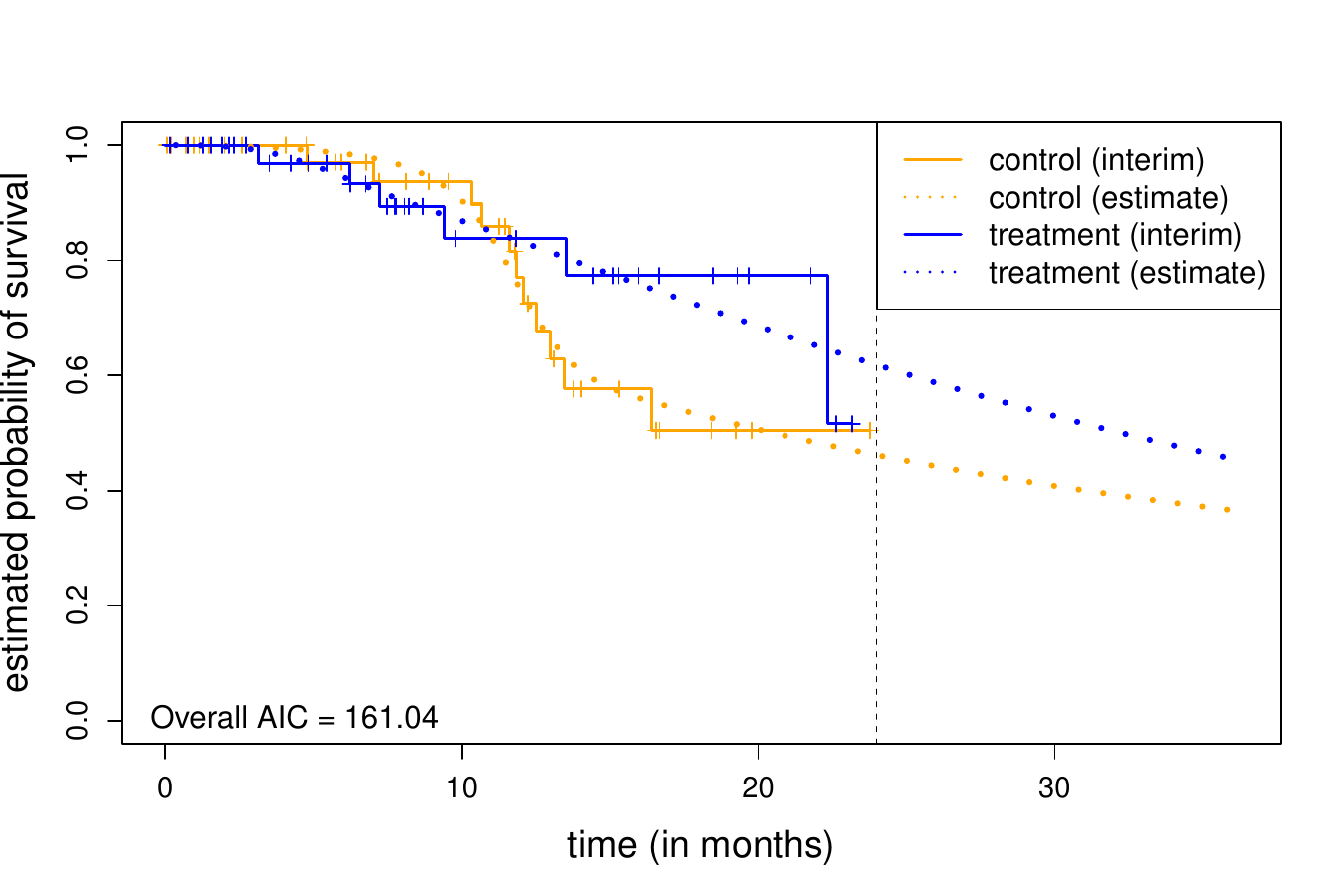}
    \caption{$p=2$, odds scale}
\end{subfigure}
\hfill
\begin{subfigure}{0.45\textwidth}
    \includegraphics[width=\textwidth]{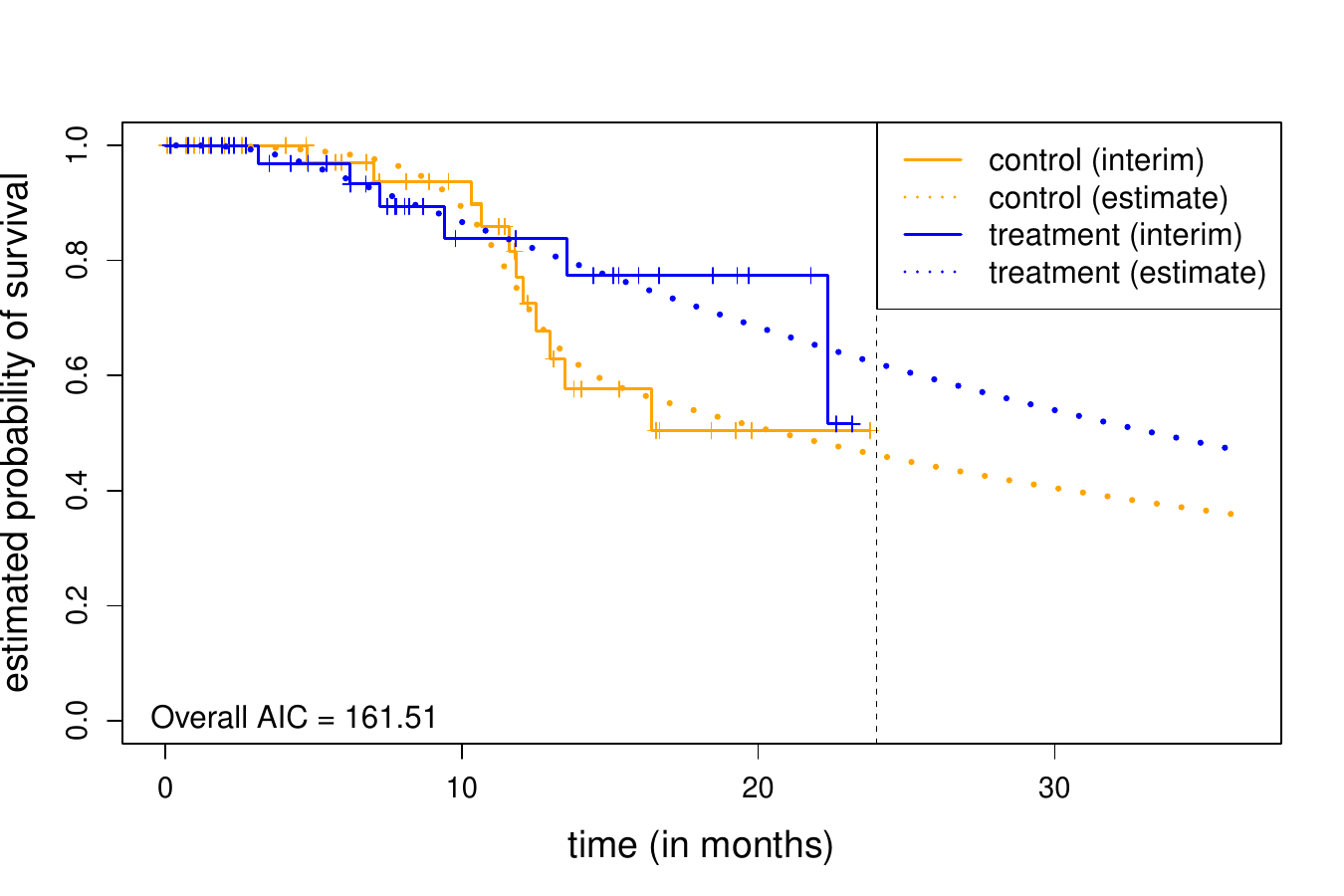}
    \caption{$p=2$, normal scale}
\end{subfigure}

\caption{Fits of Royston-Parmar spline models to interim data.}
\label{supp-fig:interim_fits}
\end{figure}

\end{landscape}

\begin{landscape}

\begin{figure}

%First row (p=0)
\centering
\begin{subfigure}{0.45\textwidth}
    \includegraphics[width=\textwidth]{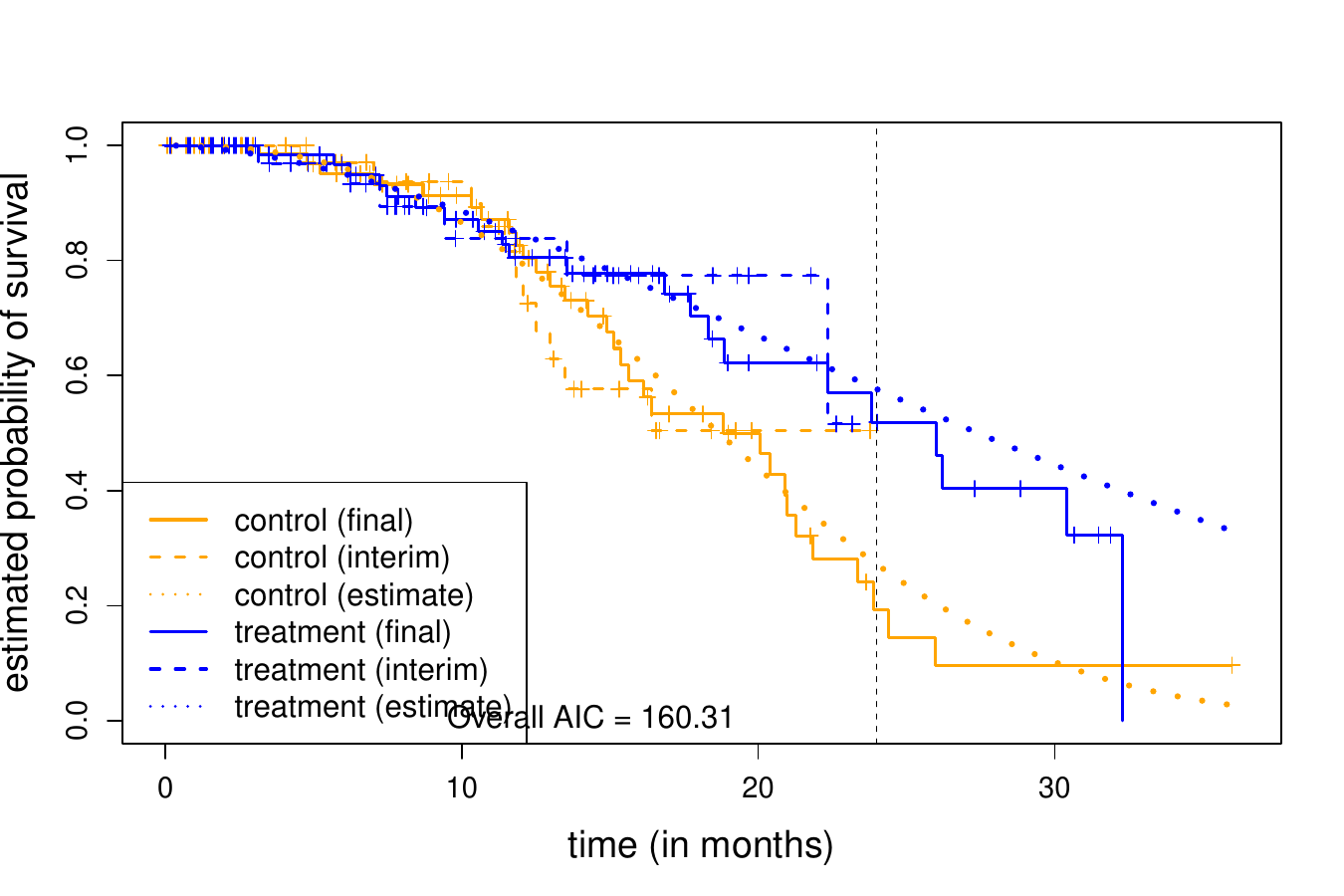}
    \caption{$p=0$, hazard scale}
\end{subfigure}
\hfill
\begin{subfigure}{0.45\textwidth}
    \includegraphics[width=\textwidth]{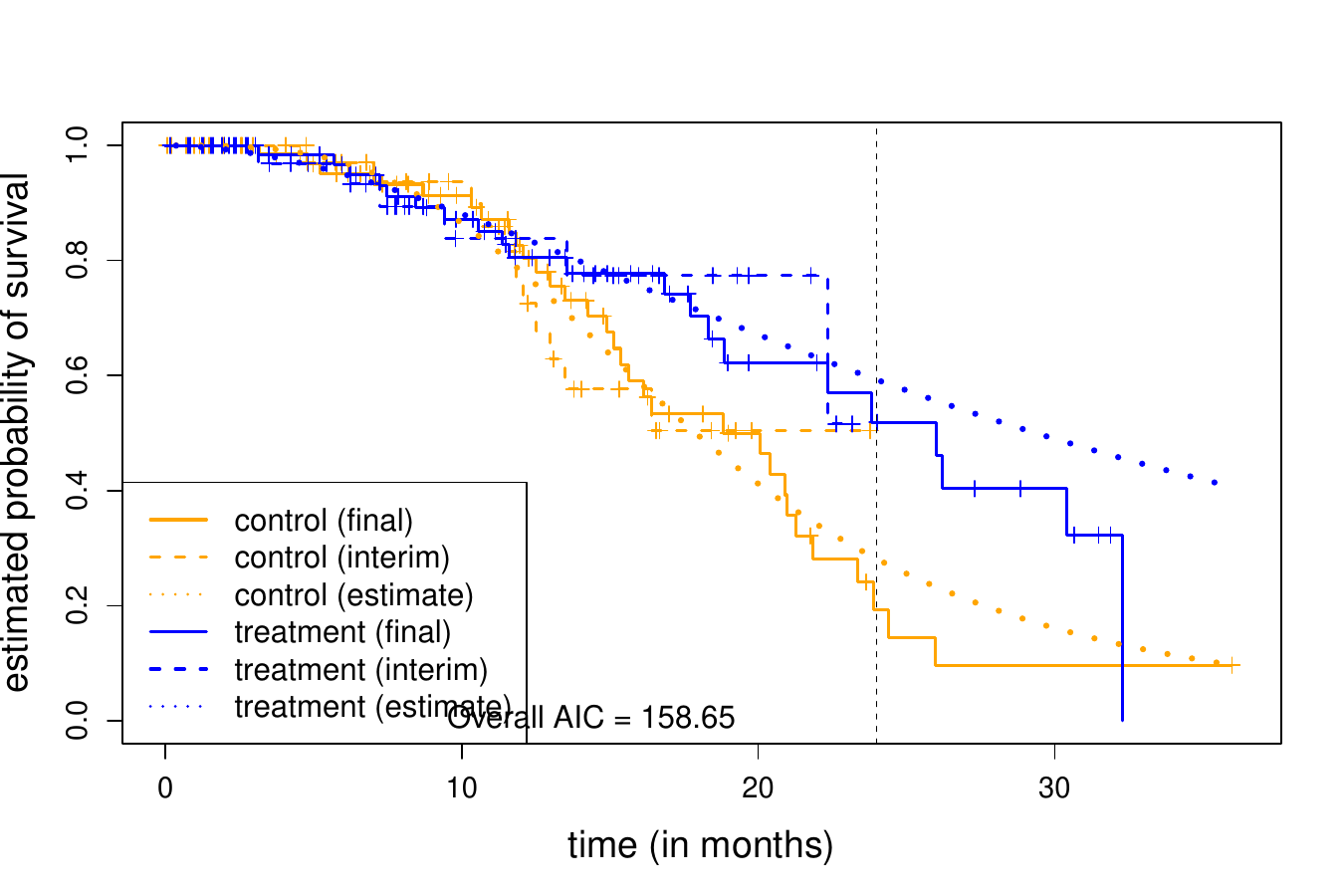}
    \caption{$p=0$, odds scale}
\end{subfigure}
\hfill
\begin{subfigure}{0.45\textwidth}
    \includegraphics[width=\textwidth]{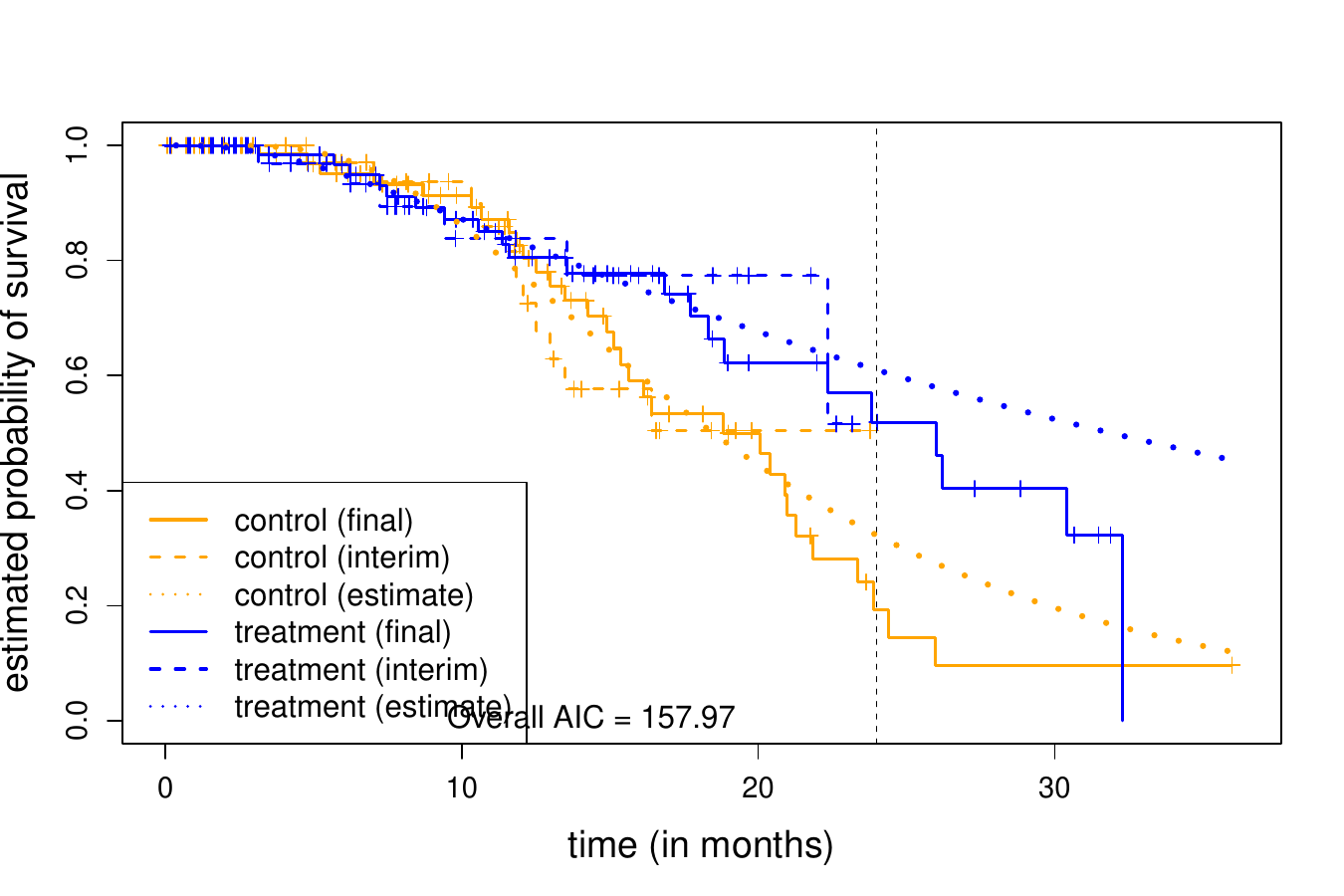}
    \caption{$p=0$, normal scale}
\end{subfigure}

%Second row (p=1)
\begin{subfigure}{0.45\textwidth}
    \includegraphics[width=\textwidth]{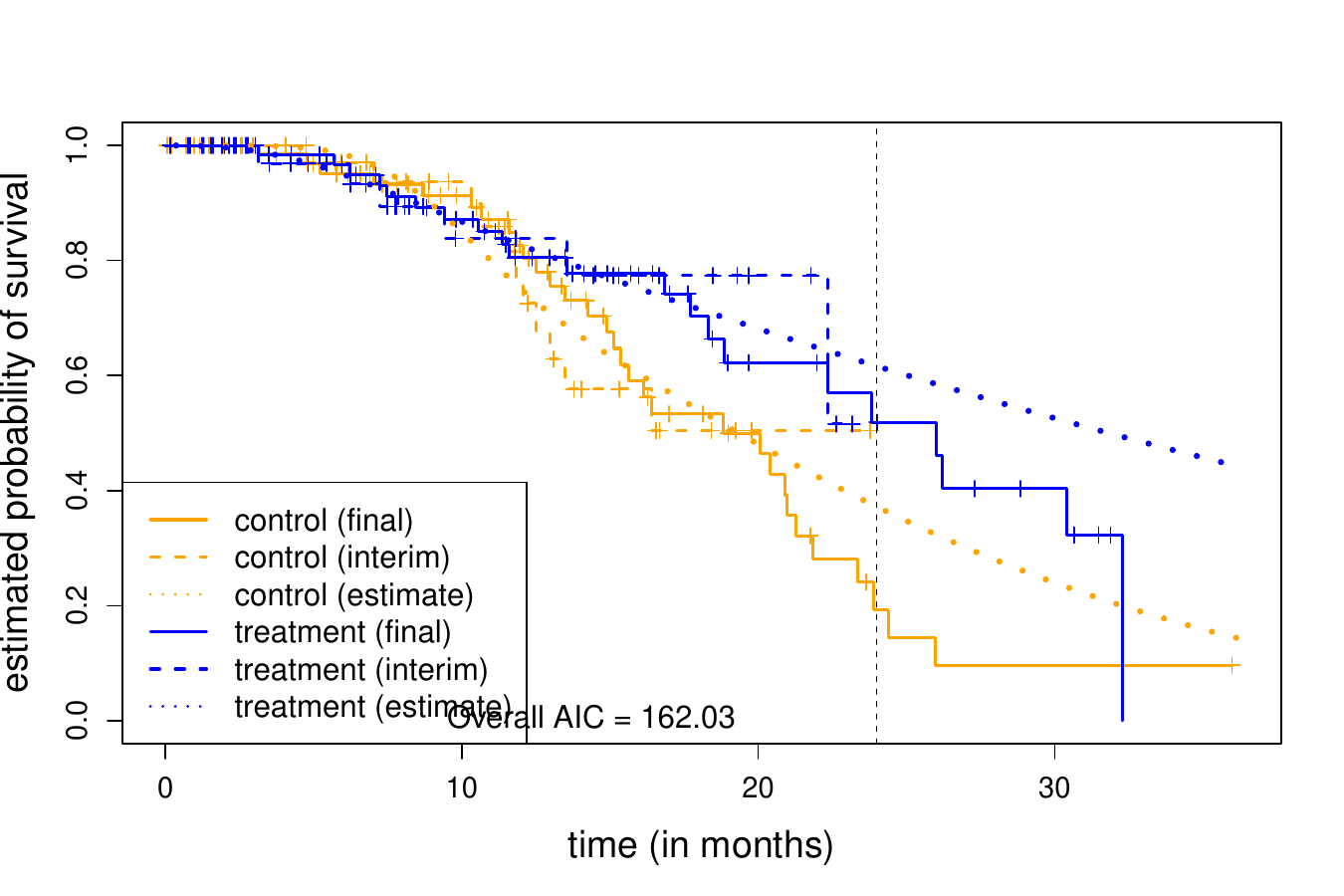}
    \caption{$p=1$, hazard scale}
\end{subfigure}
\hfill
\begin{subfigure}{0.45\textwidth}
    \includegraphics[width=\textwidth]{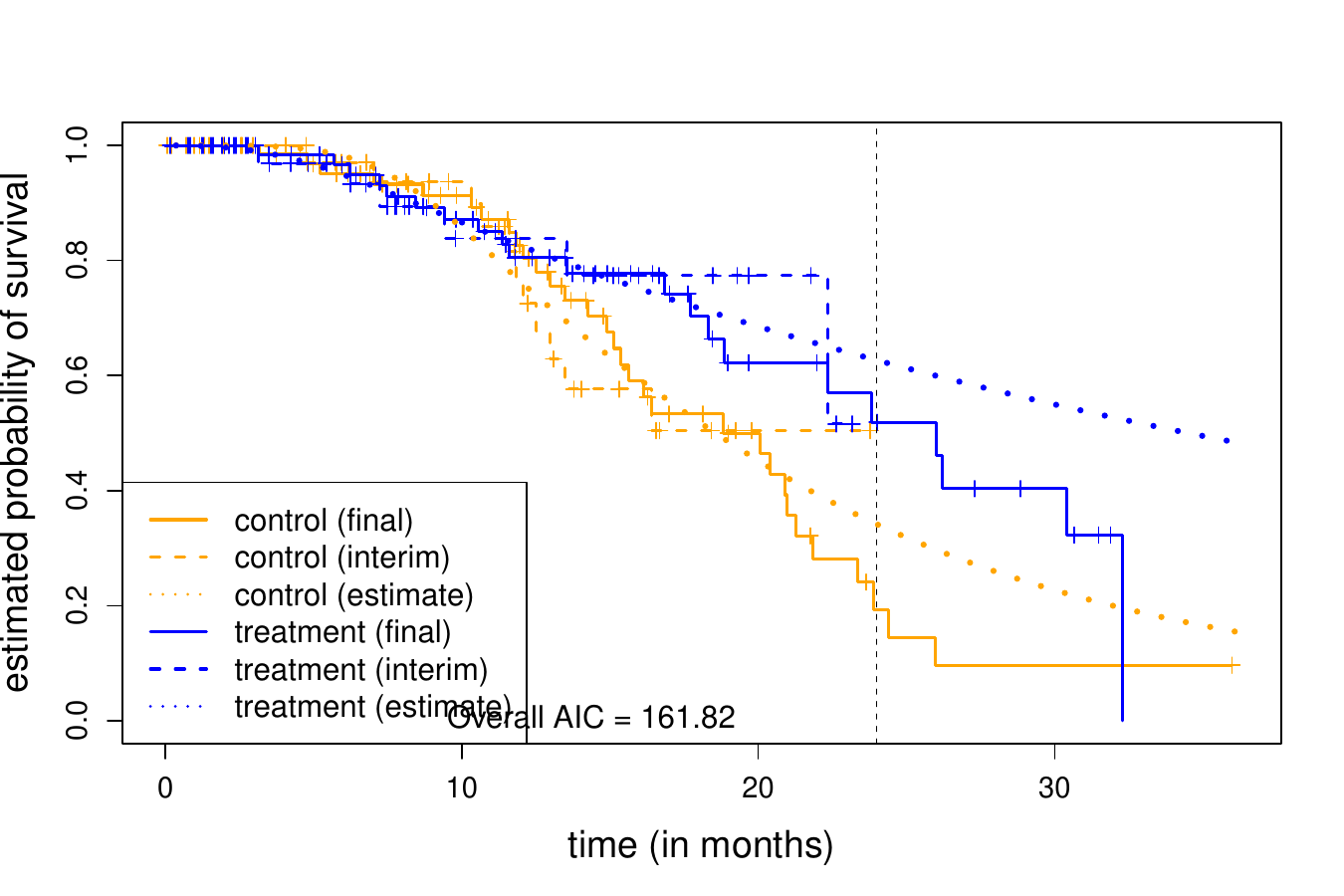}
    \caption{$p=1$, odds scale}
\end{subfigure}
\hfill
\begin{subfigure}{0.45\textwidth}
    \includegraphics[width=\textwidth]{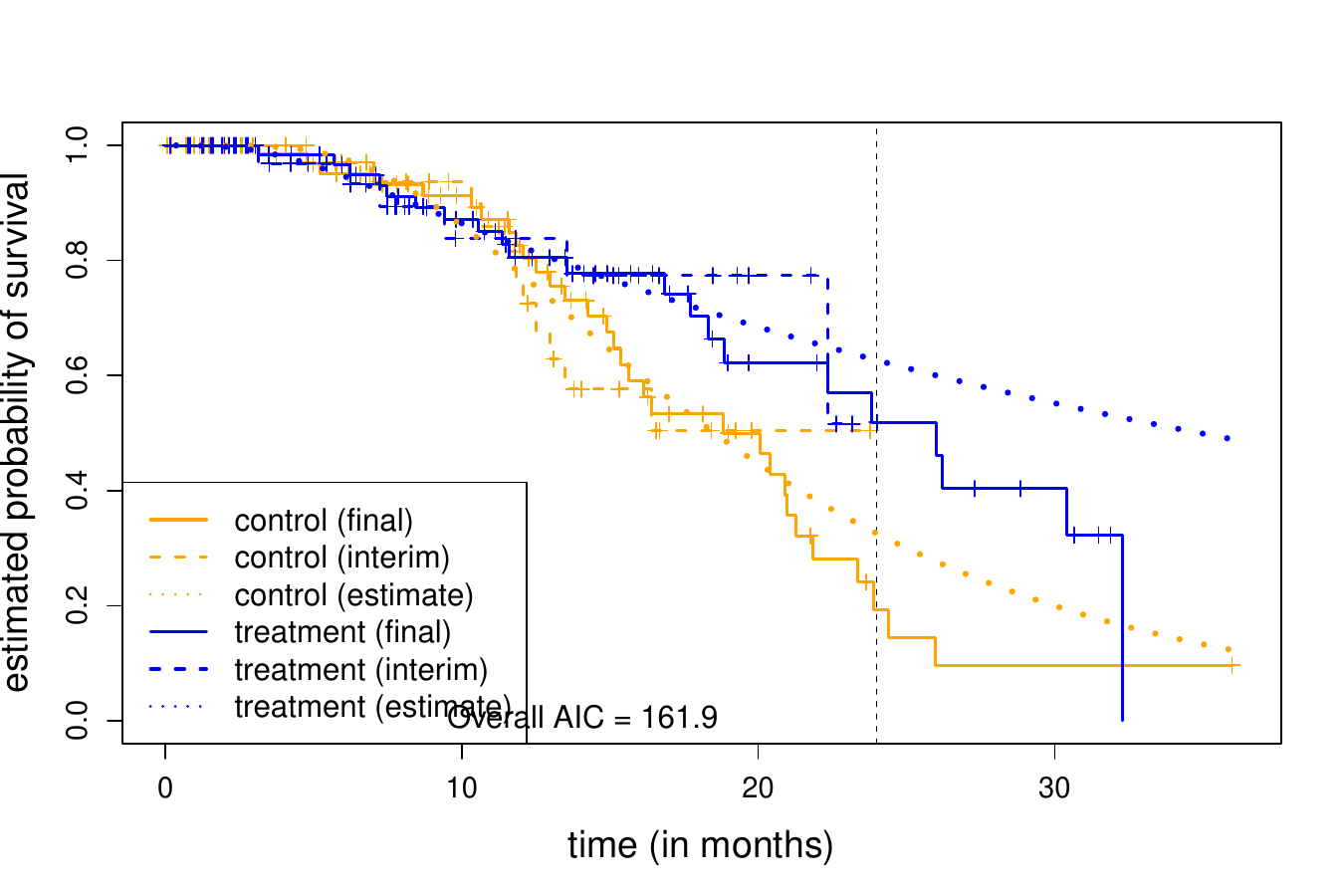}
    \caption{$p=1$, normal scale}
\end{subfigure}

%Third row (p=2)
\begin{subfigure}{0.45\textwidth}
    \includegraphics[width=\textwidth]{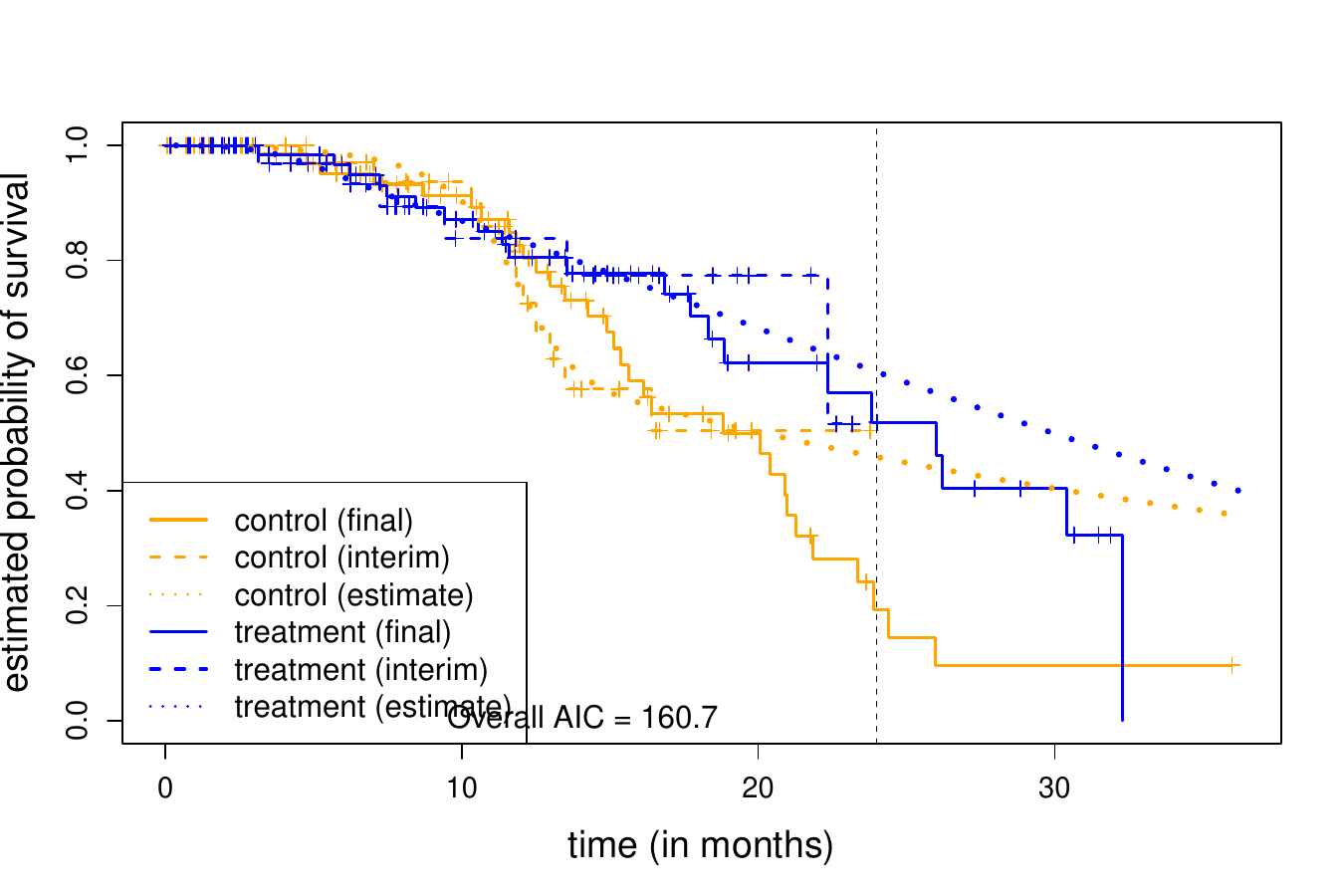}
    \caption{$p=2$, hazard scale}
\end{subfigure}
\hfill
\begin{subfigure}{0.45\textwidth}
    \includegraphics[width=\textwidth]{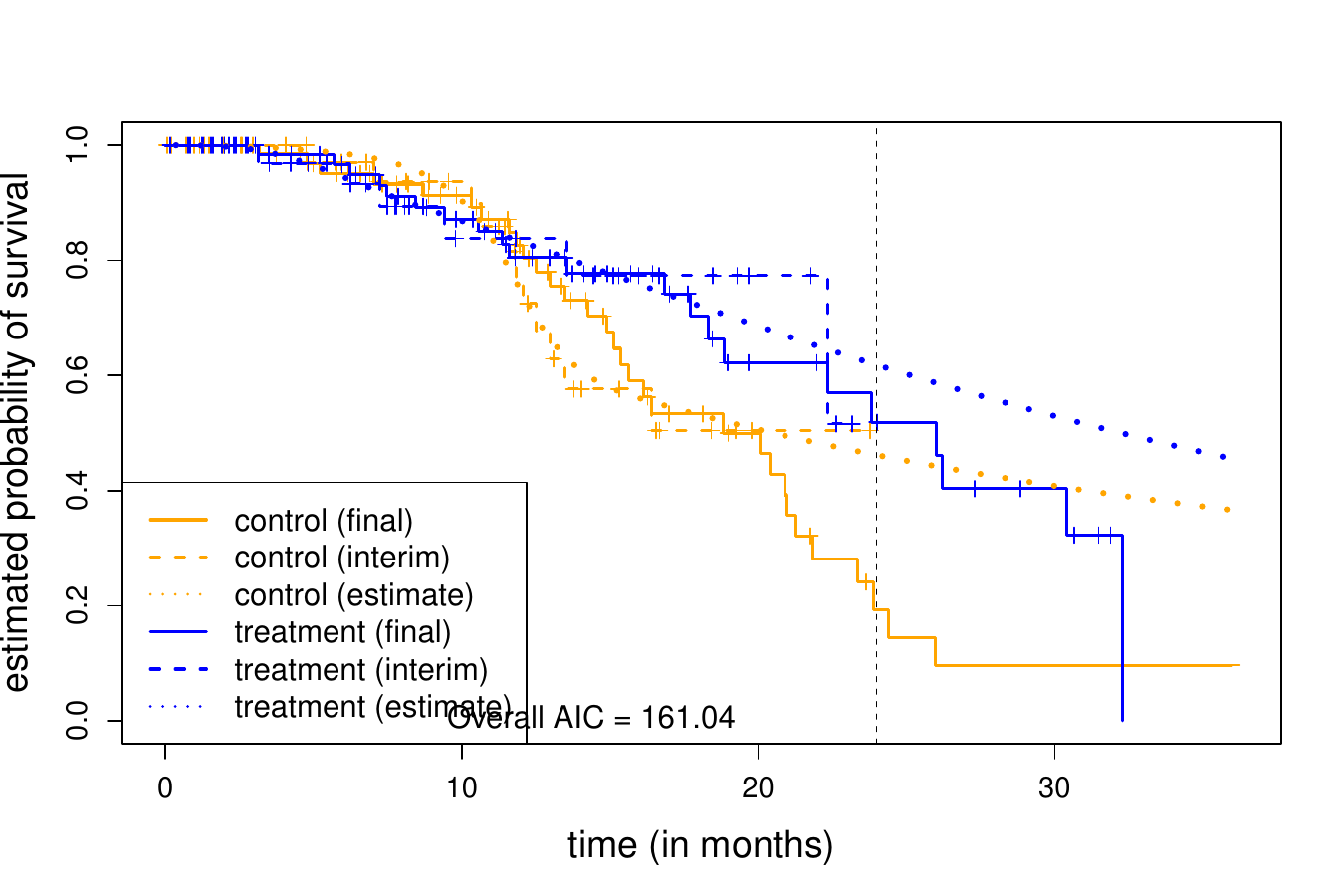}
    \caption{$p=2$, odds scale}
\end{subfigure}
\hfill
\begin{subfigure}{0.45\textwidth}
    \includegraphics[width=\textwidth]{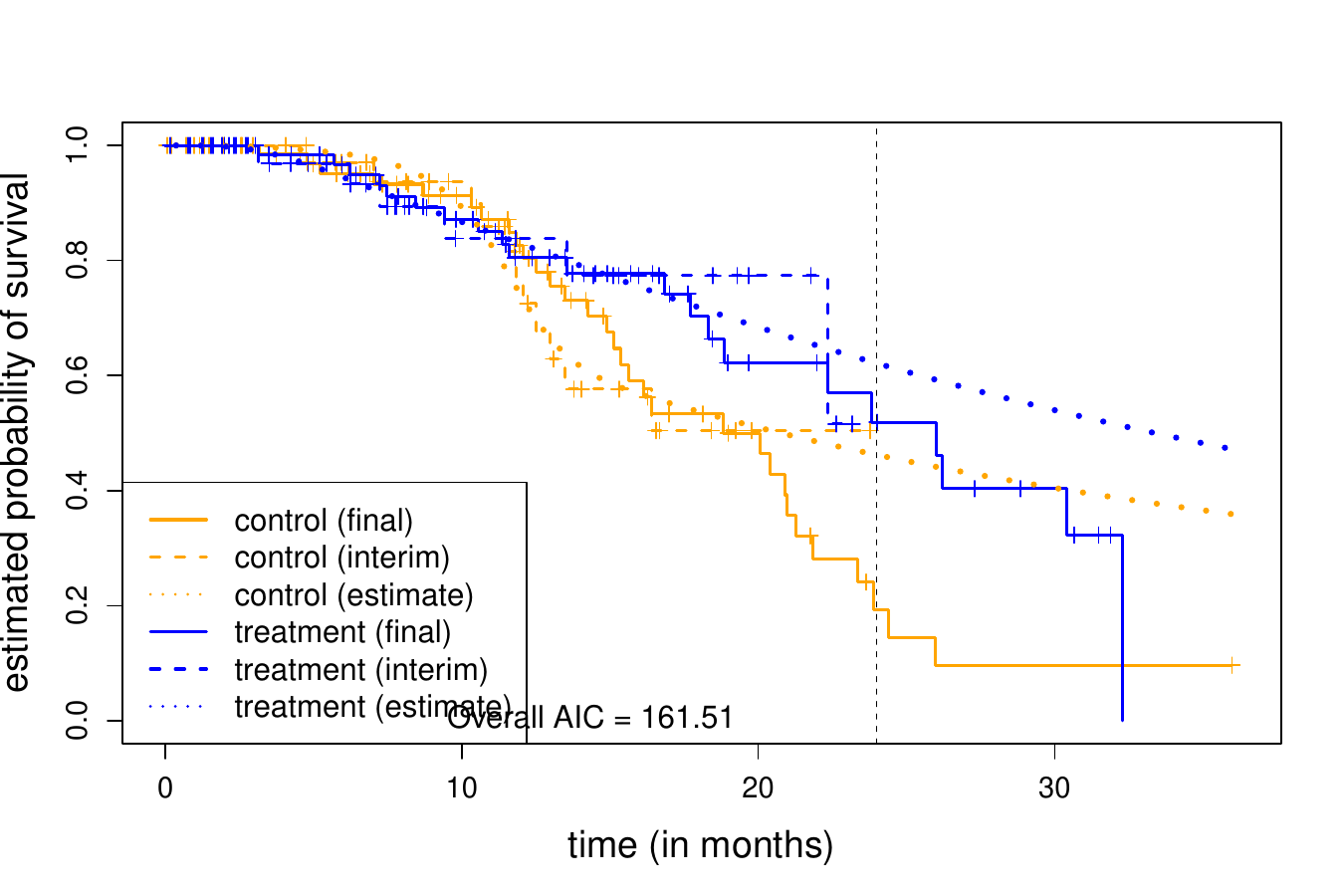}
    \caption{$p=2$, normal scale}
\end{subfigure}

\caption{Fits of Royston-Parmar spline models to interim data with additional display of the final Kaplan-Meier estimates.}
\label{supp-fig:interim_fits_ep}
\end{figure}

\end{landscape}

After administrative censoring at calendar time $t_1=24$ was applied to the reconstructed and simulated data, the Kaplan-Meier estimates in Figure \ref{supp-fig:km_interim} can be obtained. 
Royston-Parmar splines were fitted to this data for each group separately using the function \texttt{flexsurvspline} from the R package \texttt{flexsurv} \cite{Jackson:2016}. The Nelder-Mead optimization method has been chosen. Models with the number of interior knots $p \in \{0,1,2\}$ and on all three available scales (hazard, odds, normal) were fitted. For each fit, the combined AIC was computed. The values can be found in Table \ref{supp-table:rp_aic}. The lowest AIC is achieved for $p=0$ on the normal scale. This results in a log-normal distribution \cite{Royston:2002}. It would also be possible to choose a different $p$ and scale for the two groups. However, we restricted ourselves to the application of the same modeling parameters for both groups.
On the next page, in Figure \ref{supp-fig:interim_fits}, the fitted Royston-Parmar spline models for all considered configurations are shown. On the page following afterwards (Figure \ref{supp-fig:interim_fits_ep}), the Kaplan-Meier estimate based on the complete data is also shown in order to display the extrapolation performance.
\begin{table}[h]
    \centering
    \begin{tabular}{|c|c|c|c|c|}
         \cline{3-5}
         \multicolumn{2}{c|}{} & \multicolumn{3}{c|}{scale}\\
         \cline{3-5}
         \multicolumn{2}{c|}{}&hazard&odds&normal\\
         \hline
         \multirow{3}{*}{$p$}&0&160.31&158.65&157.97\\
         \cline{2-5}
         &1&162.03&161.82&162.03\\
         \cline{2-5}
         &2&160.70&161.04&161.51\\
         \hline
    \end{tabular}
    \caption{AIC values for various Royston-Parmar spline models when fitted to the interim data}
    \label{supp-table:rp_aic}
\end{table}

\subsection{Conditional power calculations}
Here, we present the results of the conditional power calculations. We consider
\begin{itemize}
    \item 8 different mdir combination tests for the first stage (as presented in Table \ref{table:example_p} in the main manuscript); the sets of weights used in these tests will be indexed as follows
    \begin{description}
        \item[$\mathcal{Q}_{\text{mdir},1}=$] $\{(0,0),(1,0),(0,1),(1,1)\}$
        \item[$\mathcal{Q}_{\text{mdir},2}=$] $\{(0,0),(1,0),(0,1)\}$
        \item[$\mathcal{Q}_{\text{mdir},3}=$] $\{(0,0),(1,0),(1,1)\}$
        \item[$\mathcal{Q}_{\text{mdir},4}=$] $\{(0,0),(0,1),(1,1)\}$
        \item[$\mathcal{Q}_{\text{mdir},5}=$] $\{(0,0),(1,0)\}$
        \item[$\mathcal{Q}_{\text{mdir},6}=$] $\{(0,0),(0,1)\}$
        \item[$\mathcal{Q}_{\text{mdir},7}=$] $\{(0,0),(1,1)\}$
        \item[$\mathcal{Q}_{\text{mdir},8}=$] $\{(0,0)\}$
    \end{description}
    \item 9 different parameter constellations to fit Royston-Parmar splines to the interim data (number of knots $p \in \{0,1,2\}$, hazard, odds or normal scale)
    \item 4 different single-weighted tests for the second stage
\end{itemize}
For each combination of the mdir combination tests in the first stage and parameter constellation of the Royston-Parmar spline mode, the weighted test achieving the highest conditional power is marked in bold. Obviously, the choice does not depend on the first stage test statistic, as it is a monotone function of the standardized drift that is computed for each of the weighted tests in the second stage.\\
It is remarkable that for most of the choices of parameters for the Royston-Parmar splines (6 out of 9), the $(1,1)$-weighted test is favoured. This test actually has the best performance (see main manuscript). However, the test with the worst performance (weight $(1,0)$) is also chosen once. In the remainig two cases, the $(0,1)$-weighted test is favoured.

\begin{landscape}

\begin{table}[h]
    \centering
    \begin{tabular}{|c|c||c|c|c|c||c|c|c|c||c|c|c|c|}
         \cline{3-14}
         \multicolumn{2}{c|}{} & \multicolumn{4}{c||}{hazard scale} & \multicolumn{4}{c||}{odds scale} & \multicolumn{4}{c||}{normal scale}\\
         \cline{3-14}
         \hline
         \multicolumn{2}{|c|}{2nd stage weight}&$(0,0)$&$(1,0)$&$(0,1)$&$(1,1)$&$(0,0)$&$(1,0)$&$(0,1)$&$(1,1)$&$(0,0)$&$(1,0)$&$(0,1)$&$(1,1)$\\
         \hline
         \hline
         \multirow{8}{*}{$p=0$}&$\mathcal{Q}_{\text{mdir},1}$ & 0.8089& 0.6396& \textbf{0.8868}& 0.87& 0.7537& 0.6233& 0.8105& \textbf{0.8295}& 0.7052& 0.5649& 0.7955& \textbf{0.8033}\\
         &$\mathcal{Q}_{\text{mdir},2}$ & 0.7653& 0.5821& \textbf{0.8553}& 0.8356& 0.704& 0.5651& 0.7671& \textbf{0.7888}& 0.6514& 0.5052& 0.7503& \textbf{0.7591}\\
         &$\mathcal{Q}_{\text{mdir},3}$& 0.8307& 0.6702& \textbf{0.9019}& 0.8868& 0.7791& 0.6543& 0.8321& \textbf{0.8497}& 0.7331& 0.5972& 0.8182& \textbf{0.8255}\\ 
         &$\mathcal{Q}_{\text{mdir},4}$ & 0.8191& 0.6538& \textbf{0.8939}& 0.8779& 0.7655& 0.6376& 0.8206& \textbf{0.839}& 0.7182& 0.5798& 0.8061& \textbf{0.8137}\\ 
         &$\mathcal{Q}_{\text{mdir},5}$ & 0.7851& 0.6076& \textbf{0.8698}& 0.8513& 0.7263& 0.5908& 0.7867& \textbf{0.8073}& 0.6754& 0.5314& 0.7707& \textbf{0.7791}\\ 
         &$\mathcal{Q}_{\text{mdir},6}$ & 0.7411& 0.5519& \textbf{0.8371}& 0.8158& 0.6769& 0.5347& 0.743& \textbf{0.7659}& 0.6226& 0.4746& 0.7253& \textbf{0.7345}\\ 
         &$\mathcal{Q}_{\text{mdir},7}$ & 0.8384& 0.6814& \textbf{0.9072}& 0.8927& 0.7882& 0.6658& 0.8398& \textbf{0.8569}& 0.7433& 0.6093& 0.8264& \textbf{0.8334}\\ 
         &$\mathcal{Q}_{\text{mdir},8}$ & 0.7605& 0.5759& \textbf{0.8517}& 0.8316& 0.6985& 0.5588& 0.7622& \textbf{0.7842}& 0.6455& 0.4989& 0.7453& \textbf{0.7541}\\ 
         \hline
         \hline
         \multirow{8}{*}{$p=1$}&$\mathcal{Q}_{\text{mdir},1}$ & 0.598& 0.4835& 0.6715& \textbf{0.6932}& 0.679& 0.5493& 0.7718& \textbf{0.7878}& 0.74& 0.5875& \textbf{0.8501}& 0.85\\ 
         &$\mathcal{Q}_{\text{mdir},2}$ & 0.539& 0.4241& 0.6156& \textbf{0.6386}& 0.6235& 0.4895& 0.7239& \textbf{0.7417}& 0.6891& 0.5282& \textbf{0.8123}& 0.8123\\ 
         &$\mathcal{Q}_{\text{mdir},3}$ & 0.6297& 0.5166& 0.7009& \textbf{0.7217}& 0.7081& 0.5819& 0.7961& \textbf{0.811}& 0.7662& 0.6194& \textbf{0.8686}& 0.8685\\ 
         &$\mathcal{Q}_{\text{mdir},4}$ & 0.6126& 0.4987& 0.6852& \textbf{0.7065}& 0.6925& 0.5643& 0.7831& \textbf{0.7987}& 0.7522& 0.6022& \textbf{0.8588}& 0.8587\\ 
         &$\mathcal{Q}_{\text{mdir},5}$ & 0.565& 0.45& 0.6405& \textbf{0.663}& 0.6482& 0.5157& 0.7455& \textbf{0.7625}& 0.7119& 0.5543& \textbf{0.8295}& 0.8295\\ 
         &$\mathcal{Q}_{\text{mdir},6}$ & 0.5084& 0.3942& 0.5859& \textbf{0.6095}& 0.594& 0.4589& 0.6977& \textbf{0.7163}& 0.6615& 0.4976& \textbf{0.791}& 0.7909\\ 
         &$\mathcal{Q}_{\text{mdir},7}$ & 0.6415& 0.5291& 0.7117& \textbf{0.7321}& 0.7188& 0.5941& 0.8048& \textbf{0.8194}& 0.7757& 0.6313& \textbf{0.8751}& 0.8751\\ 
         &$\mathcal{Q}_{\text{mdir},8}$ & 0.5327& 0.4179& 0.6095& \textbf{0.6327}& 0.6175& 0.4832& 0.7186& \textbf{0.7366}& 0.6835& 0.5219& \textbf{0.8081}& 0.808\\ 
         \hline
         \hline
         \multirow{8}{*}{$p=2$}&$\mathcal{Q}_{\text{mdir},1}$ & 0.1992& \textbf{0.222}& 0.1167& 0.191& 0.2345& 0.2411& 0.1687& \textbf{0.2442}& 0.2532& 0.2496& 0.2033& \textbf{0.2743}\\ 
         &$\mathcal{Q}_{\text{mdir},2}$ & 0.1599& \textbf{0.1799}& 0.0898& 0.1528& 0.1909& 0.1968& 0.1336& \textbf{0.1996}& 0.2076& 0.2044& 0.1636& \textbf{0.2266}\\ 
         &$\mathcal{Q}_{\text{mdir},3}$ & 0.2231& \textbf{0.2474}& 0.1338& 0.2143& 0.2607& 0.2677& 0.1904& \textbf{0.271}& 0.2805& 0.2767& 0.2276& \textbf{0.3026}\\ 
         &$\mathcal{Q}_{\text{mdir},4}$ & 0.21& \textbf{0.2334}& 0.1243& 0.2015& 0.2463& 0.2531& 0.1785& \textbf{0.2563}& 0.2655& 0.2618& 0.2143& \textbf{0.2872}\\ 
         &$\mathcal{Q}_{\text{mdir},5}$ & 0.1764& \textbf{0.1977}& 0.101& 0.1688& 0.2094& 0.2156& 0.1483& \textbf{0.2185}& 0.227& 0.2236& 0.1803& \textbf{0.2469}\\ 
         &$\mathcal{Q}_{\text{mdir},6}$ & 0.1419& \textbf{0.1604}& 0.078& 0.1354& 0.1707& 0.1762& 0.1177& \textbf{0.1788}& 0.1864& 0.1833& 0.1453& \textbf{0.2042}\\ 
         &$\mathcal{Q}_{\text{mdir},7}$ & 0.2326& \textbf{0.2574}& 0.1407& 0.2236& 0.2709& 0.2781& 0.199& \textbf{0.2814}& 0.2911& 0.2872& 0.2371& \textbf{0.3136}\\ 
         &$\mathcal{Q}_{\text{mdir},8}$ & 0.1561& \textbf{0.1758}& 0.0873& 0.1491& 0.1867& 0.1925& 0.1302& \textbf{0.1952}& 0.2032& 0.2& 0.1597& \textbf{0.2219}\\ 
         \hline
    \end{tabular}
    \caption{Conditional power for different choices of $\mathcal{Q}_{\text{mdir}}$ single weighted log-rank test for the second stage test for different parameter configurations of the Royston-Parmar spline model}
    \label{supp-table:cond_power}
\end{table}

\end{landscape}

\subsection{Dependence from simulated recruitment data}\label{supp-subsec:sim_dependence}
The results in Section \ref{sec:example} and in the previous subsections of course depend on the simulated recruitment data for patients with uncensored event time data.\\
Here, we want to assess the dependence from this simulation. Therefore, we repeat the procedure 10,000 times. Each time, interim data is analysed with the 8 different mdir combination tests mentioned above and 9 different Royston-Parmar spline models are fitted to the data. The second stage data is analysed with one of the for different single-weighted log-rank tests listed above. In Supplementary Table \ref{supp-table:power_comb}, the empirical power of the respective combinations of first- and second-stage tests is shown.

\begin{table}[h]
    \centering
    \begin{tabular}{|l||c|c|c|c|}
         \hline
         \multirow{2}{*}{first stage test} & \multicolumn{4}{c|}{second stage test}\\
         \cline{2-5}
         &$(0,0)$&$(0,1)$&$(1,0)$&$(1,1)$\\
         \hline
         $\mathcal{Q}_{\text{mdir},1}$&0.5492&0.3827&0.3596&0.8997\\
         $\mathcal{Q}_{\text{mdir},2}$&0.5297&0.3782&0.3478&0.8859\\
         $\mathcal{Q}_{\text{mdir},3}$&0.5994&0.4228&0.4032&0.9147\\
         $\mathcal{Q}_{\text{mdir},4}$&0.5380&0.3594&0.3355&0.8942\\
         $\mathcal{Q}_{\text{mdir},5}$&0.5768&0.4169&0.3924&0.9044\\
         $\mathcal{Q}_{\text{mdir},6}$&0.0769&0.0042&0.1570&0.6228\\
         $\mathcal{Q}_{\text{mdir},7}$&0.5915&0.4040&0.3800&0.9108\\
         $\mathcal{Q}_{\text{mdir},8}$&0.0890&0.0081&0.1770&0.7065\\
         \hline
    \end{tabular}
    \caption{Empirical power of the simulation study for all combinations of first stage combination testing procedures and single-weighted log-rank testing procedures in the second stage.}
    \label{supp-table:power-comb}
\end{table}
Please note that the combination of $\mathcal{Q}_{\text{mdir},8}$ in the first stage and the $(0,0)$-weighted log-rank test in the second stage basically constitutes a two-stage standard log-rank test as in \cite{Wassmer:2006}. In about 9\% of all simulation runs, this leads to a rejection although the originally applied simple one-stage standard log-rank leads to a rejection.\\
Power can strongly be increased if a different test is chosen in the first stage and the $(1,1)$-weighted log-rank test is chosen for the second stage. However, this is a retrospective assessment and one cannot guarantee a better choice for the first-stage test. Nevertheless, if a deviation from proportional hazards is anticipated, it is reasonable to consider a different test for the first stage.\\
For the second stage, we consider a choice based on an extrapolation with Royston-Parmar splines, a model selection based on AIC values and a consecutive conditional power calculation. In Supplementary Table \ref{supp-table:aic_winner}, it is shown how often the 9 Royston-Parmar spline models used in this example so far are chosen based on the AIC values.
\begin{table}[h]
    \centering
    \begin{tabular}{|c|c|c|c|c|}
         \cline{3-5}
         \multicolumn{2}{c|}{} & \multicolumn{3}{c|}{scale}\\
         \cline{3-5}
         \multicolumn{2}{c|}{}&hazard&odds&normal\\
         \hline
         \multirow{3}{*}{$p$}&0&0.1869&0.1741&0.4637\\
         \cline{2-5}
         &1&0.0258&0.0032&0.0541\\
         \cline{2-5}
         &2&0.0824&0.0011&0.0087\\
         \hline
    \end{tabular}
    \caption{Relative frequency with which the various Royston-Parmar spline models are selected based on the AIC}
    \label{supp-table:aic_winner}
\end{table}
Based on these choices, the second stage tests considered here, are chosen with the following frequencies if $\mathcal{Q}_{\text{mdir},8}$ is applied in the first stage: $(0,0)$: 0.1755; $(1,0)$: 0.4293; $(0,1)$: 0.1258; $(1,1)$: 0.2694. It seems like the chosen model prefers the $(1,0)$-weighted test although the performance for the weight $(1,1)$ is much better according to Supplementary Table \ref{supp-table:power-comb}.\\
Finally, we can evaluate the performance of this procedure in terms of the power. This needs to be done separately for each first stage test statistic. The results can be found in Supplementary Table \ref{supp-table:final_power_example}

\begin{table}[h]
    \centering
    \begin{tabular}{|l|c c c c c c c c|}
         \hline
         first stage test & $\mathcal{Q}_{\text{mdir},1}$ & $\mathcal{Q}_{\text{mdir},2}$ & $\mathcal{Q}_{\text{mdir},3}$ & $\mathcal{Q}_{\text{mdir},4}$ & $\mathcal{Q}_{\text{mdir},5}$ & $\mathcal{Q}_{\text{mdir},6}$ & $\mathcal{Q}_{\text{mdir},7}$ & $\mathcal{Q}_{\text{mdir},8}$\\
         \hline
         empirical power & 0.4802 & 0.4699 & 0.5103 & 0.4613 & 0.5027 & 0.2836 & 0.4915 & 0.3057\\
         \hline
    \end{tabular}
    \caption{Relative frequency with which the various Royston-Parmar spline models are selected based on the AIC}
    \label{supp-table:final_power_example}
\end{table}
One can see that the power can be increased by about 6 percentage points even if the standard log-rank test is chosen in the first stage. If the first stage test statistic is chosen appropriately, it can even raise to about 50\%.\\
In Supplementary Figure \ref{supp-fig:cp_distribution_example}, we show the empirical distribution of the maximal conditional power among the four tests for the second stage, computed based on the Royston-Parmar spline model with the lowest AIC. We restrict ourselves to the case that the weights in $\mathcal{Q}_{\text{mdir},1}$ have been chosen for the first stage. Correpsponding histograms for other choices for the first stage are very similar. We encounter a well-known problem in conditional power calculation in adaptive designs. This is characterised by the fact that the distribution of this variable tends towards extremes, i.e. very large and very small values \cite{Bauer:2006}. It could therefore make sense to consider alternative concepts here.
\begin{figure}[ht]
    \centering
    \includegraphics[width = .66\textwidth]{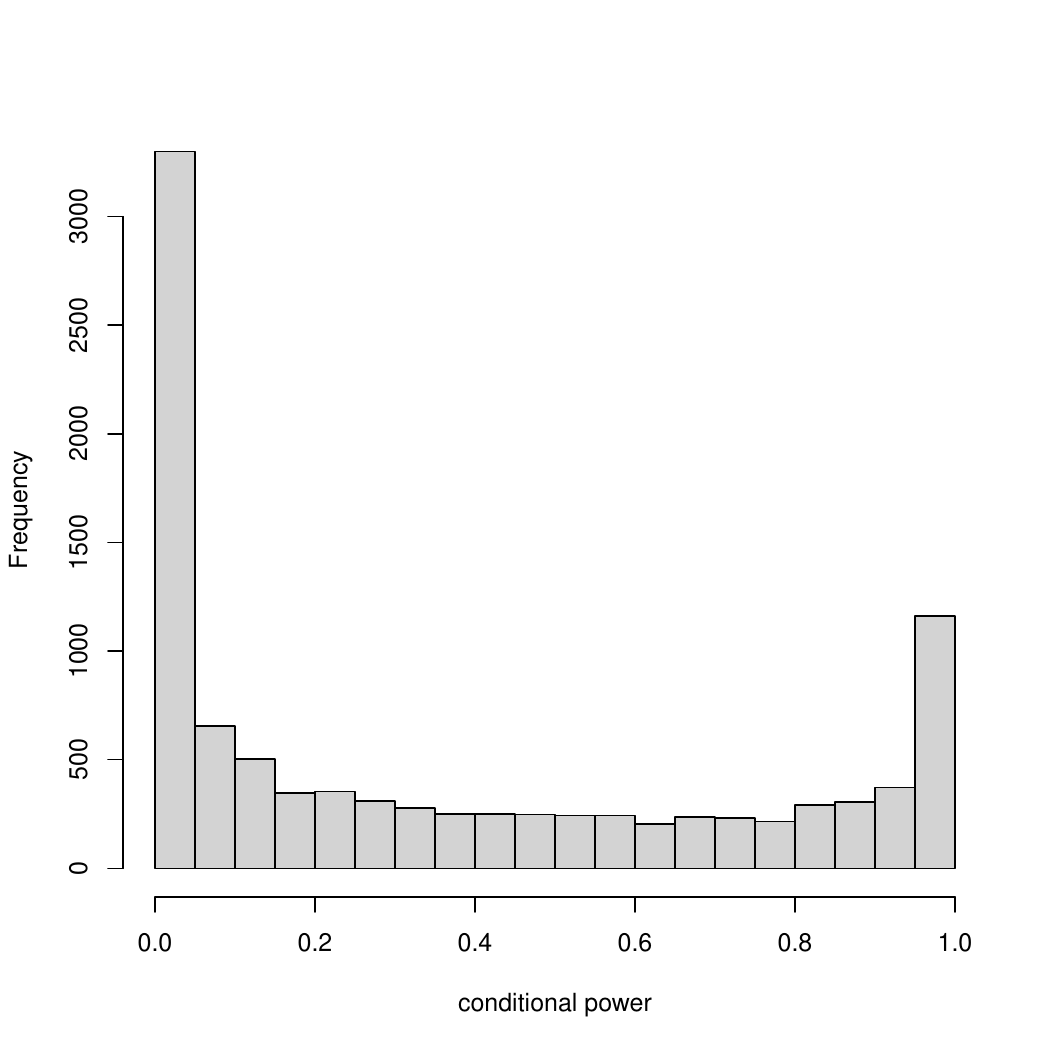}
    \caption{Empirical distribution of the maximal conditional power that can be obtained for the four candidate tests for the second stage. The first stage test is an mdir combination test with weights in $\mathcal{Q}_{\text{mdir},1}$. The Royston-Parmar spline model is chosen based on the AIC.}
    \label{supp-fig:cp_distribution_example}
\end{figure}

\subsection{Application of modestly weighted log-rank tests}\label{supp-subsec:modest_weights}

\begin{figure}[h]
    \centering
        \includegraphics[width = \textwidth]{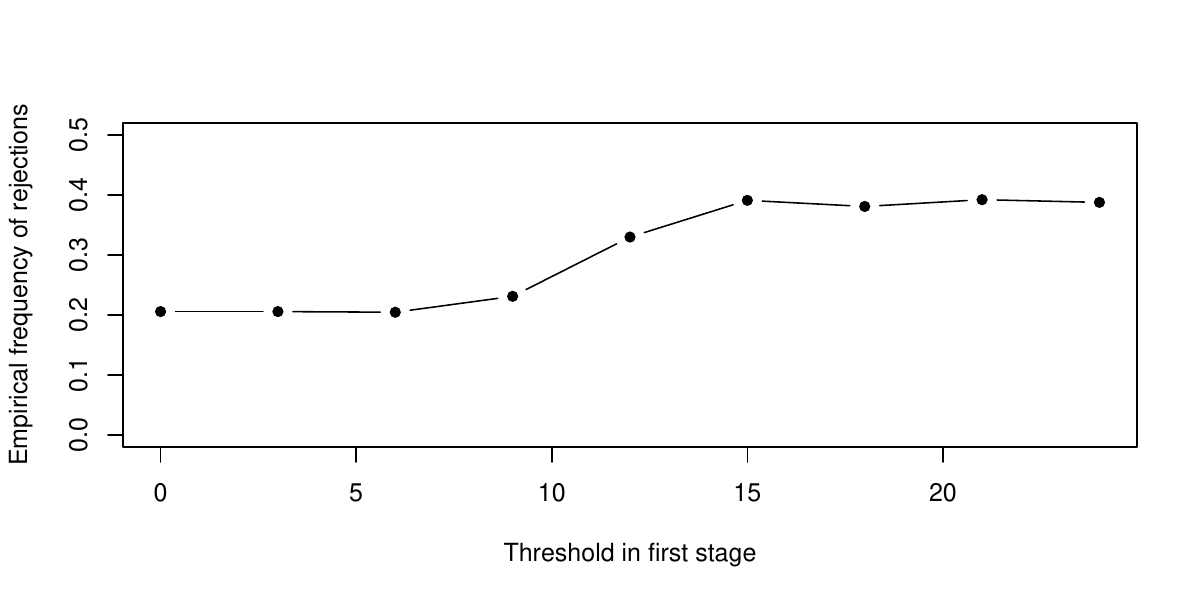}
        \caption{Empirical power for an adpaitve design with a modestly weighted log-rank test in the first stage and a modestly weighted log-rank test in the second stage that is chosen based on conditional power consideration based on Royston-Parmar spline models. 9 different modestly weighted log-rank tests are considered. The threshold for the (fixed) test in the first stage is given by the x-axis.}
        \label{supp-fig:modest_power}
\end{figure}

As an additional analysis, we conduct a similar analysis to the one of Section \ref{supp-subsec:sim_dependence} with a different class of weights. The class of Fleming-Harrington weight has been criticized for its undesirable properties when applied in a one-sided testing procedure \cite{Magirr:2023}. This applies equally when combining them in a combination testing procedure. In this context, Magirr and Burman suggested a different class of weights, which they termed "modest weights" \cite{Magirr:2019}.\\
The class of modest weights is parametrized by some threshold time $s^{\star} \geq 0$ and given by
\begin{equation}\label{eq:fh_weights}
	\hat{Q}(t,s)=w_{\text{modest}, s^{\star}}(\hat{S}(t,s-))\coloneqq \frac{1}{\max\left( \hat{S}(t,s-), \hat{S}(t,s^{\star}-) \right)}.
\end{equation}
As the weights of Fleming and Harrington with $\rho > 0$ and $\gamma = 0$, this function is increasing in the second argument. Differently from these functions, it is bounded from below and it stays constant after the second argument exceeds the threshold $s^{\star}$. As previously shown, are an appropriate tool to detect late effects \cite{Magirr:2019}. Notably, for $s^{\star}=0$ the modestly weighted log-rank test is the same as the standard log-rank test as it is constantly equal to 1. The modest weights also fulfill the assumptions of Theorem \ref{supp-thm:asymptotic_equivalence} and can hence also be incorporated into our framework.\\
Here, we consider designs where a single modestly weighted log-rank test is applied in the first stage and the second test stage is chosen among the modestly weighted tests with the thresholds $s^{\star} \in \{0,3,6,9,12,15,18,21,24\}$. The modest weights could also be used in a combination testing procedure as e.g. the \textit{mdir} test. However, we do not consider this option here.\\
As previously, 10,000 simulation runs are made where the unreconstructable recruitment dates are simulated. Conditional power calculations for the 9 different modestly weighted tests are made from the same 9 different Royston-Parmar spline models as above. The best spline model is chosen based on the AIC and the second stage test is chosen as the modestly weighted test with the highest conditional power according to this spline model. The results in terms of the empirical power of the procedure in dependence from the (fixed) test in the first stage are displayed in Figure \ref{supp-fig:modest_power}.\\
For comparison, the empirical rejection rate for the two-stage procedure which applies the standard log-rank test in both stages (i.e. $s^{\star}=0$) is 9.31\%. It is obvious that the selection procedure increases the rejection rate, even if the standard log-rank test is chosen for the first stage. The increase is even more articulate if a better-suited test is chosen in the first stage. 

\newpage

\section{Additional simulation results}
In this section, we provide additional results to the simulation study in Section \ref{sec:sim_study} of the main manuscript.

\subsection{Empirical type I error rates}
Here, we present the empirical type I error rates that emerge from fixed (in particular non-adaptive) combinations of testing procedures in the two stages. This shall demonstrate that our selection procedure (see Table \ref{table:type_1_error_rates}) induces no additional type I error rate inflation.\\
For the sake of presentability, we use the following numbering system for the various tests:
\begin{enumerate}[label = \arabic*:]
    \item Standard log-rank test
    \item Weighted log-rank test with Fleming-Harrington weight $w^{(0,1)} \circ \hat{F}$
    \item Weighted log-rank test with Fleming-Harrington weight $w^{(0,2)} \circ \hat{F}$
    \item Weighted log-rank test with Fleming-Harrington weight $w^{(0,3)} \circ \hat{F}$
    \item Weighted log-rank test with Fleming-Harrington weight $w^{(1,1)} \circ \hat{F}$
    \item Weighted log-rank test with Fleming-Harrington weight $w^{(1,0)} \circ \hat{F}$
    \item Weighted log-rank test with Fleming-Harrington weight $w^{(2,0)} \circ \hat{F}$
    \item Weighted log-rank test with Fleming-Harrington weight $w^{(3,0)} \circ \hat{F}$
    \item \textit{mdir} combination test based on the set of weights $\{w^{(0,0)} \circ \hat{F}, w^{(1,0)} \circ \hat{F}\}$
    \item \textit{mdir} combination test based on the set of weights $\{w^{(0,0)} \circ \hat{F}, w^{(1,1)} \circ \hat{F}\}$
    \item \textit{mdir} combination test based on the set of weights $\{w^{(0,0)} \circ \hat{F}, w^{(0,1)} \circ \hat{F}\}$
    \item \textit{mdir} combination test based on the set of weights $\{w^{(0,0)} \circ \hat{F}, w^{(1,0)} \circ \hat{F}, w^{(0,1)} \circ \hat{F}\}$
    \item \textit{mdir} combination test based on the set of weights $\{w^{(0,0)} \circ \hat{F}, w^{(1,0)} \circ \hat{F}, w^{(1,1)} \circ \hat{F}\}$
    \item \textit{mdir} combination test based on the set of weights $\{w^{(0,0)} \circ \hat{F}, w^{(1,1)} \circ \hat{F}, w^{(0,1)} \circ \hat{F}\}$
    \item \textit{mdir} combination test based on the set of weights $\{w^{(0,0)} \circ \hat{F}, w^{(1,0)} \circ \hat{F}, w^{(1,1)} \circ \hat{F}, w^{(0,1)} \circ \hat{F}\}$

\end{enumerate}

\begin{landscape}

\begin{table}
\footnotesize
    \centering
    \begin{tabular}{|l|c|c|c|c|c|c|c|c|c|c|c|c|c|c|c|}
    \hline
    &$1$&$2$&$3$&$4$&$5$&$6$&$7$&$8$&$9$&$10$&$11$&$12$&$13$&$14$&$15$\\
    \hline
    $ 1 $&$ 0.028 $&$ 0.029 $&$ 0.0302 $&$ 0.0317 $&$ 0.0274 $&$ 0.027 $&$ 0.0263 $&$ 0.0259 $&$ 0.0241 $&$ 0.0255 $&$ 0.0259 $&$ 0.0254 $&$ 0.0255 $&$ 0.0262 $&$ 0.0251 $\\
    \hline
    $ 2 $&$ 0.0301 $&$ 0.0302 $&$ 0.0316 $&$ 0.0323 $&$ 0.0308 $&$ 0.0292 $&$ 0.0289 $&$ 0.0282 $&$ 0.0279 $&$ 0.028 $&$ 0.0275 $&$ 0.0274 $&$ 0.0275 $&$ 0.0273 $&$ 0.0278 $\\
    \hline
    $ 3 $&$ 0.0313 $&$ 0.0318 $&$ 0.032 $&$ 0.0337 $&$ 0.0311 $&$ 0.0306 $&$ 0.0305 $&$ 0.0305 $&$ 0.0279 $&$ 0.0289 $&$ 0.0294 $&$ 0.0278 $&$ 0.0275 $&$ 0.0287 $&$ 0.028 $\\
    \hline
    $ 4 $&$ 0.0315 $&$ 0.0319 $&$ 0.0328 $&$ 0.034 $&$ 0.0317 $&$ 0.0297 $&$ 0.0305 $&$ 0.0312 $&$ 0.0286 $&$ 0.0298 $&$ 0.0293 $&$ 0.0287 $&$ 0.0294 $&$ 0.0299 $&$ 0.0284 $\\
    \hline
    $ 5 $&$ 0.0292 $&$ 0.0292 $&$ 0.0292 $&$ 0.0301 $&$ 0.0286 $&$ 0.0273 $&$ 0.0272 $&$ 0.0278 $&$ 0.0262 $&$ 0.0267 $&$ 0.0272 $&$ 0.0275 $&$ 0.0264 $&$ 0.0268 $&$ 0.0265 $\\
    \hline
    $ 6 $&$ 0.0269 $&$ 0.0273 $&$ 0.0302 $&$ 0.0313 $&$ 0.0269 $&$ 0.0263 $&$ 0.025 $&$ 0.025 $&$ 0.0256 $&$ 0.0253 $&$ 0.0259 $&$ 0.0264 $&$ 0.0252 $&$ 0.0263 $&$ 0.0258 $\\
    \hline
    $ 7 $&$ 0.0267 $&$ 0.0281 $&$ 0.0297 $&$ 0.0315 $&$ 0.0272 $&$ 0.0254 $&$ 0.0245 $&$ 0.0245 $&$ 0.0249 $&$ 0.0255 $&$ 0.0255 $&$ 0.0255 $&$ 0.0252 $&$ 0.0253 $&$ 0.0259 $\\
    \hline
    $ 8 $&$ 0.0269 $&$ 0.0277 $&$ 0.03 $&$ 0.0317 $&$ 0.0269 $&$ 0.0253 $&$ 0.0252 $&$ 0.0244 $&$ 0.0249 $&$ 0.0257 $&$ 0.0257 $&$ 0.0259 $&$ 0.0253 $&$ 0.0261 $&$ 0.0262 $\\
    \hline
    $ 9 $&$ 0.0258 $&$ 0.0268 $&$ 0.0274 $&$ 0.0283 $&$ 0.0255 $&$ 0.025 $&$ 0.0236 $&$ 0.0237 $&$ 0.0227 $&$ 0.0236 $&$ 0.0232 $&$ 0.0234 $&$ 0.0233 $&$ 0.0236 $&$ 0.0234 $\\
    \hline
    $ 10 $&$ 0.0267 $&$ 0.0275 $&$ 0.0278 $&$ 0.0285 $&$ 0.0261 $&$ 0.0257 $&$ 0.0253 $&$ 0.0256 $&$ 0.0231 $&$ 0.0241 $&$ 0.0248 $&$ 0.0242 $&$ 0.0241 $&$ 0.0251 $&$ 0.0246 $\\
    \hline
    $ 11 $&$ 0.0253 $&$ 0.0258 $&$ 0.0288 $&$ 0.0304 $&$ 0.0262 $&$ 0.0251 $&$ 0.0241 $&$ 0.0235 $&$ 0.0237 $&$ 0.0239 $&$ 0.0234 $&$ 0.0234 $&$ 0.0234 $&$ 0.0229 $&$ 0.0236 $\\
    \hline
    $ 12 $&$ 0.0257 $&$ 0.0272 $&$ 0.0282 $&$ 0.0291 $&$ 0.0265 $&$ 0.0252 $&$ 0.024 $&$ 0.0243 $&$ 0.0234 $&$ 0.0238 $&$ 0.0237 $&$ 0.0232 $&$ 0.0236 $&$ 0.0246 $&$ 0.0237 $\\
    \hline
    $ 13 $&$ 0.0263 $&$ 0.0268 $&$ 0.0276 $&$ 0.0275 $&$ 0.0267 $&$ 0.0257 $&$ 0.0245 $&$ 0.0254 $&$ 0.0232 $&$ 0.0241 $&$ 0.0242 $&$ 0.0241 $&$ 0.0238 $&$ 0.025 $&$ 0.024 $\\
    \hline
    $ 14 $&$ 0.0256 $&$ 0.0268 $&$ 0.0282 $&$ 0.0299 $&$ 0.0266 $&$ 0.0262 $&$ 0.0245 $&$ 0.0244 $&$ 0.0238 $&$ 0.0248 $&$ 0.0242 $&$ 0.0237 $&$ 0.0241 $&$ 0.0245 $&$ 0.0246 $\\
    \hline
    $ 15 $&$ 0.0258 $&$ 0.027 $&$ 0.0274 $&$ 0.0287 $&$ 0.0266 $&$ 0.0244 $&$ 0.0242 $&$ 0.0249 $&$ 0.0227 $&$ 0.0235 $&$ 0.0239 $&$ 0.0231 $&$ 0.0232 $&$ 0.0252 $&$ 0.0232 $\\
    \hline
    \end{tabular}
    \caption{Empirical type I error rates for fixed combinations of stagewise tests with 50 patients in each group. Rows correspond to first-stage tests and columns correspond to second-stage tests.}
    \label{supp-table:t1e_all_n50}
\end{table}

\begin{table}
\footnotesize
    \centering
    \begin{tabular}{|l|c|c|c|c|c|c|c|c|c|c|c|c|c|c|c|}
    \hline
    &$1$&$2$&$3$&$4$&$5$&$6$&$7$&$8$&$9$&$10$&$11$&$12$&$13$&$14$&$15$\\
    \hline
    $ 1 $&$ 0.0283 $&$ 0.0273 $&$ 0.0278 $&$ 0.0272 $&$ 0.0285 $&$ 0.0281 $&$ 0.0283 $&$ 0.0264 $&$ 0.0259 $&$ 0.0282 $&$ 0.0279 $&$ 0.0279 $&$ 0.0266 $&$ 0.0285 $&$ 0.0275 $\\
    \hline
    $ 2 $&$ 0.0264 $&$ 0.0274 $&$ 0.0258 $&$ 0.0258 $&$ 0.0269 $&$ 0.0261 $&$ 0.0267 $&$ 0.026 $&$ 0.0248 $&$ 0.0257 $&$ 0.0258 $&$ 0.0269 $&$ 0.0265 $&$ 0.0267 $&$ 0.027 $\\
    \hline
    $ 3 $&$ 0.0272 $&$ 0.0258 $&$ 0.0271 $&$ 0.028 $&$ 0.0275 $&$ 0.0271 $&$ 0.0254 $&$ 0.0256 $&$ 0.0259 $&$ 0.0274 $&$ 0.0271 $&$ 0.0288 $&$ 0.0277 $&$ 0.0277 $&$ 0.029 $\\
    \hline
    $ 4 $&$ 0.0281 $&$ 0.0264 $&$ 0.0273 $&$ 0.0283 $&$ 0.0284 $&$ 0.0275 $&$ 0.0258 $&$ 0.0261 $&$ 0.0278 $&$ 0.0287 $&$ 0.0286 $&$ 0.0285 $&$ 0.029 $&$ 0.0283 $&$ 0.0288 $\\
    \hline
    $ 5 $&$ 0.026 $&$ 0.0275 $&$ 0.027 $&$ 0.0273 $&$ 0.0272 $&$ 0.0266 $&$ 0.0276 $&$ 0.0265 $&$ 0.0256 $&$ 0.0262 $&$ 0.026 $&$ 0.0274 $&$ 0.0265 $&$ 0.0277 $&$ 0.0267 $\\
    \hline
    $ 6 $&$ 0.0276 $&$ 0.0266 $&$ 0.0267 $&$ 0.0263 $&$ 0.0283 $&$ 0.0294 $&$ 0.0277 $&$ 0.0277 $&$ 0.025 $&$ 0.0277 $&$ 0.0282 $&$ 0.0276 $&$ 0.0262 $&$ 0.0285 $&$ 0.027 $\\
    \hline
    $ 7 $&$ 0.0278 $&$ 0.0262 $&$ 0.0266 $&$ 0.0266 $&$ 0.0283 $&$ 0.0303 $&$ 0.0294 $&$ 0.0278 $&$ 0.0264 $&$ 0.0268 $&$ 0.0294 $&$ 0.0277 $&$ 0.0266 $&$ 0.0289 $&$ 0.0278 $\\
    \hline
    $ 8 $&$ 0.0276 $&$ 0.026 $&$ 0.0259 $&$ 0.0264 $&$ 0.0279 $&$ 0.03 $&$ 0.0282 $&$ 0.0287 $&$ 0.0264 $&$ 0.0276 $&$ 0.0287 $&$ 0.0285 $&$ 0.0258 $&$ 0.0279 $&$ 0.0272 $\\
    \hline
    $ 9 $&$ 0.0261 $&$ 0.0251 $&$ 0.026 $&$ 0.0253 $&$ 0.0272 $&$ 0.026 $&$ 0.0258 $&$ 0.0244 $&$ 0.0243 $&$ 0.0264 $&$ 0.0265 $&$ 0.027 $&$ 0.0263 $&$ 0.0269 $&$ 0.0267 $\\
    \hline
    $ 10 $&$ 0.0268 $&$ 0.0269 $&$ 0.0261 $&$ 0.0246 $&$ 0.0283 $&$ 0.0265 $&$ 0.0263 $&$ 0.0245 $&$ 0.0245 $&$ 0.0278 $&$ 0.0266 $&$ 0.0265 $&$ 0.0262 $&$ 0.0273 $&$ 0.0258 $\\
    \hline
    $ 11 $&$ 0.0256 $&$ 0.0263 $&$ 0.0262 $&$ 0.0257 $&$ 0.027 $&$ 0.0279 $&$ 0.0272 $&$ 0.0263 $&$ 0.0239 $&$ 0.0261 $&$ 0.0275 $&$ 0.0266 $&$ 0.0248 $&$ 0.0279 $&$ 0.0252 $\\
    \hline
    $ 12 $&$ 0.027 $&$ 0.0249 $&$ 0.0264 $&$ 0.0267 $&$ 0.0276 $&$ 0.0272 $&$ 0.0269 $&$ 0.0247 $&$ 0.0241 $&$ 0.0264 $&$ 0.0272 $&$ 0.0271 $&$ 0.0256 $&$ 0.0276 $&$ 0.027 $\\
    \hline
    $ 13 $&$ 0.0264 $&$ 0.0261 $&$ 0.026 $&$ 0.0252 $&$ 0.0271 $&$ 0.0263 $&$ 0.0262 $&$ 0.0256 $&$ 0.0244 $&$ 0.0272 $&$ 0.0262 $&$ 0.0272 $&$ 0.0264 $&$ 0.0269 $&$ 0.0267 $\\
    \hline
    $ 14 $&$ 0.0273 $&$ 0.0263 $&$ 0.0266 $&$ 0.0258 $&$ 0.0291 $&$ 0.0284 $&$ 0.0275 $&$ 0.0256 $&$ 0.025 $&$ 0.0281 $&$ 0.0283 $&$ 0.0278 $&$ 0.0264 $&$ 0.0283 $&$ 0.0273 $\\
    \hline
    $ 15 $&$ 0.0271 $&$ 0.0264 $&$ 0.0258 $&$ 0.0257 $&$ 0.0276 $&$ 0.0277 $&$ 0.0258 $&$ 0.0251 $&$ 0.0237 $&$ 0.0269 $&$ 0.0275 $&$ 0.0272 $&$ 0.0256 $&$ 0.0277 $&$ 0.0261 $\\
    \hline
    \end{tabular}
    \caption{Empirical type I error rates for fixed combinations of stagewise tests with 100 patients in each group. Rows correspond to first-stage tests and columns correspond to second-stage tests.}
    \label{supp-table:t1e_all_n100}
\end{table}

\begin{table}
\footnotesize
    \centering
    \begin{tabular}{|l|c|c|c|c|c|c|c|c|c|c|c|c|c|c|c|}
    \hline
    &$1$&$2$&$3$&$4$&$5$&$6$&$7$&$8$&$9$&$10$&$11$&$12$&$13$&$14$&$15$\\
    \hline
    $ 1 $&$ 0.0228 $&$ 0.0211 $&$ 0.0214 $&$ 0.0218 $&$ 0.0232 $&$ 0.024 $&$ 0.0236 $&$ 0.0238 $&$ 0.0224 $&$ 0.0231 $&$ 0.0233 $&$ 0.0231 $&$ 0.023 $&$ 0.0231 $&$ 0.0231 $\\
    \hline
    $ 2 $&$ 0.0257 $&$ 0.0243 $&$ 0.0234 $&$ 0.0237 $&$ 0.0249 $&$ 0.0254 $&$ 0.024 $&$ 0.0261 $&$ 0.025 $&$ 0.0255 $&$ 0.0254 $&$ 0.0256 $&$ 0.0246 $&$ 0.0241 $&$ 0.0251 $\\
    \hline
    $ 3 $&$ 0.0262 $&$ 0.0252 $&$ 0.0251 $&$ 0.0249 $&$ 0.026 $&$ 0.0258 $&$ 0.0262 $&$ 0.0274 $&$ 0.0255 $&$ 0.0252 $&$ 0.026 $&$ 0.0256 $&$ 0.0247 $&$ 0.0255 $&$ 0.0259 $\\
    \hline
    $ 4 $&$ 0.0261 $&$ 0.026 $&$ 0.0255 $&$ 0.026 $&$ 0.0268 $&$ 0.0264 $&$ 0.0264 $&$ 0.0276 $&$ 0.0271 $&$ 0.0262 $&$ 0.027 $&$ 0.0272 $&$ 0.026 $&$ 0.0266 $&$ 0.027 $\\
    \hline
    $ 5 $&$ 0.0237 $&$ 0.0221 $&$ 0.0222 $&$ 0.0229 $&$ 0.0242 $&$ 0.0248 $&$ 0.0246 $&$ 0.0247 $&$ 0.0225 $&$ 0.0238 $&$ 0.0229 $&$ 0.0235 $&$ 0.0224 $&$ 0.0242 $&$ 0.0239 $\\
    \hline
    $ 6 $&$ 0.0244 $&$ 0.0216 $&$ 0.0216 $&$ 0.0218 $&$ 0.0237 $&$ 0.0239 $&$ 0.0237 $&$ 0.0237 $&$ 0.0225 $&$ 0.0245 $&$ 0.0245 $&$ 0.0233 $&$ 0.0222 $&$ 0.0237 $&$ 0.0243 $\\
    \hline
    $ 7 $&$ 0.024 $&$ 0.0234 $&$ 0.0228 $&$ 0.0227 $&$ 0.0239 $&$ 0.0236 $&$ 0.024 $&$ 0.0223 $&$ 0.0235 $&$ 0.0248 $&$ 0.0241 $&$ 0.0243 $&$ 0.023 $&$ 0.0242 $&$ 0.0234 $\\
    \hline
    $ 8 $&$ 0.0242 $&$ 0.0224 $&$ 0.0225 $&$ 0.0228 $&$ 0.0233 $&$ 0.0236 $&$ 0.0226 $&$ 0.0228 $&$ 0.0232 $&$ 0.024 $&$ 0.0239 $&$ 0.0238 $&$ 0.0235 $&$ 0.0239 $&$ 0.0228 $\\
    \hline
    $ 9 $&$ 0.0237 $&$ 0.0229 $&$ 0.0221 $&$ 0.0226 $&$ 0.0236 $&$ 0.0245 $&$ 0.0244 $&$ 0.0246 $&$ 0.0227 $&$ 0.0242 $&$ 0.0238 $&$ 0.024 $&$ 0.0227 $&$ 0.024 $&$ 0.0241 $\\
    \hline
    $ 10 $&$ 0.0219 $&$ 0.0217 $&$ 0.021 $&$ 0.0212 $&$ 0.0233 $&$ 0.0237 $&$ 0.0237 $&$ 0.0241 $&$ 0.0225 $&$ 0.0223 $&$ 0.023 $&$ 0.0227 $&$ 0.0218 $&$ 0.023 $&$ 0.0227 $\\
    \hline
    $ 11 $&$ 0.0231 $&$ 0.0218 $&$ 0.0221 $&$ 0.0222 $&$ 0.0225 $&$ 0.0248 $&$ 0.0236 $&$ 0.0241 $&$ 0.0218 $&$ 0.0238 $&$ 0.0238 $&$ 0.0237 $&$ 0.0223 $&$ 0.0239 $&$ 0.0237 $\\
    \hline
    $ 12 $&$ 0.0234 $&$ 0.0213 $&$ 0.0213 $&$ 0.0224 $&$ 0.0237 $&$ 0.0239 $&$ 0.0247 $&$ 0.0241 $&$ 0.0217 $&$ 0.0239 $&$ 0.0239 $&$ 0.0231 $&$ 0.022 $&$ 0.0239 $&$ 0.0235 $\\
    \hline
    $ 13 $&$ 0.0237 $&$ 0.0228 $&$ 0.0216 $&$ 0.0228 $&$ 0.0245 $&$ 0.0235 $&$ 0.0253 $&$ 0.025 $&$ 0.0233 $&$ 0.024 $&$ 0.0238 $&$ 0.0235 $&$ 0.023 $&$ 0.0245 $&$ 0.0238 $\\
    \hline
    $ 14 $&$ 0.023 $&$ 0.0217 $&$ 0.021 $&$ 0.0219 $&$ 0.0244 $&$ 0.0251 $&$ 0.0245 $&$ 0.0239 $&$ 0.0228 $&$ 0.0246 $&$ 0.0237 $&$ 0.0228 $&$ 0.0226 $&$ 0.0228 $&$ 0.023 $\\
    \hline
    $ 15 $&$ 0.0239 $&$ 0.0224 $&$ 0.0214 $&$ 0.0223 $&$ 0.0233 $&$ 0.0242 $&$ 0.0255 $&$ 0.0243 $&$ 0.0223 $&$ 0.0239 $&$ 0.024 $&$ 0.0235 $&$ 0.0229 $&$ 0.0243 $&$ 0.0236 $\\
    \hline
    \end{tabular}
    \caption{Empirical type I error rates for fixed combinations of stagewise tests with 200 patients in each group. Rows correspond to first-stage tests and columns correspond to second-stage tests.}
    \label{supp-table:t1e_all_n200}
\end{table}

\begin{table}
\footnotesize
    \centering
    \begin{tabular}{|l|c|c|c|c|c|c|c|c|c|c|c|c|c|c|c|}
    \hline
    &$1$&$2$&$3$&$4$&$5$&$6$&$7$&$8$&$9$&$10$&$11$&$12$&$13$&$14$&$15$\\
    \hline
    $ 1 $&$ 0.0263 $&$ 0.0286 $&$ 0.0288 $&$ 0.029 $&$ 0.0269 $&$ 0.026 $&$ 0.0254 $&$ 0.0258 $&$ 0.0277 $&$ 0.027 $&$ 0.0269 $&$ 0.0278 $&$ 0.0284 $&$ 0.028 $&$ 0.0282 $\\
    \hline
    $ 2 $&$ 0.0252 $&$ 0.0279 $&$ 0.0272 $&$ 0.0275 $&$ 0.0275 $&$ 0.0257 $&$ 0.025 $&$ 0.0247 $&$ 0.0271 $&$ 0.0264 $&$ 0.0243 $&$ 0.025 $&$ 0.0269 $&$ 0.0256 $&$ 0.0268 $\\
    \hline
    $ 3 $&$ 0.0263 $&$ 0.0255 $&$ 0.0259 $&$ 0.0254 $&$ 0.0265 $&$ 0.0248 $&$ 0.0247 $&$ 0.0239 $&$ 0.0261 $&$ 0.0262 $&$ 0.025 $&$ 0.0252 $&$ 0.0261 $&$ 0.025 $&$ 0.0262 $\\
    \hline
    $ 4 $&$ 0.0271 $&$ 0.0266 $&$ 0.0266 $&$ 0.0265 $&$ 0.0257 $&$ 0.0235 $&$ 0.0257 $&$ 0.0251 $&$ 0.0256 $&$ 0.0276 $&$ 0.0259 $&$ 0.0259 $&$ 0.0268 $&$ 0.0258 $&$ 0.0269 $\\
    \hline
    $ 5 $&$ 0.0255 $&$ 0.0277 $&$ 0.0283 $&$ 0.0279 $&$ 0.0273 $&$ 0.0263 $&$ 0.0263 $&$ 0.0254 $&$ 0.0263 $&$ 0.0257 $&$ 0.025 $&$ 0.0256 $&$ 0.0271 $&$ 0.0256 $&$ 0.0256 $\\
    \hline
    $ 6 $&$ 0.0279 $&$ 0.0298 $&$ 0.0286 $&$ 0.0291 $&$ 0.0288 $&$ 0.0263 $&$ 0.026 $&$ 0.0256 $&$ 0.0286 $&$ 0.0288 $&$ 0.0272 $&$ 0.0284 $&$ 0.0294 $&$ 0.0284 $&$ 0.0284 $\\
    \hline
    $ 7 $&$ 0.0276 $&$ 0.0286 $&$ 0.0279 $&$ 0.0281 $&$ 0.0285 $&$ 0.0278 $&$ 0.0269 $&$ 0.0267 $&$ 0.0282 $&$ 0.0285 $&$ 0.0278 $&$ 0.028 $&$ 0.0291 $&$ 0.0284 $&$ 0.0287 $\\
    \hline
    $ 8 $&$ 0.0275 $&$ 0.0277 $&$ 0.0271 $&$ 0.0276 $&$ 0.0284 $&$ 0.027 $&$ 0.027 $&$ 0.0258 $&$ 0.0282 $&$ 0.0278 $&$ 0.0268 $&$ 0.0269 $&$ 0.0277 $&$ 0.0272 $&$ 0.0276 $\\
    \hline
    $ 9 $&$ 0.0259 $&$ 0.0285 $&$ 0.0288 $&$ 0.0286 $&$ 0.0278 $&$ 0.0261 $&$ 0.0271 $&$ 0.0261 $&$ 0.0264 $&$ 0.0266 $&$ 0.0256 $&$ 0.0269 $&$ 0.0273 $&$ 0.0262 $&$ 0.0274 $\\
    \hline
    $ 10 $&$ 0.026 $&$ 0.0274 $&$ 0.0276 $&$ 0.0276 $&$ 0.0264 $&$ 0.0258 $&$ 0.0262 $&$ 0.0256 $&$ 0.0267 $&$ 0.0265 $&$ 0.0251 $&$ 0.0269 $&$ 0.0266 $&$ 0.0259 $&$ 0.0266 $\\
    \hline
    $ 11 $&$ 0.0274 $&$ 0.0287 $&$ 0.0285 $&$ 0.0289 $&$ 0.0286 $&$ 0.0269 $&$ 0.0264 $&$ 0.0258 $&$ 0.0281 $&$ 0.0274 $&$ 0.0267 $&$ 0.0276 $&$ 0.0284 $&$ 0.0269 $&$ 0.0266 $\\
    \hline
    $ 12 $&$ 0.0253 $&$ 0.0276 $&$ 0.0281 $&$ 0.0272 $&$ 0.0263 $&$ 0.0258 $&$ 0.0264 $&$ 0.0269 $&$ 0.0258 $&$ 0.0267 $&$ 0.0252 $&$ 0.0261 $&$ 0.0267 $&$ 0.0254 $&$ 0.0258 $\\
    \hline
    $ 13 $&$ 0.0253 $&$ 0.0274 $&$ 0.0275 $&$ 0.0273 $&$ 0.0265 $&$ 0.025 $&$ 0.0266 $&$ 0.0264 $&$ 0.0259 $&$ 0.0263 $&$ 0.0252 $&$ 0.0256 $&$ 0.0271 $&$ 0.0261 $&$ 0.0266 $\\
    \hline
    $ 14 $&$ 0.0265 $&$ 0.028 $&$ 0.0278 $&$ 0.0275 $&$ 0.0265 $&$ 0.0258 $&$ 0.0254 $&$ 0.0262 $&$ 0.0276 $&$ 0.0266 $&$ 0.0254 $&$ 0.0265 $&$ 0.0274 $&$ 0.0255 $&$ 0.0267 $\\
    \hline
    $ 15 $&$ 0.0255 $&$ 0.0278 $&$ 0.0272 $&$ 0.027 $&$ 0.0274 $&$ 0.0257 $&$ 0.0261 $&$ 0.0259 $&$ 0.0268 $&$ 0.0265 $&$ 0.025 $&$ 0.0262 $&$ 0.0265 $&$ 0.0261 $&$ 0.0258 $\\
    \hline
    \end{tabular}
    \caption{Empirical type I error rates for fixed combinations of stagewise tests with 500 patients in each group. Rows correspond to first-stage tests and columns correspond to second-stage tests.}
    \label{supp-table:t1e_all_n500}
\end{table}

\end{landscape}

\subsection{Power comparisons}\label{supp-subsec:power}

\subsubsection{Additional deviation types}\label{supp-subsubsec:add_types}

\begin{figure}[h!]
	\centering
	\begin{subfigure}[t]{0.49\textwidth}
		{\Large \textsf{\textbf{A}}}
	\end{subfigure}
	\begin{subfigure}[t]{0.49\textwidth}
		{\Large \textsf{\textbf{B}}}
	\end{subfigure}
	\begin{subfigure}[tb]{0.49\textwidth}
		\includegraphics[width=\linewidth]{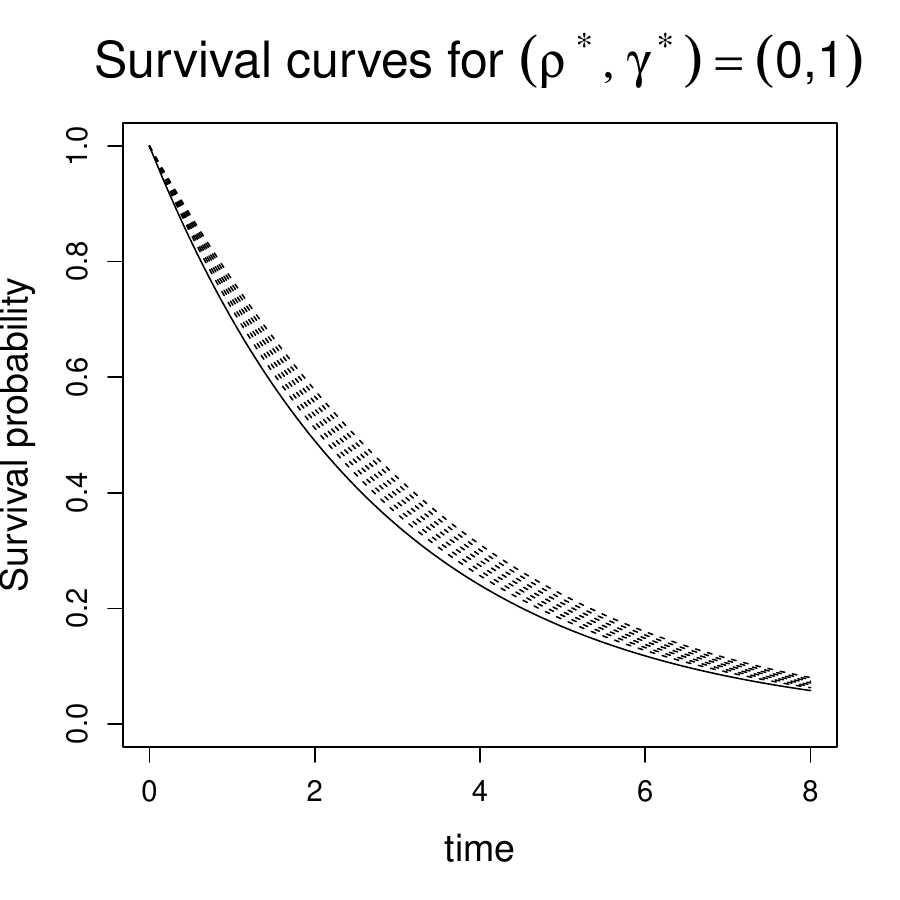}
	\end{subfigure}
	\begin{subfigure}[tb]{0.49\textwidth}
		\includegraphics[width=\linewidth]{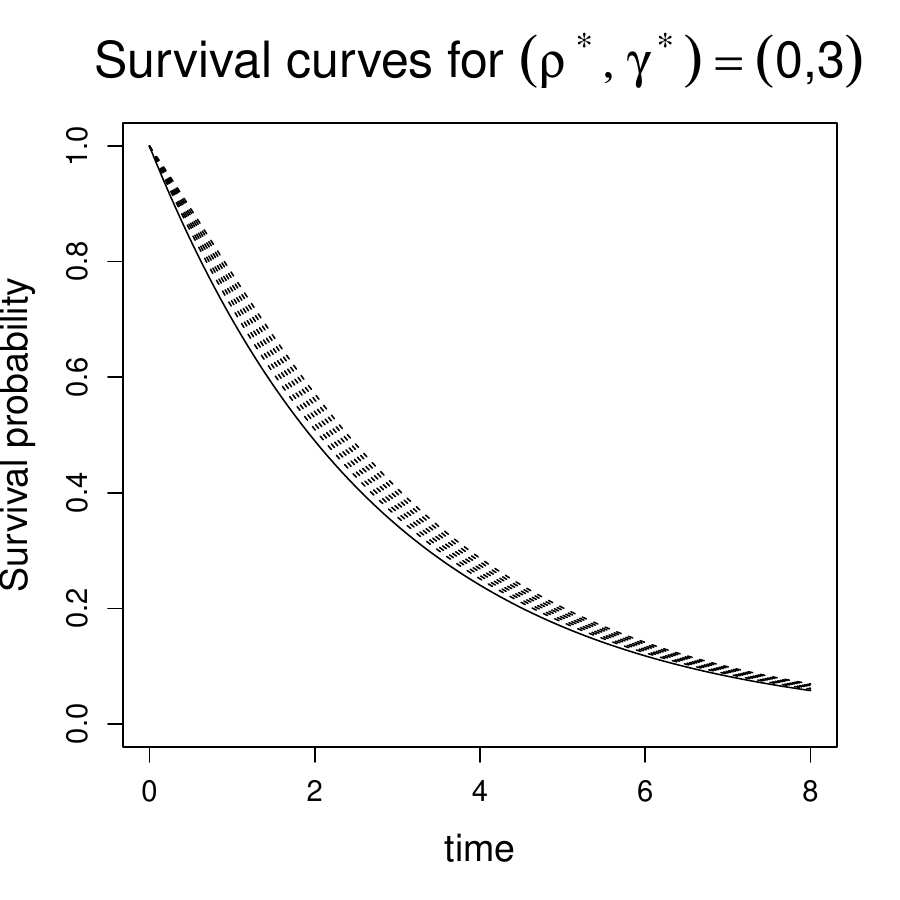}
	\end{subfigure}
    \begin{subfigure}[t]{0.49\textwidth}
		{\Large \textsf{\textbf{C}}}
	\end{subfigure}
    \begin{subfigure}[t]{0.49\textwidth}
		{\Large \textsf{\textbf{D}}}
	\end{subfigure}
    \begin{subfigure}[tb]{0.49\textwidth}
		\includegraphics[width=\linewidth]{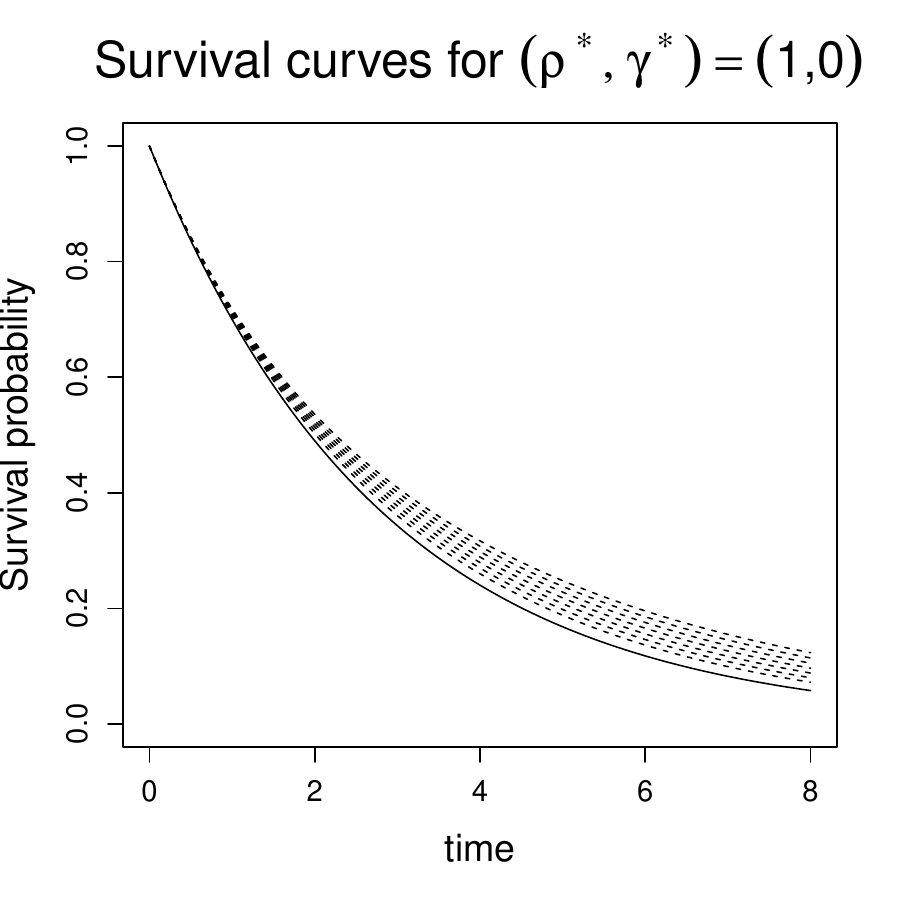}
	\end{subfigure}
    \begin{subfigure}[tb]{0.49\textwidth}
		\includegraphics[width=\linewidth]{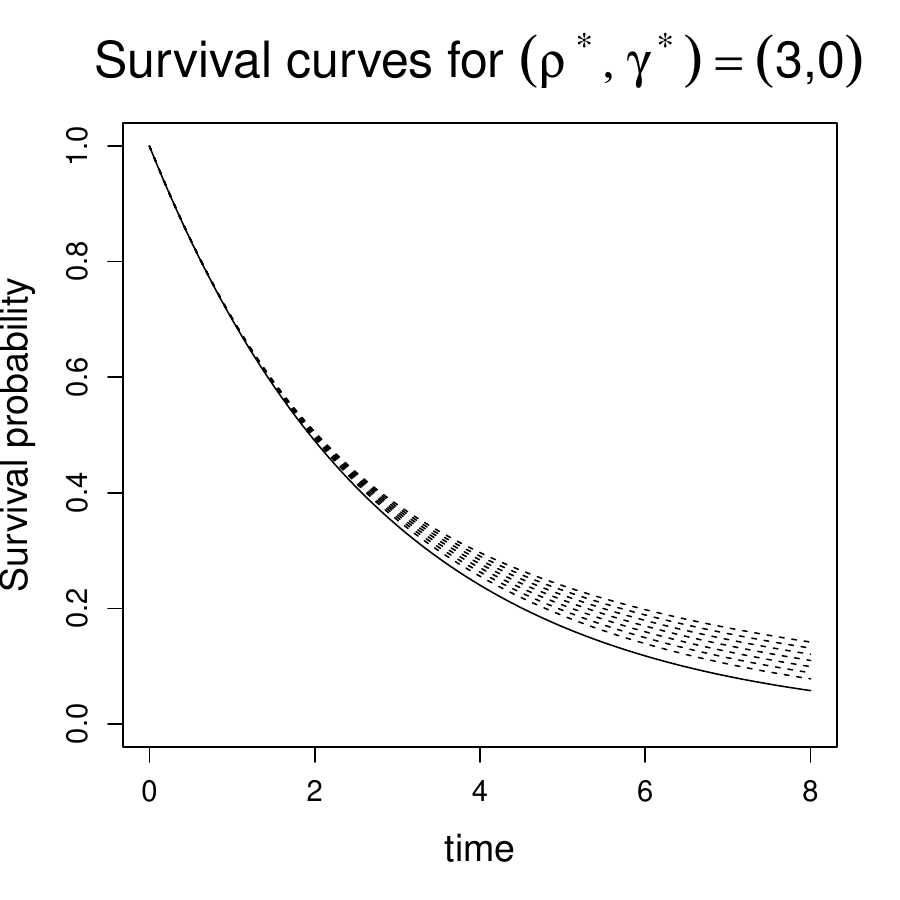}
	\end{subfigure}
	\caption{Survival curves for three types of deviation of the distribution in the experimental group from the distribution in the control group. The survival curve in the control group is given by the solid line. Dashed lines are survival curves in the experimental group for the seven effect sizes $\{0.4\cdot\theta_0, 0.6\cdot\theta_0, 0.8\cdot\theta_0, \theta_0, 1.2\cdot\theta_0, 1.4\cdot\theta_0, 1.6\cdot\theta_0\}$. A) Survival curves in the slightly early effect case $((\rho^{\star},\gamma^{\star}) = (0,1))$ B) Survival curves in the very early effect case $((\rho^{\star},\gamma^{\star}) = (0,3))$ C) Survival curves in the slightly late effect case $((\rho^{\star},\gamma^{\star}) = (1,0))$ B) Survival curves in the very late effect case $((\rho^{\star},\gamma^{\star}) = (3,0))$}
	\label{supp-fig:add_survival curves}
\end{figure}

\begin{figure}[h]
	\centering
	\begin{subfigure}[t]{0.5\textwidth}
		{\Large \textsf{\textbf{A}}}
	\end{subfigure}
	\begin{subfigure}[t]{0.49\textwidth}
		{\Large \textsf{\textbf{B}}}
	\end{subfigure}
	\begin{subfigure}[tb]{0.5\textwidth}
		\includegraphics[width=\linewidth]{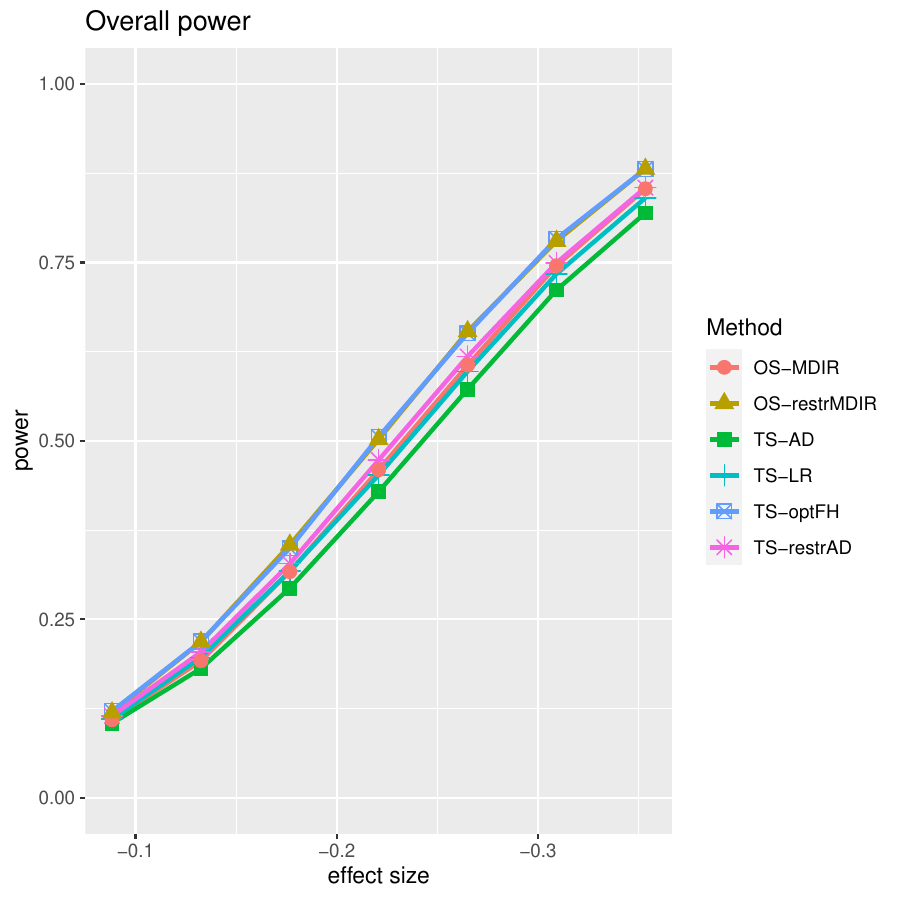}
	\end{subfigure}
	\begin{subfigure}[tb]{0.49\textwidth}
		\includegraphics[width=\linewidth]{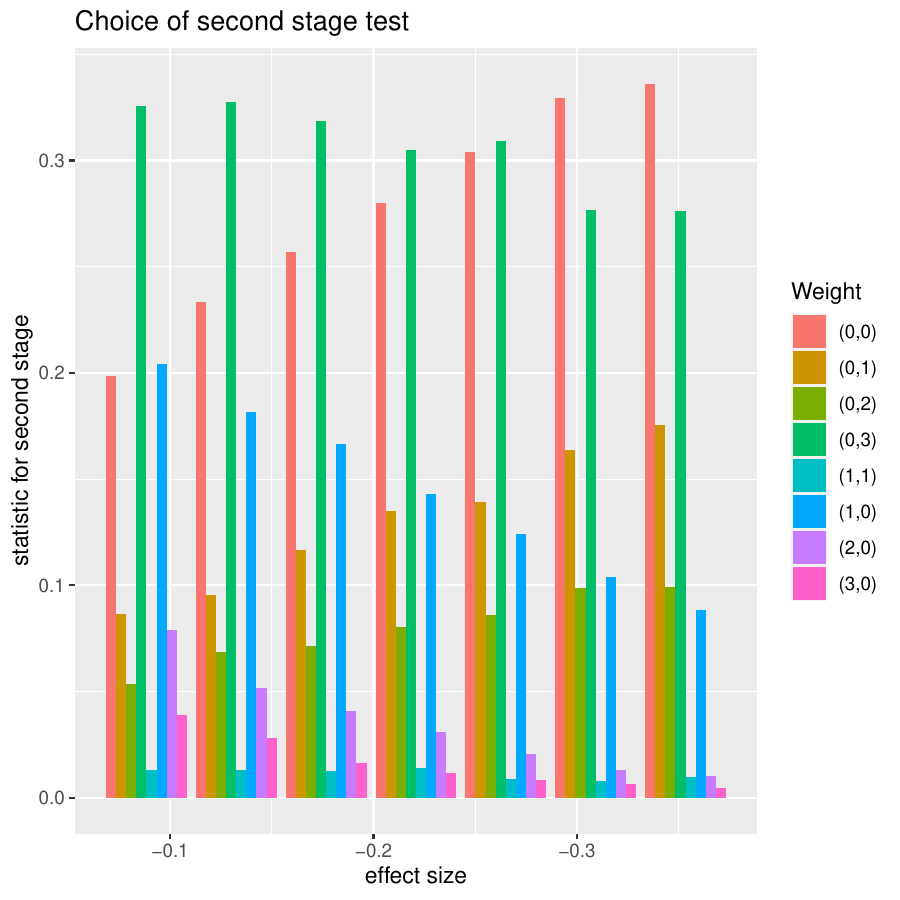}
	\end{subfigure}
	\caption{A) Power curve for six testing procedures in case of proportional hazards $((\rho^{\star}, \gamma^{\star})=(0,1))$. Please note that the two procedures TS-AD and TS-optFH as well as the two procedures OS-MDIR and OS-restrMDIR coincide in this case. B) Relative frequencies of the choice of single weighted tests for the second stage for the testing procedure TS-AD.}
	\label{supp-fig:power_slightly_early}
\end{figure}

\begin{figure}[h]
	\centering
	\begin{subfigure}[t]{0.5\textwidth}
		{\Large \textsf{\textbf{A}}}
	\end{subfigure}
	\begin{subfigure}[t]{0.49\textwidth}
		{\Large \textsf{\textbf{B}}}
	\end{subfigure}
	\begin{subfigure}[tb]{0.5\textwidth}
		\includegraphics[width=\linewidth]{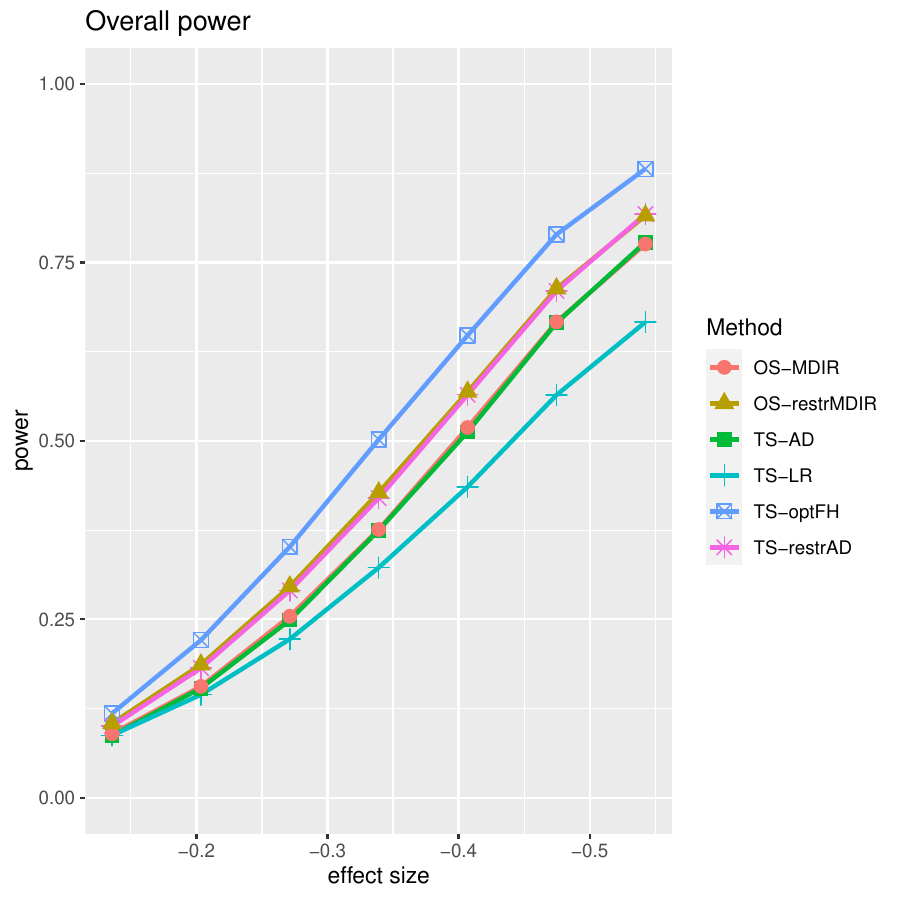}
	\end{subfigure}
	\begin{subfigure}[tb]{0.49\textwidth}
		\includegraphics[width=\linewidth]{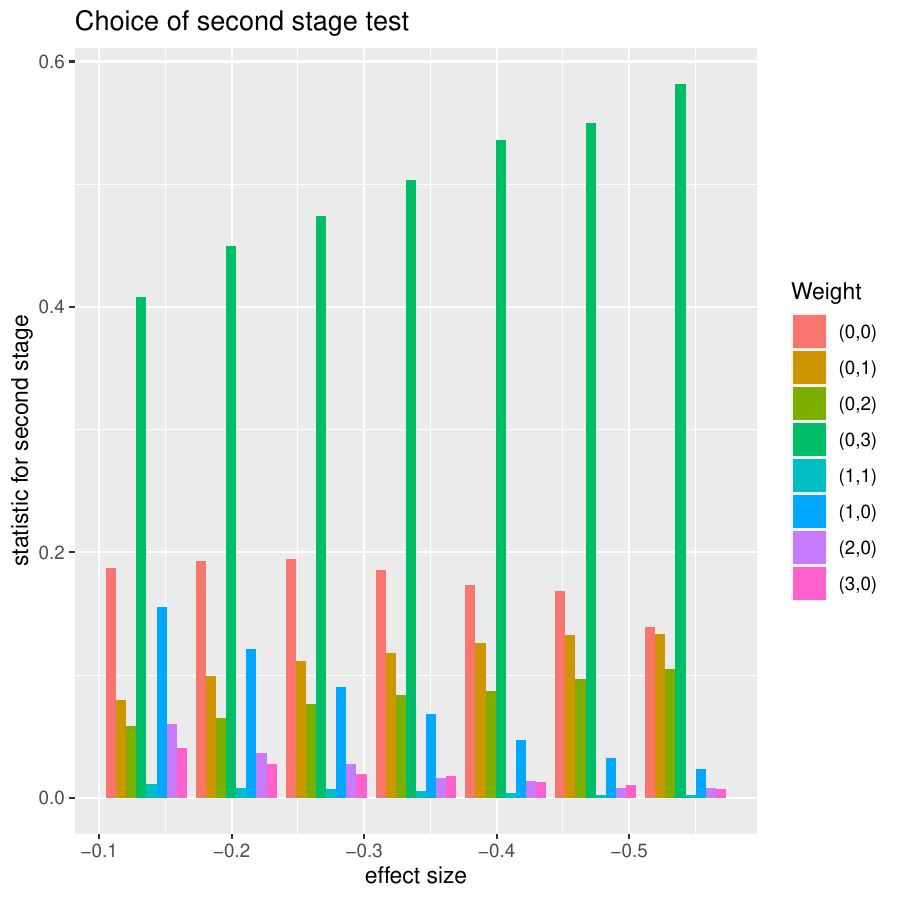}
	\end{subfigure}
	\caption{A) Power curve for six testing procedures in case of proportional hazards $((\rho^{\star}, \gamma^{\star})=(0,3))$. Please note that the two procedures TS-AD and TS-optFH as well as the two procedures OS-MDIR and OS-restrMDIR coincide in this case. B) Relative frequencies of the choice of single weighted tests for the second stage for the testing procedure TS-AD.}
	\label{supp-fig:power_very_early}
\end{figure}

\begin{figure}[h]
	\centering
	\begin{subfigure}[t]{0.5\textwidth}
		{\Large \textsf{\textbf{A}}}
	\end{subfigure}
	\begin{subfigure}[t]{0.49\textwidth}
		{\Large \textsf{\textbf{B}}}
	\end{subfigure}
	\begin{subfigure}[tb]{0.5\textwidth}
		\includegraphics[width=\linewidth]{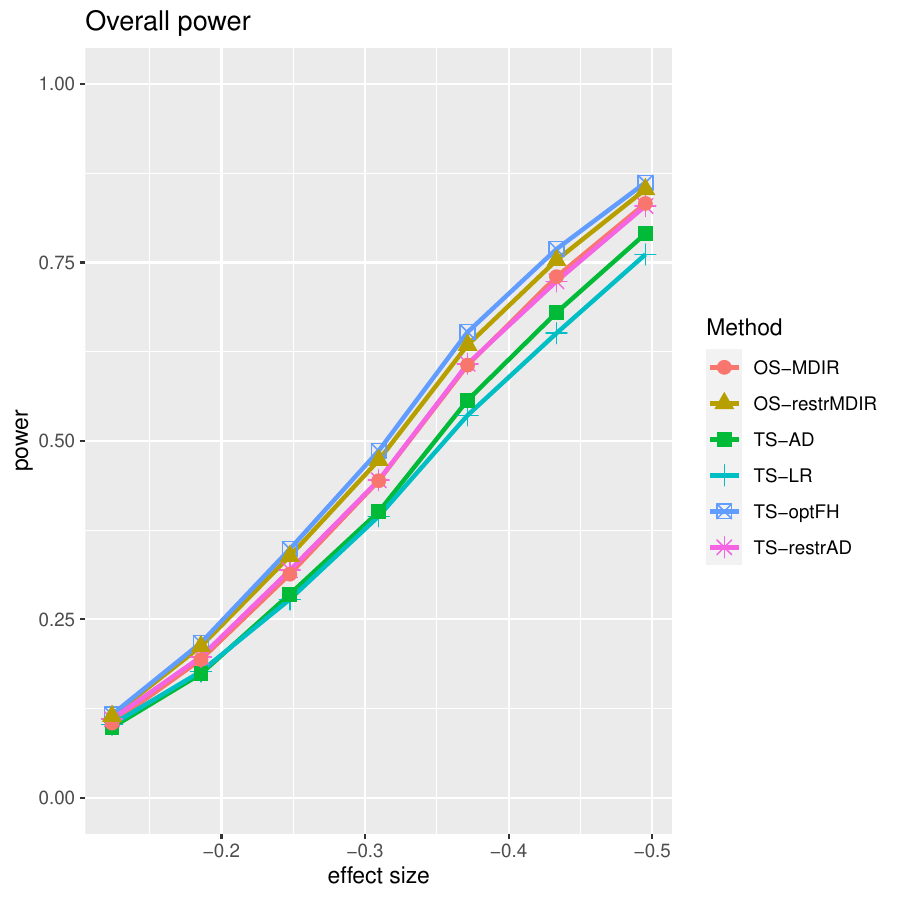}
	\end{subfigure}
	\begin{subfigure}[tb]{0.49\textwidth}
		\includegraphics[width=\linewidth]{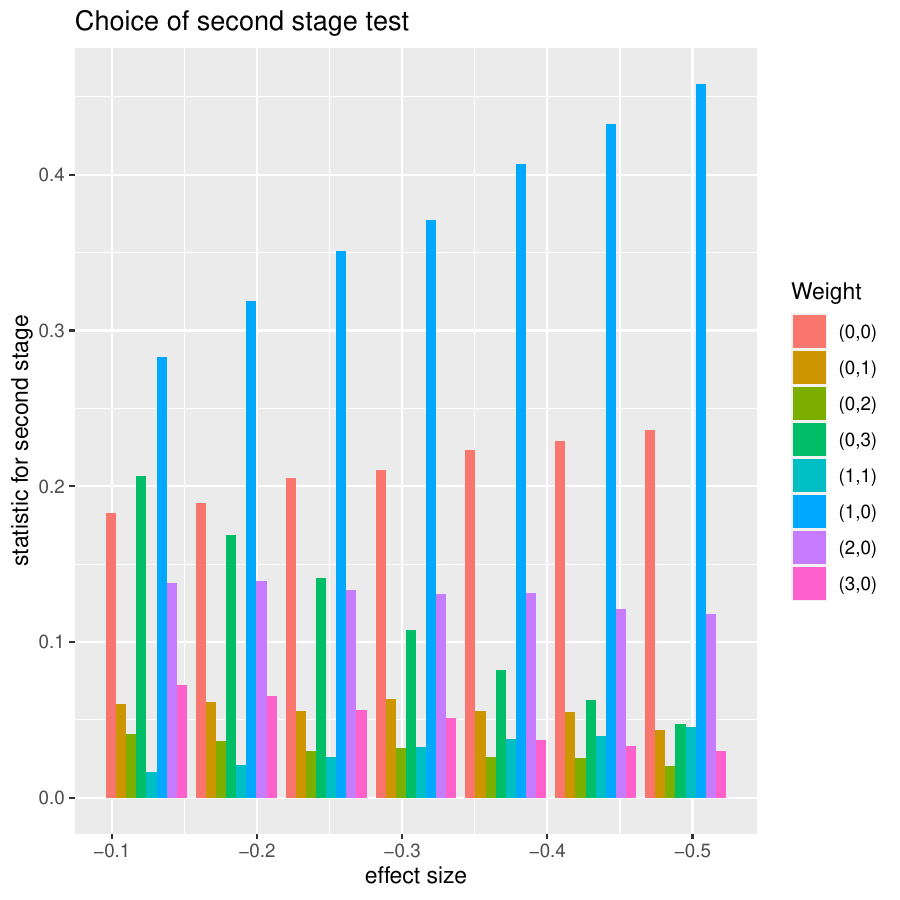}
	\end{subfigure}
	\caption{A) Power curve for six testing procedures in case of proportional hazards $((\rho^{\star}, \gamma^{\star})=(1,0))$. Please note that the two procedures TS-AD and TS-optFH as well as the two procedures OS-MDIR and OS-restrMDIR coincide in this case. B) Relative frequencies of the choice of single weighted tests for the second stage for the testing procedure TS-AD.}
	\label{supp-fig:power_slightly_late}
\end{figure}

\begin{figure}[h]
	\centering
	\begin{subfigure}[t]{0.5\textwidth}
		{\Large \textsf{\textbf{A}}}
	\end{subfigure}
	\begin{subfigure}[t]{0.49\textwidth}
		{\Large \textsf{\textbf{B}}}
	\end{subfigure}
	\begin{subfigure}[tb]{0.5\textwidth}
		\includegraphics[width=\linewidth]{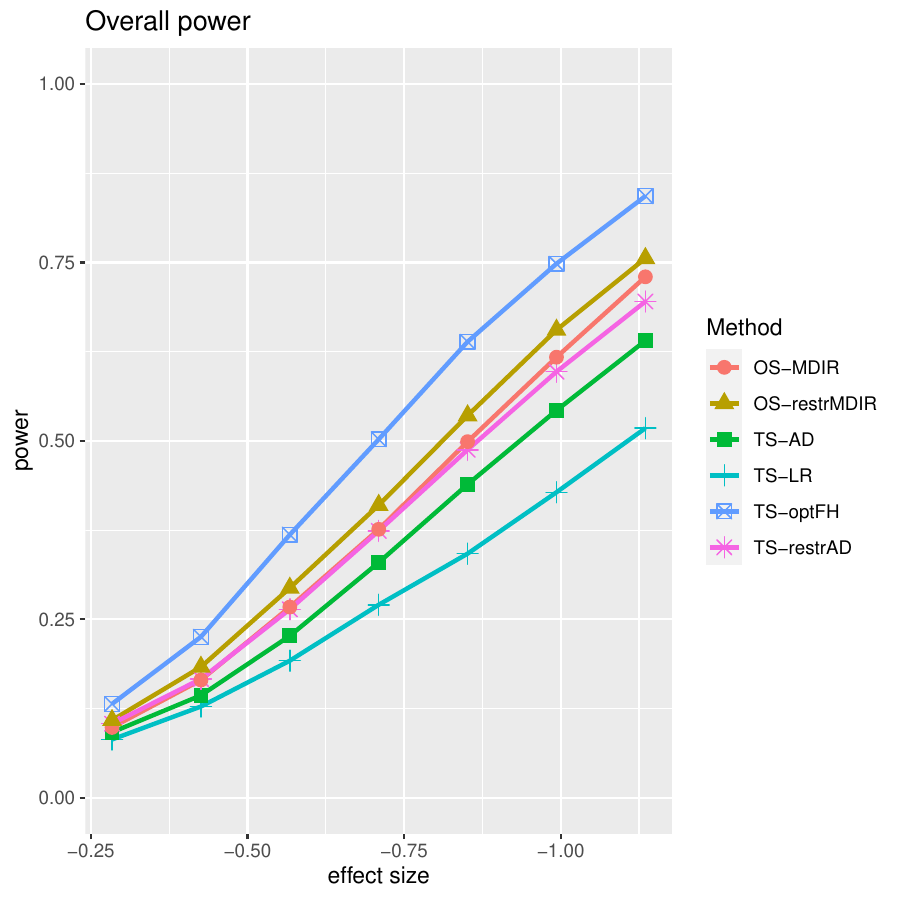}
	\end{subfigure}
	\begin{subfigure}[tb]{0.49\textwidth}
		\includegraphics[width=\linewidth]{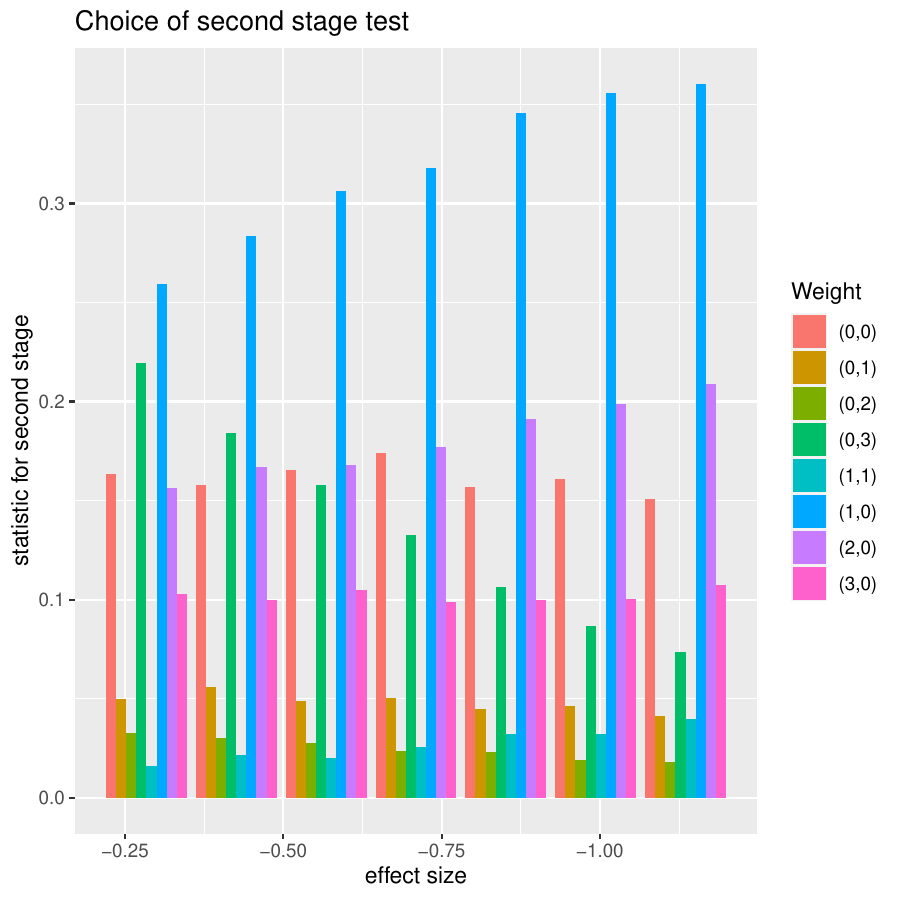}
	\end{subfigure}
	\caption{A) Power curve for six testing procedures in case of proportional hazards $((\rho^{\star}, \gamma^{\star})=(3,0))$. Please note that the two procedures TS-AD and TS-optFH as well as the two procedures OS-MDIR and OS-restrMDIR coincide in this case. B) Relative frequencies of the choice of single weighted tests for the second stage for the testing procedure TS-AD.}
	\label{supp-fig:power_very_late}
\end{figure}

In this subsection, we supplement the results from Section \ref{subsec:power} of the main manuscript by those results for the settings $(\rho^{\star}, \gamma^{\star}) \in \{(0,1), (0,3), (1,0), (3,0)\}$. The corresponding survival curves are shown in Figure \ref{supp-fig:add_survival curves}.\\
\clearpage
For these scenarios, the power curves and the bar plots showing the choices made by the procedure \textbf{TS-AD} for the second stage test can be found in Figures \ref{supp-fig:power_slightly_early}-\ref{supp-fig:power_very_late}. For slightly early and late effects (i.e. $(\rho^{\star}, \gamma^{\star}) = (0,1)$ or $(\rho^{\star}, \gamma^{\star}) = (1,0)$, respectively) we can see that all procedures perform very similarly. For strong early and late effects (i.e. $(\rho^{\star}, \gamma^{\star}) = (0,3)$ or $(\rho^{\star}, \gamma^{\star}) = (3,0)$, respectively), there is a marked gap between the optimal two-stage test and the two-stage standard log-rank test. The one-stage combination testing procedures and the adaptive procedures fill this gap. Once again, we can see that the two restricted procedures with pre-chosen sets of weights perform better than the unrestricted ones. It should be noted that our selection procedure tends to favor the log-rank test with weights $w^{(0,3)} \circ \hat{F}$ in the case of early effects and the log-rank test with weights $w^{(1,0)} \circ \hat{F}$ in the case of late effects. This shows that there is certainly room for improvement in this selection process.\\

\clearpage

\subsubsection{Model choice based on AIC}\label{supp-subsubsec:model_choice}

Here we show which of the 9 Royston-Parmar spline models is chosen based on their AIC value. As already mentioned, we can see that the AIC selection mechanism prefers models with 0 interior knots for our simulation scenarios. In particular, the model on the hazard scale is preferred. This spline model induces Weibull distributions for both groups.

\begin{table}[h]
\footnotesize
    \centering
    \begin{tabular}{|l||c|c|c|c|c|c|c|c|c|}
    \hline
    \multirow{2}{*}{Effect size}&\multicolumn{3}{c|}{$p=0$}&\multicolumn{3}{c|}{$p=1$}&\multicolumn{3}{c|}{$p=2$}\\
    \cline{2-10}
    &hazard&odds&normal&hazard&odds&normal&hazard&odds&normal\\
    \hline
    $ 0.4 \theta_0 $&$ 0.7892 $&$ 0.0224 $&$ 0.0001 $&$ 0.0799 $&$ 0.0243 $&$ 0.0227 $&$ 0.0233 $&$ 0.0141 $&$ 0.0241$\\
    \hline 
    $ 0.6 \theta_0 $&$ 0.7807 $&$ 0.0252 $&$ 0.0001 $&$ 0.0816 $&$ 0.0273 $&$ 0.0254 $&$ 0.022 $&$ 0.0143 $&$ 0.0235$\\
    \hline 
    $ 0.8 \theta_0 $&$ 0.7728 $&$ 0.0276 $&$ 0.0001 $&$ 0.0806 $&$ 0.0243 $&$ 0.0263 $&$ 0.0255 $&$ 0.0166 $&$ 0.0262$\\
    \hline 
    $\theta_0 $&$ 0.7774 $&$ 0.0284 $&$ 0.0001 $&$ 0.0788 $&$ 0.0288 $&$ 0.0253 $&$ 0.0236 $&$ 0.0141 $&$ 0.0234$\\
    \hline 
    $ 1.2 \theta_0 $&$ 0.777 $&$ 0.0308 $&$ 0 $&$ 0.0755 $&$ 0.0291 $&$ 0.0203 $&$ 0.0247 $&$ 0.015 $&$ 0.0276$\\
    \hline 
    $ 1.4 \theta_0 $&$ 0.7689 $&$ 0.038 $&$ 0.0001 $&$ 0.0754 $&$ 0.0299 $&$ 0.0223 $&$ 0.0235 $&$ 0.0153 $&$ 0.0267$\\
    \hline 
    $ 1.6 \theta_0 $&$ 0.7745 $&$ 0.0368 $&$ 0.0001 $&$ 0.0747 $&$ 0.0264 $&$ 0.0244 $&$ 0.0223 $&$ 0.0163 $&$ 0.0245$\\
    \hline 
    \end{tabular}
    \caption{For each of the seven effect sizes, this table displays which of the nine Royston-Parmar spline models is rated as the best based on the AIC. As above, $p$ refers to the number of inner knots in the spline model. The empirical rates refer to the total quantity of runs without early stopping of the corresponding simulated trial. This table refers to the scenario $(\rho^{\star}, \gamma^{\star}) = (0,0)$}
    \label{supp-table:aic_choice_00}
\end{table}

\begin{table}[h]
\footnotesize
    \centering
    \begin{tabular}{|l||c|c|c|c|c|c|c|c|c|}
    \hline
    \multirow{2}{*}{Effect size}&\multicolumn{3}{c|}{$p=0$}&\multicolumn{3}{c|}{$p=1$}&\multicolumn{3}{c|}{$p=2$}\\
    \cline{2-10}
    &hazard&odds&normal&hazard&odds&normal&hazard&odds&normal\\
    \hline
    $ 0.4 \theta_0 $&$ 0.763 $&$ 0.0314 $&$ 0.0001 $&$ 0.0827 $&$ 0.0262 $&$ 0.0321 $&$ 0.0219 $&$ 0.0169 $&$ 0.0257$\\
    \hline 
    $ 0.6 \theta_0 $&$ 0.7614 $&$ 0.0362 $&$ 0 $&$ 0.0799 $&$ 0.0291 $&$ 0.0347 $&$ 0.0231 $&$ 0.0131 $&$ 0.0226$\\
    \hline 
    $ 0.8 \theta_0 $&$ 0.7483 $&$ 0.0412 $&$ 0 $&$ 0.0836 $&$ 0.0291 $&$ 0.0338 $&$ 0.0239 $&$ 0.0148 $&$ 0.0253$\\
    \hline 
    $ 1 \theta_0 $&$ 0.7332 $&$ 0.0479 $&$ 0.0001 $&$ 0.0874 $&$ 0.032 $&$ 0.0365 $&$ 0.0238 $&$ 0.0149 $&$ 0.0242$\\
    \hline 
    $ 1.2 \theta_0 $&$ 0.7192 $&$ 0.054 $&$ 0.0005 $&$ 0.0906 $&$ 0.0338 $&$ 0.0373 $&$ 0.0221 $&$ 0.0156 $&$ 0.027$\\
    \hline 
    $ 1.4 \theta_0 $&$ 0.7104 $&$ 0.058 $&$ 0.0003 $&$ 0.0918 $&$ 0.0362 $&$ 0.0409 $&$ 0.0247 $&$ 0.0128 $&$ 0.0248$\\
    \hline 
    $ 1.6 \theta_0 $&$ 0.6998 $&$ 0.063 $&$ 0.0003 $&$ 0.0936 $&$ 0.0321 $&$ 0.0416 $&$ 0.0249 $&$ 0.0162 $&$ 0.0284$\\
    \hline 
    \end{tabular}
    \caption{For each of the seven effect sizes, this table displays which of the nine Royston-Parmar spline models is rated as the best based on the AIC. As above, $p$ refers to the number of inner knots in the spline model. The empirical rates refer to the total quantity of runs without early stopping of the corresponding simulated trial. This table refers to the scenario $(\rho^{\star}, \gamma^{\star}) = (1,0)$}
    \label{supp-table:aic_choice_10}
\end{table}

\begin{table}[h]
\footnotesize
    \centering
    \begin{tabular}{|l||c|c|c|c|c|c|c|c|c|}
    \hline
    \multirow{2}{*}{Effect size}&\multicolumn{3}{c|}{$p=0$}&\multicolumn{3}{c|}{$p=1$}&\multicolumn{3}{c|}{$p=2$}\\
    \cline{2-10}
    &hazard&odds&normal&hazard&odds&normal&hazard&odds&normal\\
    \hline
    $ 0.4 \theta_0 $&$ 0.7595 $&$ 0.0335 $&$ 0.0001 $&$ 0.0828 $&$ 0.0326 $&$ 0.035 $&$ 0.0193 $&$ 0.014 $&$ 0.0232$\\
    \hline 
    $ 0.6 \theta_0 $&$ 0.7429 $&$ 0.042 $&$ 0 $&$ 0.0836 $&$ 0.0357 $&$ 0.0368 $&$ 0.0214 $&$ 0.0134 $&$ 0.0242$\\
    \hline 
    $ 0.8 \theta_0 $&$ 0.7213 $&$ 0.0518 $&$ 0.0003 $&$ 0.0812 $&$ 0.0395 $&$ 0.0429 $&$ 0.0219 $&$ 0.0157 $&$ 0.0255$\\
    \hline 
    $ 1 \theta_0 $&$ 0.6932 $&$ 0.064 $&$ 0.0001 $&$ 0.094 $&$ 0.0391 $&$ 0.0489 $&$ 0.0227 $&$ 0.0154 $&$ 0.0226$\\
    \hline 
    $ 1.2 \theta_0 $&$ 0.6595 $&$ 0.0711 $&$ 0 $&$ 0.0985 $&$ 0.047 $&$ 0.0517 $&$ 0.0267 $&$ 0.0167 $&$ 0.0288$\\
    \hline 
    $ 1.4 \theta_0 $&$ 0.6348 $&$ 0.0833 $&$ 0.0002 $&$ 0.105 $&$ 0.048 $&$ 0.0517 $&$ 0.0268 $&$ 0.0191 $&$ 0.031$\\
    \hline 
    $ 1.6 \theta_0 $&$ 0.6007 $&$ 0.0943 $&$ 0.0001 $&$ 0.1136 $&$ 0.0533 $&$ 0.0598 $&$ 0.0297 $&$ 0.0174 $&$ 0.031$\\
    \hline 
    \end{tabular}
    \caption{For each of the seven effect sizes, this table displays which of the nine Royston-Parmar spline models is rated as the best based on the AIC. As above, $p$ refers to the number of inner knots in the spline model. The empirical rates refer to the total quantity of runs without early stopping of the corresponding simulated trial. This table refers to the scenario $(\rho^{\star}, \gamma^{\star}) = (2,0)$}
    \label{supp-table:aic_choice_20}
\end{table}

\begin{table}[h]
\footnotesize
    \centering
    \begin{tabular}{|l||c|c|c|c|c|c|c|c|c|}
    \hline
    \multirow{2}{*}{Effect size}&\multicolumn{3}{c|}{$p=0$}&\multicolumn{3}{c|}{$p=1$}&\multicolumn{3}{c|}{$p=2$}\\
    \cline{2-10}
    &hazard&odds&normal&hazard&odds&normal&hazard&odds&normal\\
    \hline
    $ 0.4 \theta_0 $&$ 0.7586 $&$ 0.0323 $&$ 0 $&$ 0.0765 $&$ 0.0351 $&$ 0.0359 $&$ 0.0223 $&$ 0.0147 $&$ 0.0246$\\
    \hline 
    $ 0.6 \theta_0 $&$ 0.7262 $&$ 0.0461 $&$ 0.0001 $&$ 0.0827 $&$ 0.0369 $&$ 0.0435 $&$ 0.0236 $&$ 0.0167 $&$ 0.0243$\\
    \hline 
    $ 0.8 \theta_0 $&$ 0.7094 $&$ 0.053 $&$ 0.0001 $&$ 0.0851 $&$ 0.0414 $&$ 0.0426 $&$ 0.0234 $&$ 0.0183 $&$ 0.0266$\\
    \hline 
    $ 1 \theta_0 $&$ 0.6718 $&$ 0.0697 $&$ 0.0001 $&$ 0.0954 $&$ 0.0451 $&$ 0.0496 $&$ 0.0233 $&$ 0.0175 $&$ 0.0274$\\
    \hline 
    $ 1.2 \theta_0 $&$ 0.6471 $&$ 0.074 $&$ 0.0003 $&$ 0.0958 $&$ 0.0522 $&$ 0.058 $&$ 0.0267 $&$ 0.02 $&$ 0.0259$\\
    \hline 
    $ 1.4 \theta_0 $&$ 0.6214 $&$ 0.0853 $&$ 0.0004 $&$ 0.1007 $&$ 0.0557 $&$ 0.0589 $&$ 0.0288 $&$ 0.0216 $&$ 0.0273$\\
    \hline 
    $ 1.6 \theta_0 $&$ 0.569 $&$ 0.1016 $&$ 0.0002 $&$ 0.1132 $&$ 0.0631 $&$ 0.0644 $&$ 0.0331 $&$ 0.0229 $&$ 0.0324$\\
    \hline 
    \end{tabular}
    \caption{For each of the seven effect sizes, this table displays which of the nine Royston-Parmar spline models is rated as the best based on the AIC. As above, $p$ refers to the number of inner knots in the spline model. The empirical rates refer to the total quantity of runs without early stopping of the corresponding simulated trial. This table refers to the scenario $(\rho^{\star}, \gamma^{\star}) = (3,0)$}
    \label{supp-table:aic_choice_30}
\end{table}

\begin{table}[h]
\footnotesize
    \centering
    \begin{tabular}{|l||c|c|c|c|c|c|c|c|c|}
    \hline
    \multirow{2}{*}{Effect size}&\multicolumn{3}{c|}{$p=0$}&\multicolumn{3}{c|}{$p=1$}&\multicolumn{3}{c|}{$p=2$}\\
    \cline{2-10}
    &hazard&odds&normal&hazard&odds&normal&hazard&odds&normal\\
    \hline
    $ 0.4 \theta_0 $&$ 0.7827 $&$ 0.0209 $&$ 0 $&$ 0.0804 $&$ 0.0256 $&$ 0.0226 $&$ 0.0246 $&$ 0.0155 $&$ 0.0276$\\
    \hline 
    $ 0.6 \theta_0 $&$ 0.7762 $&$ 0.0189 $&$ 0 $&$ 0.0864 $&$ 0.0259 $&$ 0.0215 $&$ 0.0274 $&$ 0.0163 $&$ 0.0274$\\
    \hline 
    $ 0.8 \theta_0 $&$ 0.7787 $&$ 0.0189 $&$ 0 $&$ 0.0861 $&$ 0.0244 $&$ 0.0224 $&$ 0.0259 $&$ 0.0163 $&$ 0.0273$\\
    \hline 
    $ 1 \theta_0 $&$ 0.7788 $&$ 0.0212 $&$ 0 $&$ 0.0903 $&$ 0.0249 $&$ 0.0176 $&$ 0.0252 $&$ 0.0142 $&$ 0.0279$\\
    \hline 
    $ 1.2 \theta_0 $&$ 0.7861 $&$ 0.0182 $&$ 0 $&$ 0.0875 $&$ 0.0229 $&$ 0.0183 $&$ 0.0236 $&$ 0.0176 $&$ 0.0258$\\
    \hline 
    $ 1.4 \theta_0 $&$ 0.7889 $&$ 0.0158 $&$ 0 $&$ 0.0836 $&$ 0.0249 $&$ 0.0135 $&$ 0.0298 $&$ 0.0132 $&$ 0.0303$\\
    \hline 
    $ 1.6 \theta_0 $&$ 0.7798 $&$ 0.0179 $&$ 0 $&$ 0.0913 $&$ 0.0253 $&$ 0.0137 $&$ 0.029 $&$ 0.0134 $&$ 0.0297$\\
    \hline 
    \end{tabular}
    \caption{For each of the seven effect sizes, this table displays which of the nine Royston-Parmar spline models is rated as the best based on the AIC. As above, $p$ refers to the number of inner knots in the spline model. The empirical rates refer to the total quantity of runs without early stopping of the corresponding simulated trial. This table refers to the scenario $(\rho^{\star}, \gamma^{\star}) = (0,1)$}
    \label{supp-table:aic_choice_01}
\end{table}

\begin{table}[h]
\footnotesize
    \centering
    \begin{tabular}{|l||c|c|c|c|c|c|c|c|c|}
    \hline
    \multirow{2}{*}{Effect size}&\multicolumn{3}{c|}{$p=0$}&\multicolumn{3}{c|}{$p=1$}&\multicolumn{3}{c|}{$p=2$}\\
    \cline{2-10}
    &hazard&odds&normal&hazard&odds&normal&hazard&odds&normal\\
    \hline
    $ 0.4 \theta_0 $&$ 0.7812 $&$ 0.0194 $&$ 0 $&$ 0.0863 $&$ 0.0284 $&$ 0.0201 $&$ 0.0254 $&$ 0.0143 $&$ 0.0248$\\
    \hline 
    $ 0.6 \theta_0 $&$ 0.7752 $&$ 0.0172 $&$ 0 $&$ 0.0887 $&$ 0.0294 $&$ 0.0223 $&$ 0.0238 $&$ 0.0159 $&$ 0.0275$\\
    \hline 
    $ 0.8 \theta_0 $&$ 0.7659 $&$ 0.0177 $&$ 0 $&$ 0.0969 $&$ 0.0306 $&$ 0.0208 $&$ 0.0238 $&$ 0.0165 $&$ 0.0279$\\
    \hline 
    $ 1 \theta_0 $&$ 0.7668 $&$ 0.0148 $&$ 0 $&$ 0.1012 $&$ 0.0294 $&$ 0.0204 $&$ 0.024 $&$ 0.0165 $&$ 0.0269$\\
    \hline 
    $ 1.2 \theta_0 $&$ 0.7636 $&$ 0.0145 $&$ 0 $&$ 0.1015 $&$ 0.0294 $&$ 0.0162 $&$ 0.0278 $&$ 0.0192 $&$ 0.0278$\\
    \hline 
    $ 1.4 \theta_0 $&$ 0.7588 $&$ 0.014 $&$ 0 $&$ 0.1107 $&$ 0.0309 $&$ 0.0148 $&$ 0.0268 $&$ 0.0173 $&$ 0.0267$\\
    \hline 
    $ 1.6 \theta_0 $&$ 0.7526 $&$ 0.0123 $&$ 0 $&$ 0.1194 $&$ 0.0271 $&$ 0.015 $&$ 0.0306 $&$ 0.0174 $&$ 0.0256$\\
    \hline 
    \end{tabular}
    \caption{For each of the seven effect sizes, this table displays which of the nine Royston-Parmar spline models is rated as the best based on the AIC. As above, $p$ refers to the number of inner knots in the spline model. The empirical rates refer to the total quantity of runs without early stopping of the corresponding simulated trial. This table refers to the scenario $(\rho^{\star}, \gamma^{\star}) = (0,2)$}
    \label{supp-table:aic_choice_02}
\end{table}

\begin{table}[h]
\footnotesize
    \centering
    \begin{tabular}{|l||c|c|c|c|c|c|c|c|c|}
    \hline
    \multirow{2}{*}{Effect size}&\multicolumn{3}{c|}{$p=0$}&\multicolumn{3}{c|}{$p=1$}&\multicolumn{3}{c|}{$p=2$}\\
    \cline{2-10}
    &hazard&odds&normal&hazard&odds&normal&hazard&odds&normal\\
    \hline
    $ 0.4 \theta_0 $&$ 0.7807 $&$ 0.0142 $&$ 0 $&$ 0.0861 $&$ 0.0289 $&$ 0.0226 $&$ 0.0233 $&$ 0.0168 $&$ 0.0274$\\
    \hline 
    $ 0.6 \theta_0 $&$ 0.7706 $&$ 0.0185 $&$ 0 $&$ 0.088 $&$ 0.0307 $&$ 0.0254 $&$ 0.0244 $&$ 0.0164 $&$ 0.026$\\
    \hline 
    $ 0.8 \theta_0 $&$ 0.7672 $&$ 0.0178 $&$ 0 $&$ 0.0964 $&$ 0.0309 $&$ 0.0233 $&$ 0.0229 $&$ 0.0186 $&$ 0.0229$\\
    \hline 
    $ 1 \theta_0 $&$ 0.7573 $&$ 0.0183 $&$ 0.0001 $&$ 0.0984 $&$ 0.0351 $&$ 0.0228 $&$ 0.0259 $&$ 0.0174 $&$ 0.0247$\\
    \hline 
    $ 1.2 \theta_0 $&$ 0.7571 $&$ 0.0152 $&$ 0 $&$ 0.1066 $&$ 0.0368 $&$ 0.0181 $&$ 0.0227 $&$ 0.0175 $&$ 0.0261$\\
    \hline 
    $ 1.4 \theta_0 $&$ 0.7528 $&$ 0.0159 $&$ 0 $&$ 0.1123 $&$ 0.0383 $&$ 0.0195 $&$ 0.022 $&$ 0.0154 $&$ 0.0239$\\
    \hline 
    $ 1.6 \theta_0 $&$ 0.731 $&$ 0.0151 $&$ 0 $&$ 0.1192 $&$ 0.0413 $&$ 0.0216 $&$ 0.0236 $&$ 0.0228 $&$ 0.0254$\\
    \hline 
    \end{tabular}
    \caption{For each of the seven effect sizes, this table displays which of the nine Royston-Parmar spline models is rated as the best based on the AIC. As above, $p$ refers to the number of inner knots in the spline model. The empirical rates refer to the total quantity of runs without early stopping of the corresponding simulated trial. This table refers to the scenario $(\rho^{\star}, \gamma^{\star}) = (0,3)$}
    \label{supp-table:aic_choice_03}
\end{table}

\clearpage

\subsubsection{Modelwise choice of test statistics in the second stage}\label{supp-subsubsec:modelwise_test_choice}

For the three deviation types from the main manuscript, we also show how the choice of the second stage test is distributed for each of the nine spline models. As already mentioned, models with a higher number of inner knots seem to be better suited to choose an appropriate test in late effects settings. Because of that, one should maybe restrict the set of models to a set with a higher number of inner knots if a possible late effect is anticipated.

\begin{figure}[h!]
	\centering
    \begin{subfigure}[t]{0.075\textwidth}
	
	\end{subfigure}
    \begin{subfigure}[t]{0.3\textwidth}
    \centering
		Hazard scale
        \vspace{6pt}
	\end{subfigure}
	\begin{subfigure}[t]{0.3\textwidth}
    \centering
		Odds scale
        \vspace{6pt}
	\end{subfigure}
	\begin{subfigure}[t]{0.3\textwidth}
    \centering
		Normal scale
        \vspace{6pt}
	\end{subfigure}
    \begin{subfigure}{0.075\textwidth}
        $p=0$
        \vspace{2.5cm}
	\end{subfigure}
	\begin{subfigure}[t]{0.3\textwidth}
		\includegraphics[width=\linewidth]{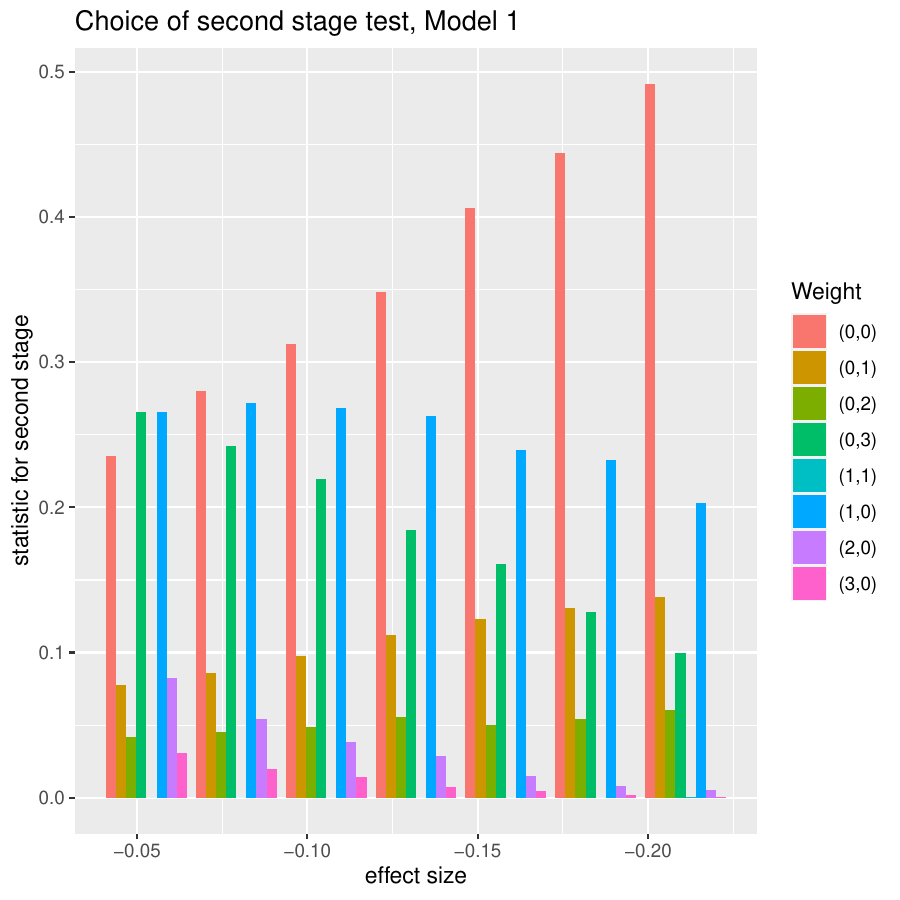}
	\end{subfigure}
	\begin{subfigure}[t]{0.3\textwidth}
		\includegraphics[width=\linewidth]{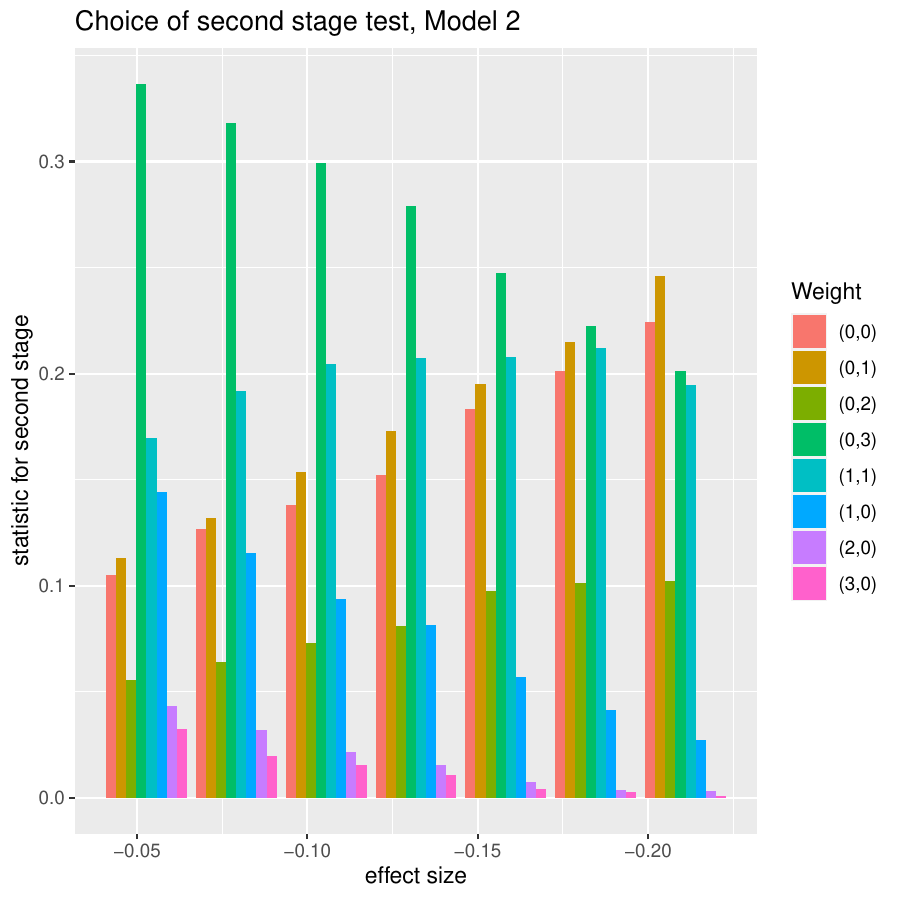}
	\end{subfigure}
	\begin{subfigure}[t]{0.3\textwidth}
		\includegraphics[width=\linewidth]{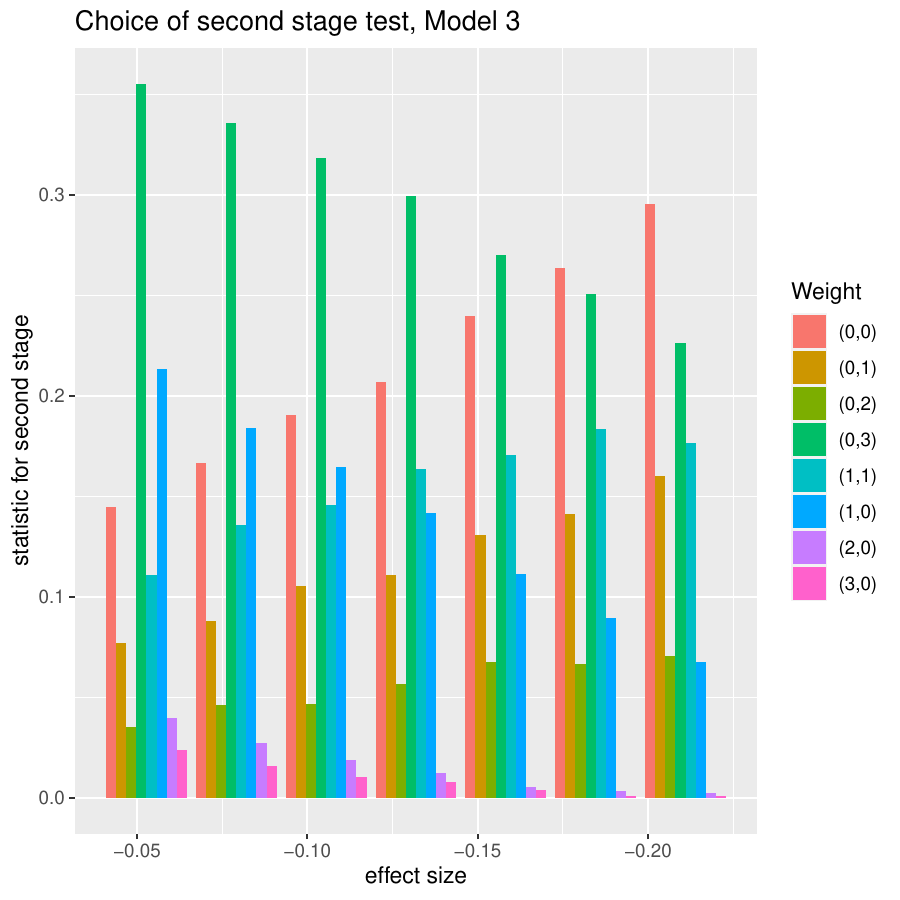}
	\end{subfigure}
    \begin{subfigure}{0.075\textwidth}
        $p=1$
        \vspace{2.5cm}
	\end{subfigure}
	\begin{subfigure}[t]{0.3\textwidth}
		\includegraphics[width=\linewidth]{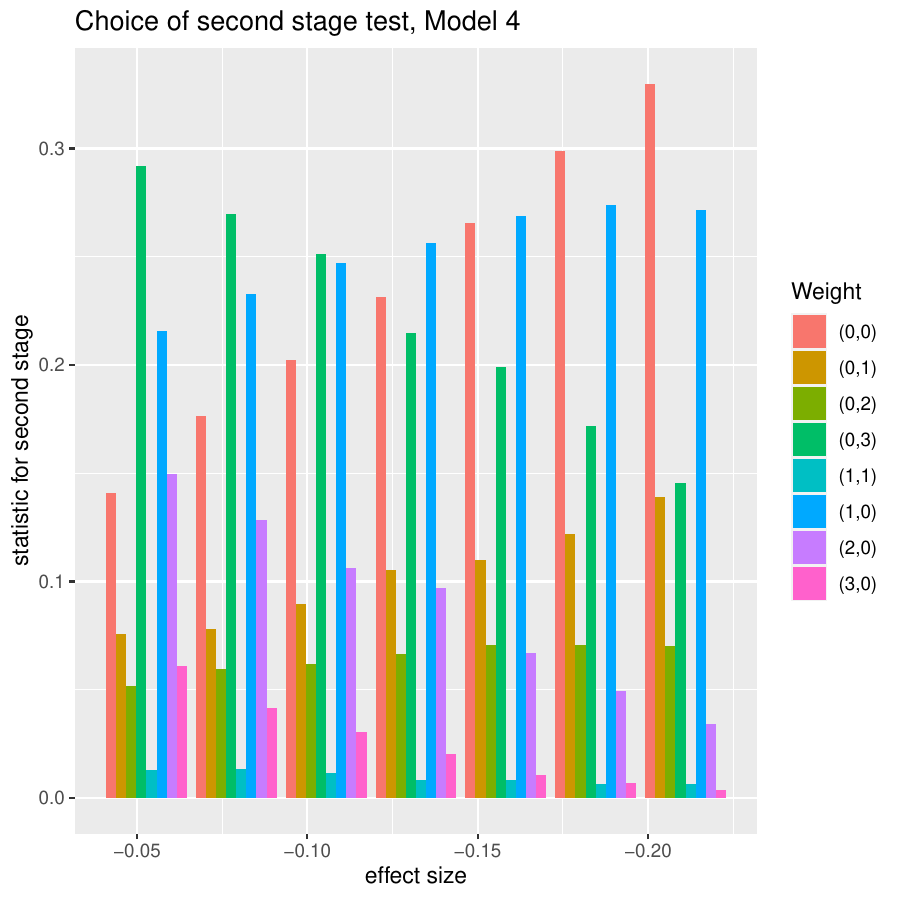}
	\end{subfigure}
    \begin{subfigure}[t]{0.3\textwidth}
		\includegraphics[width=\linewidth]{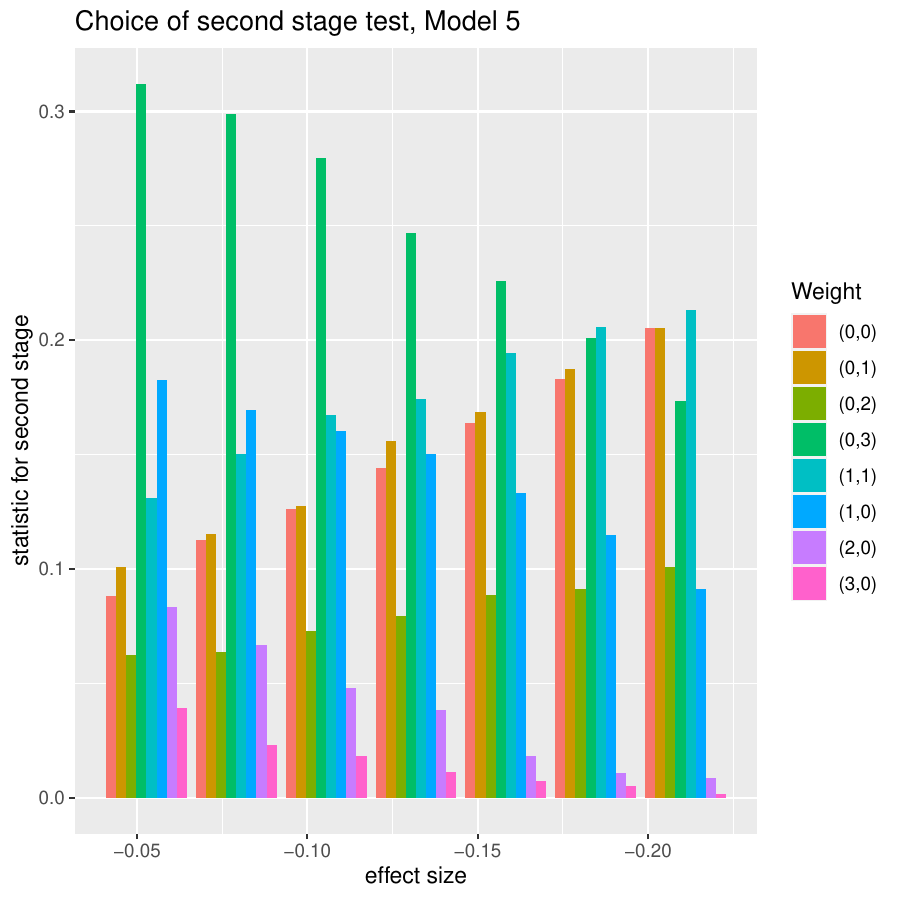}
	\end{subfigure}
    \begin{subfigure}[t]{0.3\textwidth}
		\includegraphics[width=\linewidth]{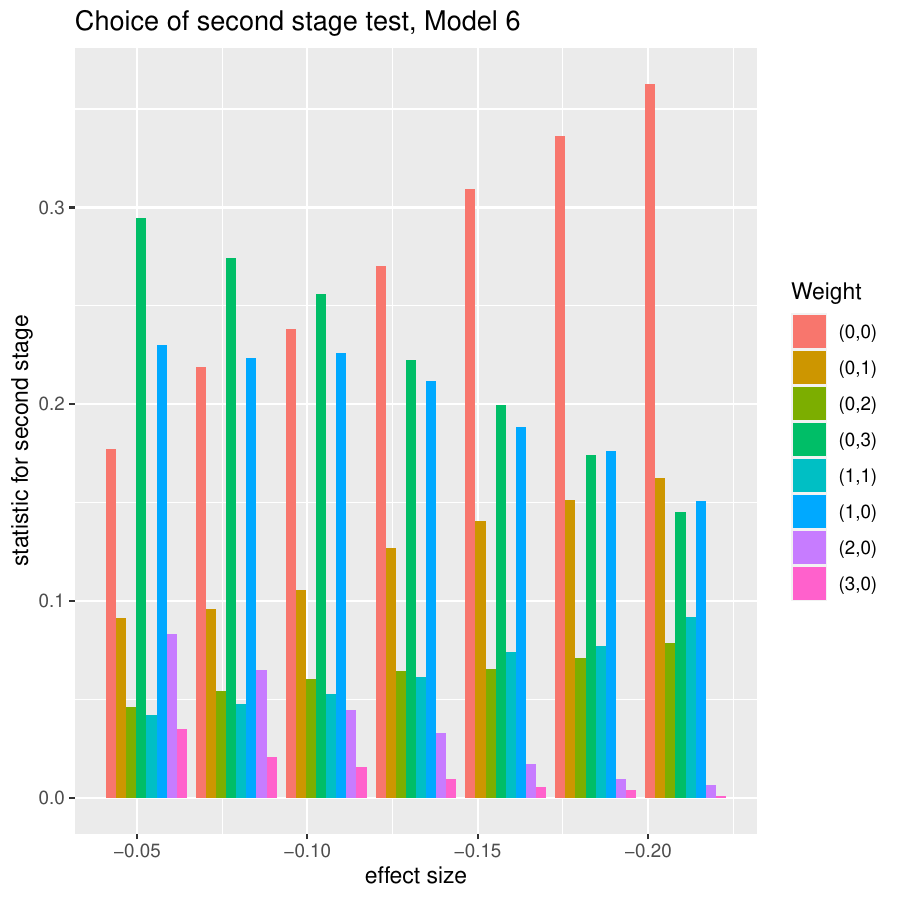}
	\end{subfigure}
    \begin{subfigure}{0.075\textwidth}
        $p=2$
        \vspace{2.5cm}
	\end{subfigure}
    \begin{subfigure}[t]{0.3\textwidth}
		\includegraphics[width=\linewidth]{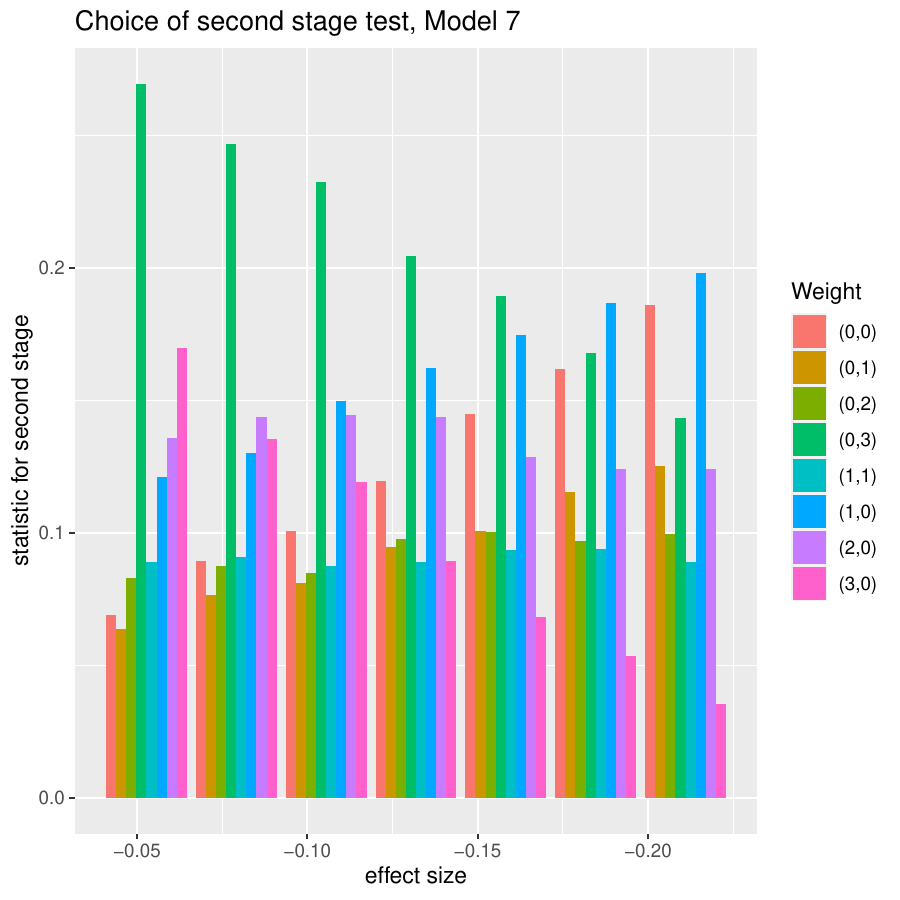}
	\end{subfigure}
    \begin{subfigure}[t]{0.3\textwidth}
		\includegraphics[width=\linewidth]{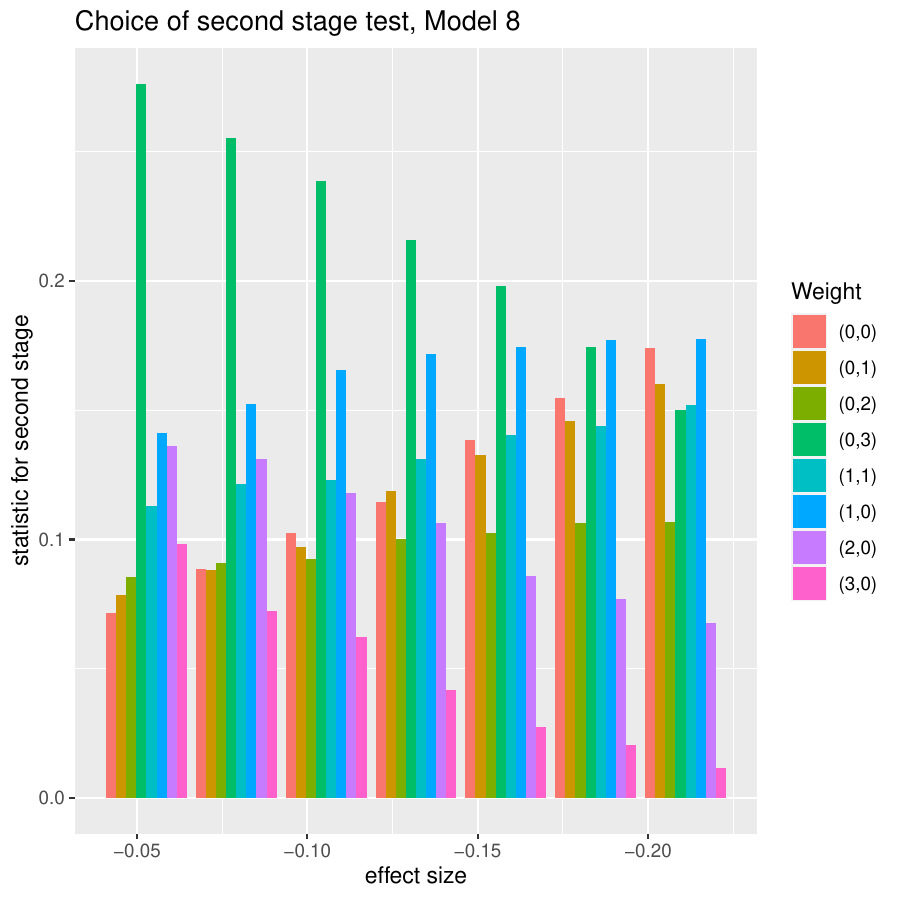}
	\end{subfigure}
    \begin{subfigure}[t]{0.3\textwidth}
		\includegraphics[width=\linewidth]{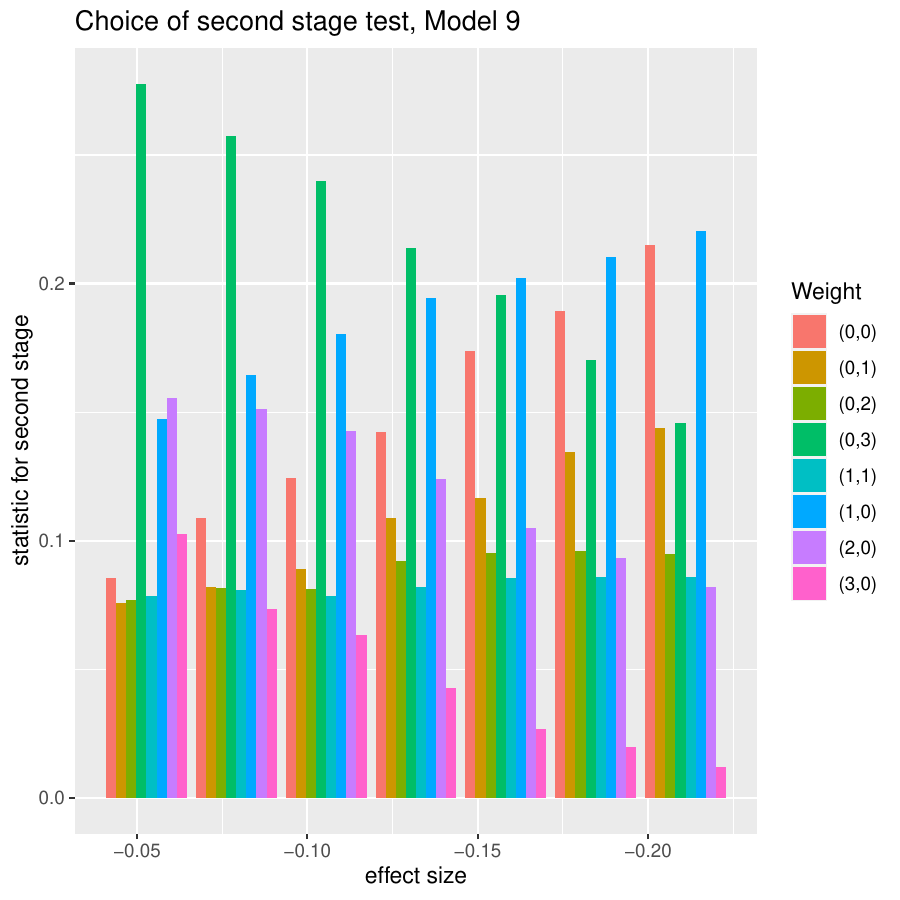}
	\end{subfigure}
	\caption{Choices of the log-rank test statistic with Fleming-Harrington weights for our nine different Royston-Parmar spline models. The rates refer to the total quantity of simulation runs in which the corresponding simulated trial proceeded to a second stage (i.e. no early termination). The plots are arranged in a grid where the columns refer to different scales and the rows to different numbers of interior knots $p$. These figures refer to the deviation type $(\rho^{\star}, \gamma^{\star})=(0,0)$, i.e. proportional hazards.}
	\label{supp-fig:modelwise_choice_0_0}
\end{figure}

\begin{figure}[h!]
	\centering
    \begin{subfigure}[t]{0.075\textwidth}
	
	\end{subfigure}
    \begin{subfigure}[t]{0.3\textwidth}
    \centering
		Hazard scale
        \vspace{6pt}
	\end{subfigure}
	\begin{subfigure}[t]{0.3\textwidth}
    \centering
		Odds scale
        \vspace{6pt}
	\end{subfigure}
	\begin{subfigure}[t]{0.3\textwidth}
    \centering
		Normal scale
        \vspace{6pt}
	\end{subfigure}
    \begin{subfigure}{0.075\textwidth}
        $p=0$
        \vspace{2.5cm}
	\end{subfigure}
	\begin{subfigure}[t]{0.3\textwidth}
		\includegraphics[width=\linewidth]{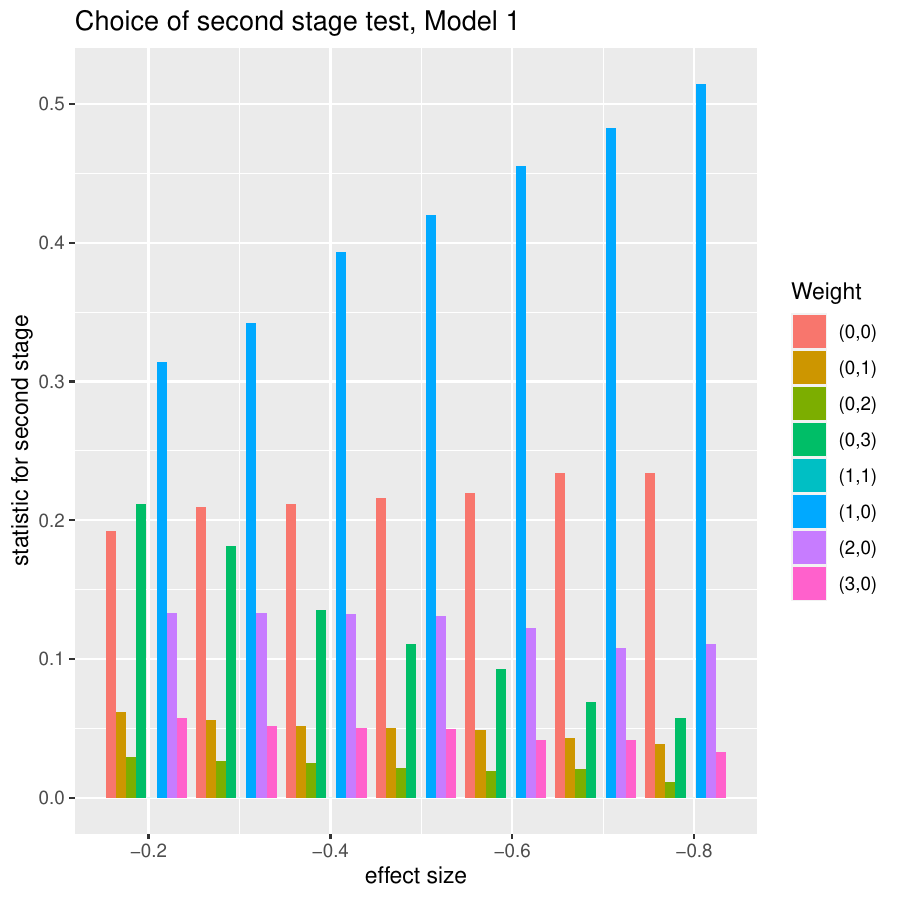}
	\end{subfigure}
	\begin{subfigure}[t]{0.3\textwidth}
		\includegraphics[width=\linewidth]{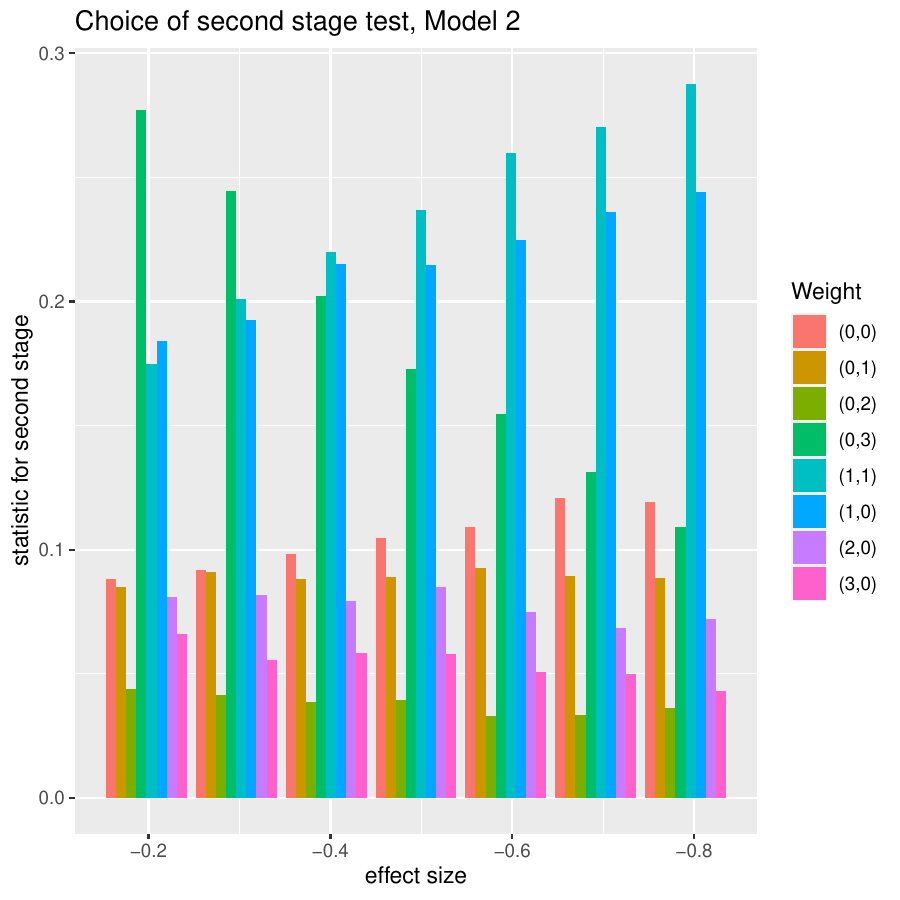}
	\end{subfigure}
	\begin{subfigure}[t]{0.3\textwidth}
		\includegraphics[width=\linewidth]{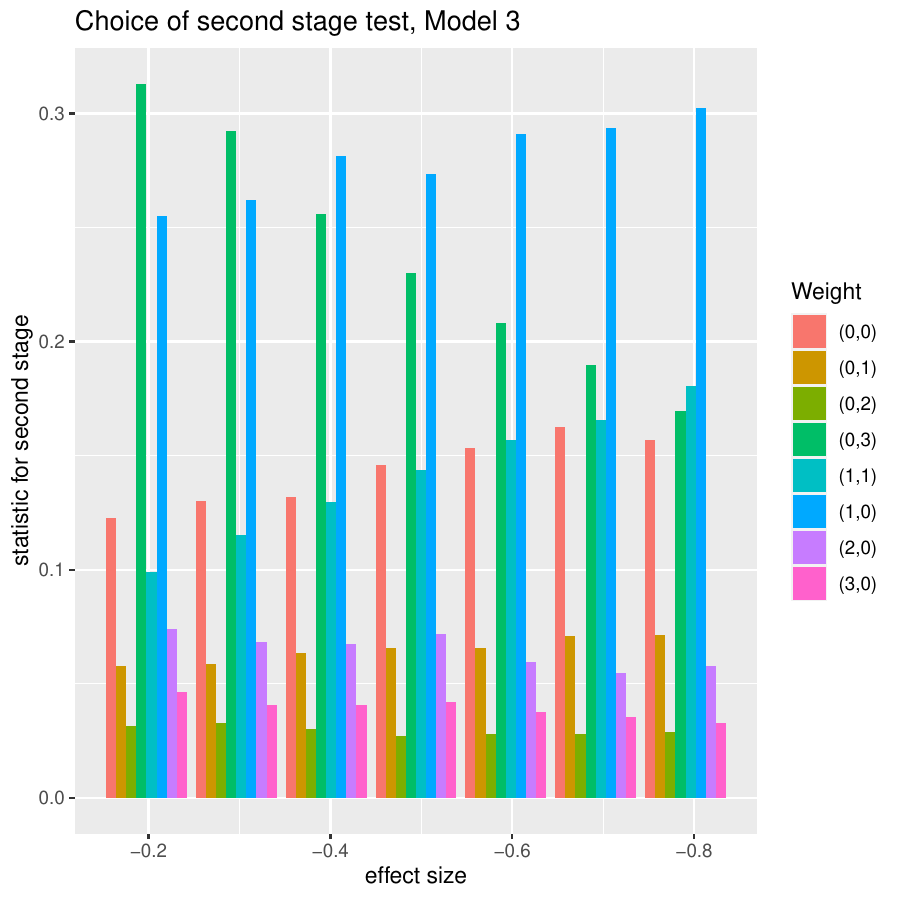}
	\end{subfigure}
    \begin{subfigure}{0.075\textwidth}
        $p=1$
        \vspace{2.5cm}
	\end{subfigure}
	\begin{subfigure}[t]{0.3\textwidth}
		\includegraphics[width=\linewidth]{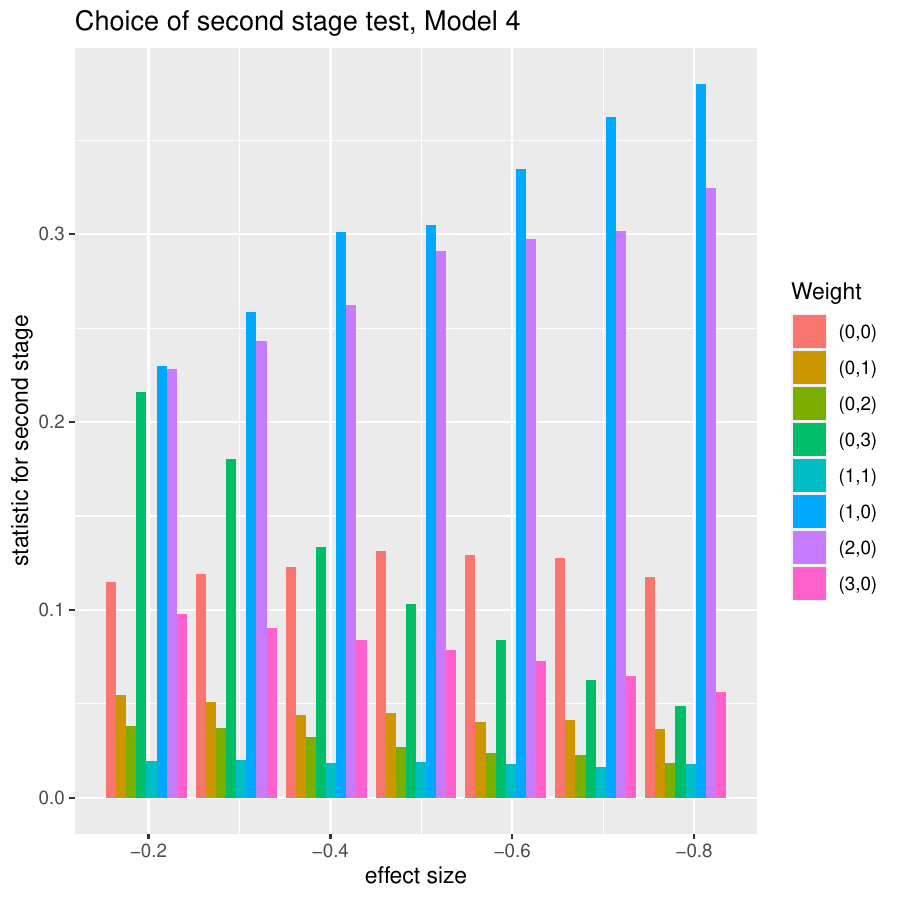}
	\end{subfigure}
    \begin{subfigure}[t]{0.3\textwidth}
		\includegraphics[width=\linewidth]{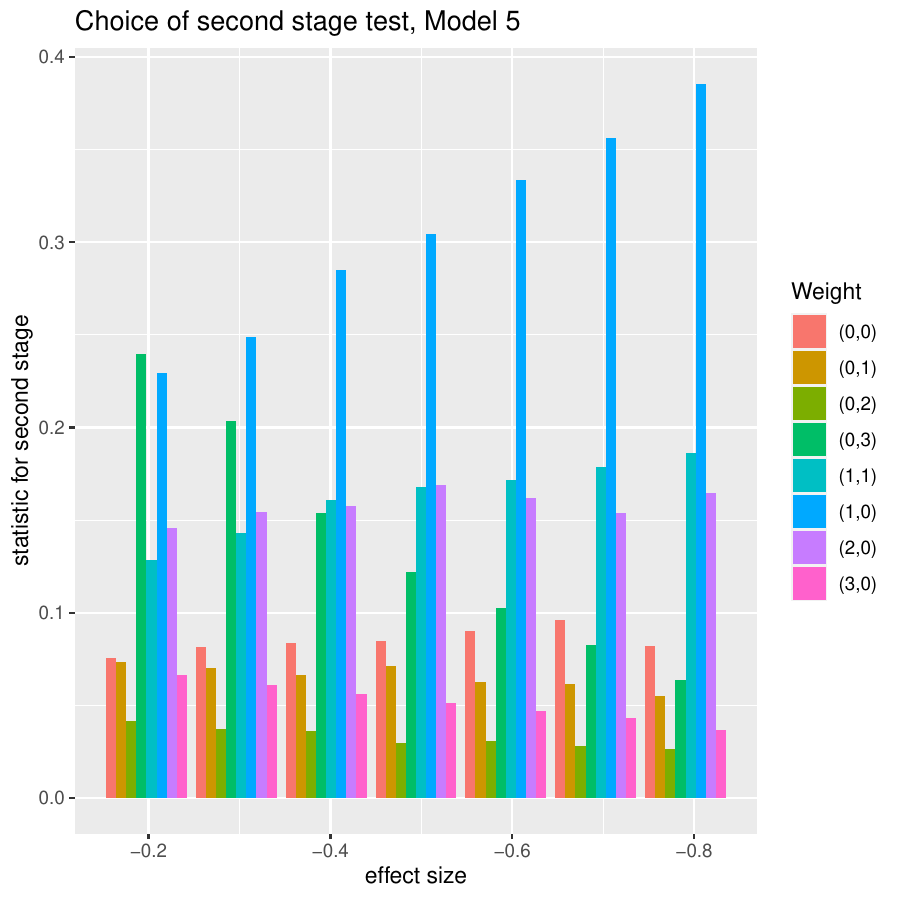}
	\end{subfigure}
    \begin{subfigure}[t]{0.3\textwidth}
		\includegraphics[width=\linewidth]{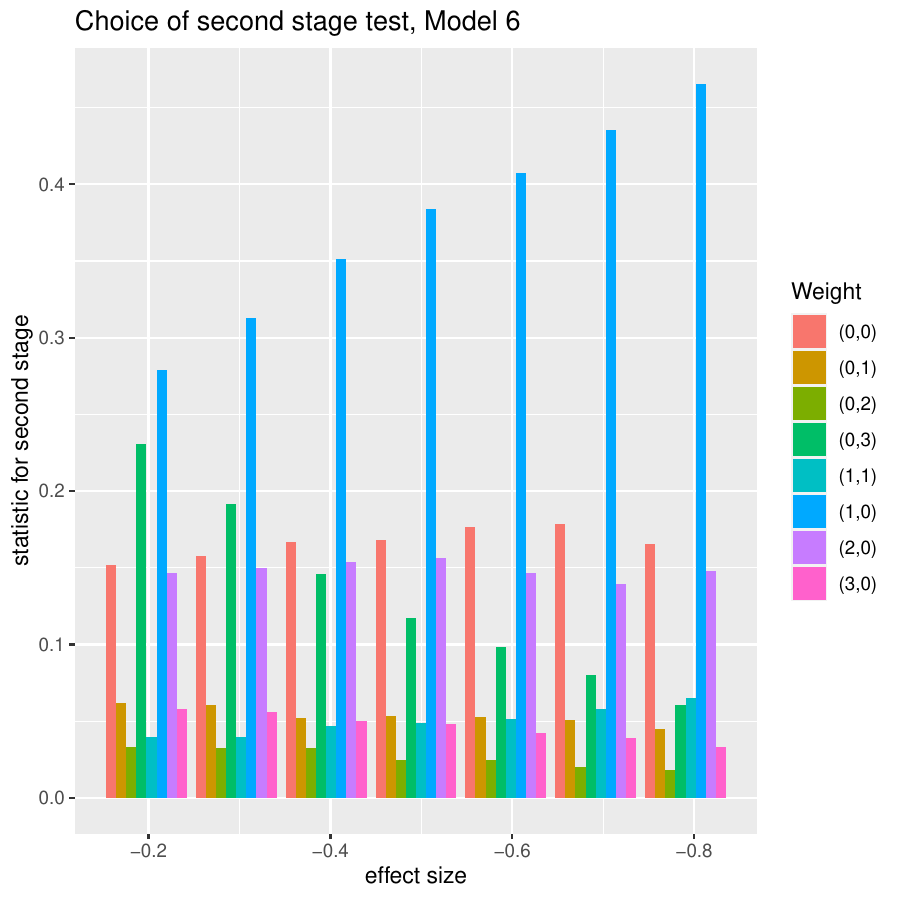}
	\end{subfigure}
    \begin{subfigure}{0.075\textwidth}
        $p=2$
        \vspace{2.5cm}
	\end{subfigure}
    \begin{subfigure}[t]{0.3\textwidth}
		\includegraphics[width=\linewidth]{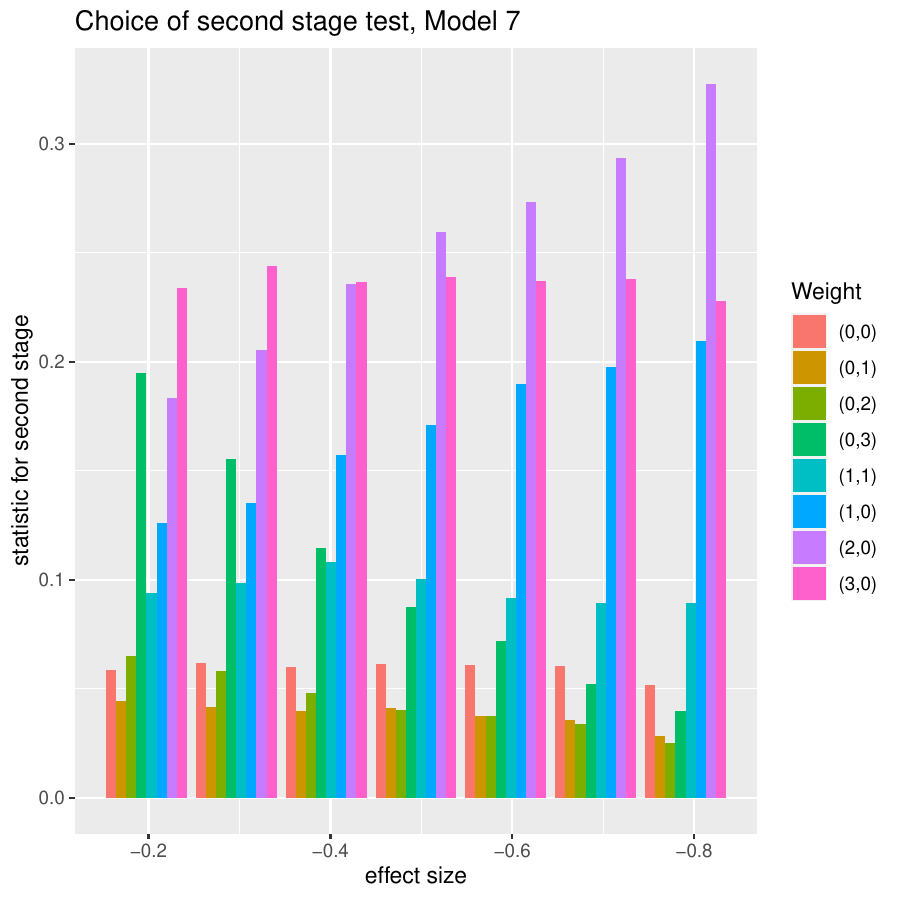}
	\end{subfigure}
    \begin{subfigure}[t]{0.3\textwidth}
		\includegraphics[width=\linewidth]{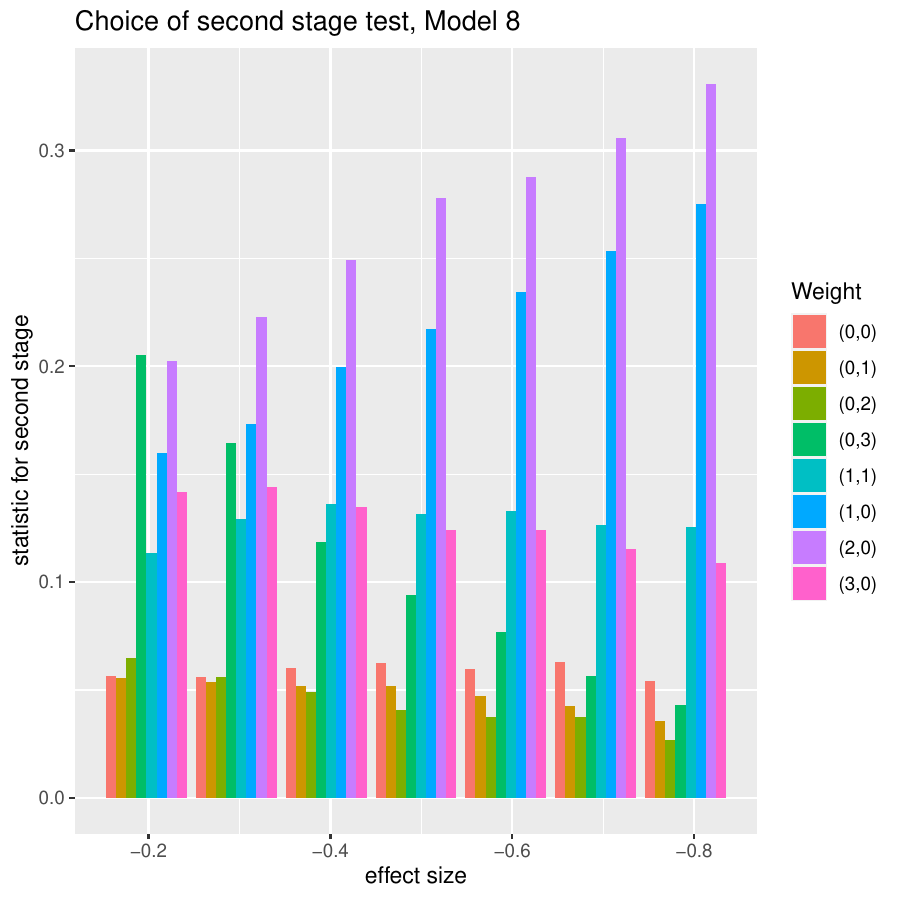}
	\end{subfigure}
    \begin{subfigure}[t]{0.3\textwidth}
		\includegraphics[width=\linewidth]{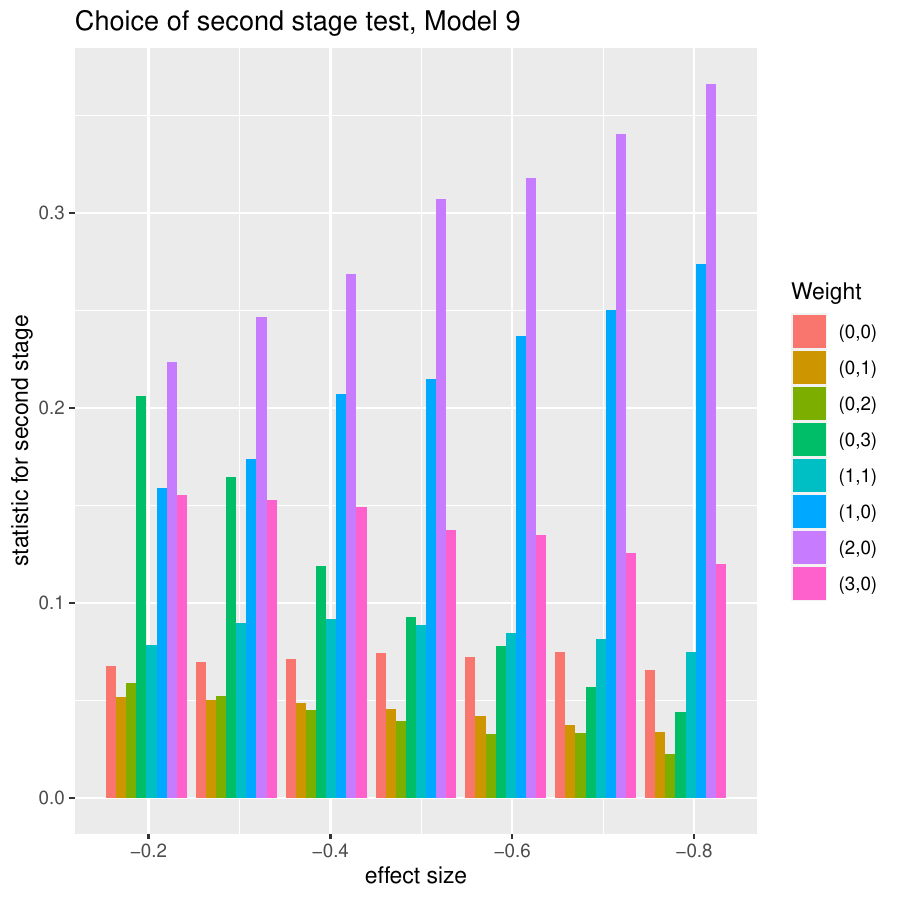}
	\end{subfigure}
	\caption{Choices of the log-rank test statistic with Fleming-Harrington weights for our nine different Royston-Parmar spline models. The rates refer to the total quantity of simulation runs in which the corresponding simulated trial proceeded to a second stage (i.e. no early termination). The plots are arranged in a grid where the columns refer to different scales and the rows to different numbers of interior knots $p$. These figures refer to the deviation type $(\rho^{\star}, \gamma^{\star})=(2,0)$, i.e. a late effects scenario.}
	\label{supp-fig:modelwise_choice_2_0}
\end{figure}

\begin{figure}[h!]
	\centering
    \begin{subfigure}[t]{0.075\textwidth}
	
	\end{subfigure}
    \begin{subfigure}[t]{0.3\textwidth}
    \centering
		Hazard scale
        \vspace{6pt}
	\end{subfigure}
	\begin{subfigure}[t]{0.3\textwidth}
    \centering
		Odds scale
        \vspace{6pt}
	\end{subfigure}
	\begin{subfigure}[t]{0.3\textwidth}
    \centering
		Normal scale
        \vspace{6pt}
	\end{subfigure}
    \begin{subfigure}{0.075\textwidth}
        $p=0$
        \vspace{2.5cm}
	\end{subfigure}
	\begin{subfigure}[t]{0.3\textwidth}
		\includegraphics[width=\linewidth]{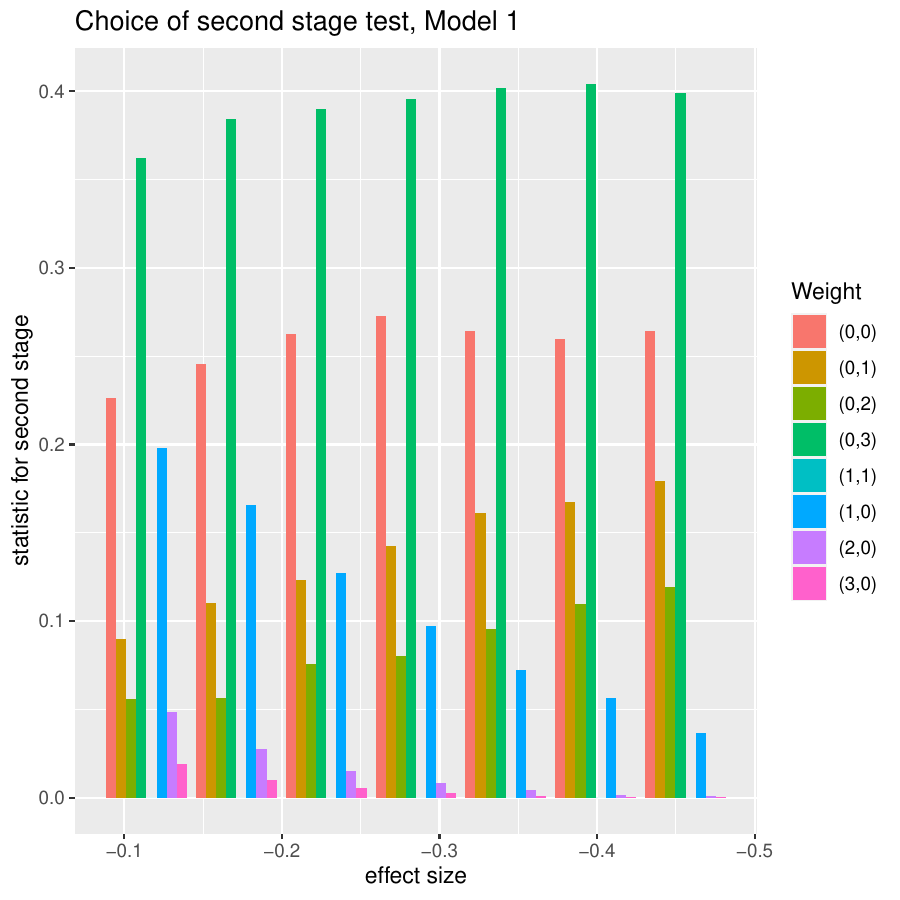}
	\end{subfigure}
	\begin{subfigure}[t]{0.3\textwidth}
		\includegraphics[width=\linewidth]{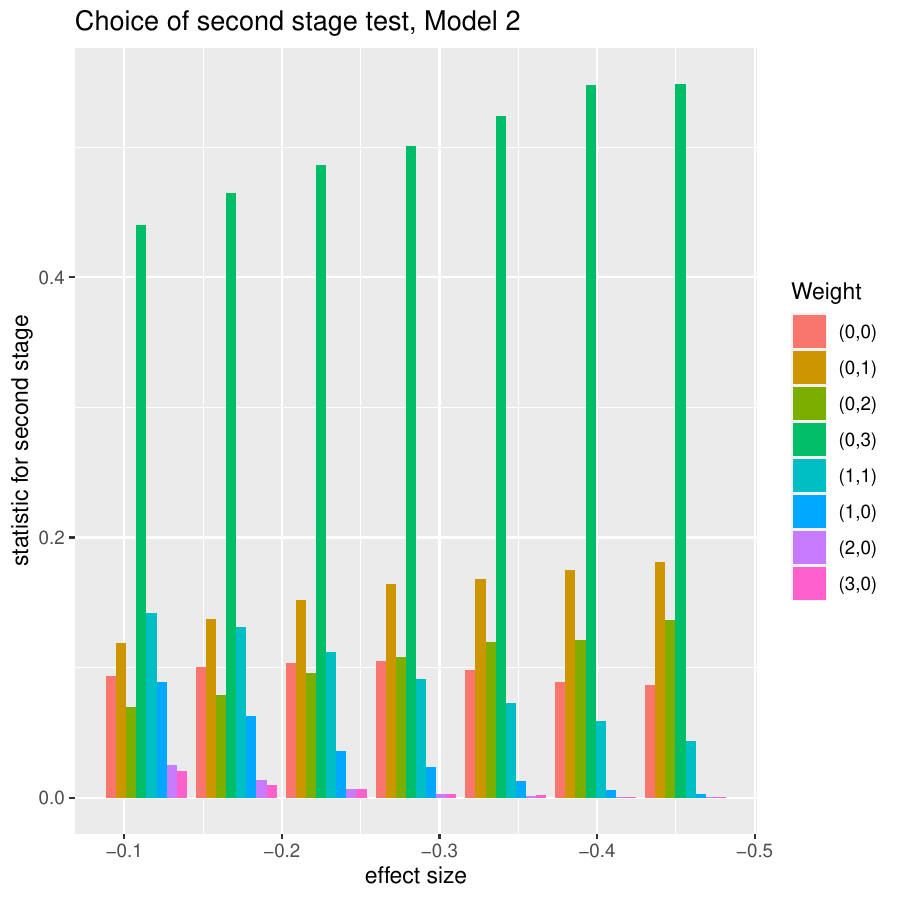}
	\end{subfigure}
	\begin{subfigure}[t]{0.3\textwidth}
		\includegraphics[width=\linewidth]{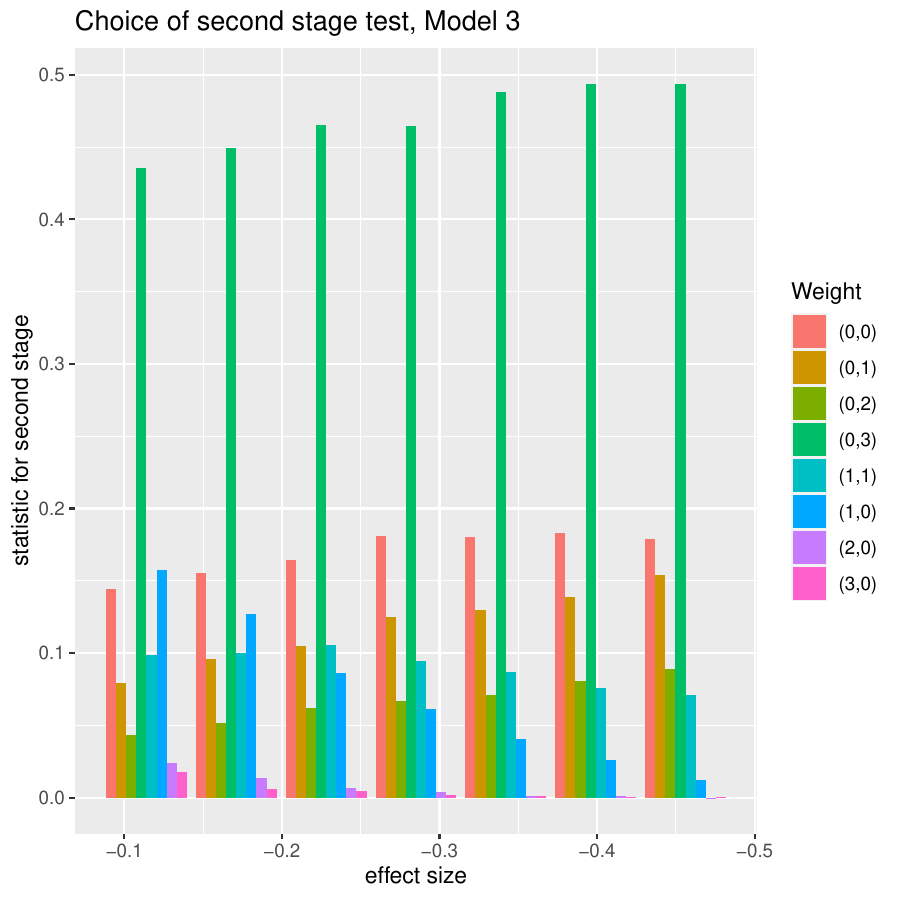}
	\end{subfigure}
    \begin{subfigure}{0.075\textwidth}
        $p=1$
        \vspace{2.5cm}
	\end{subfigure}
	\begin{subfigure}[t]{0.3\textwidth}
		\includegraphics[width=\linewidth]{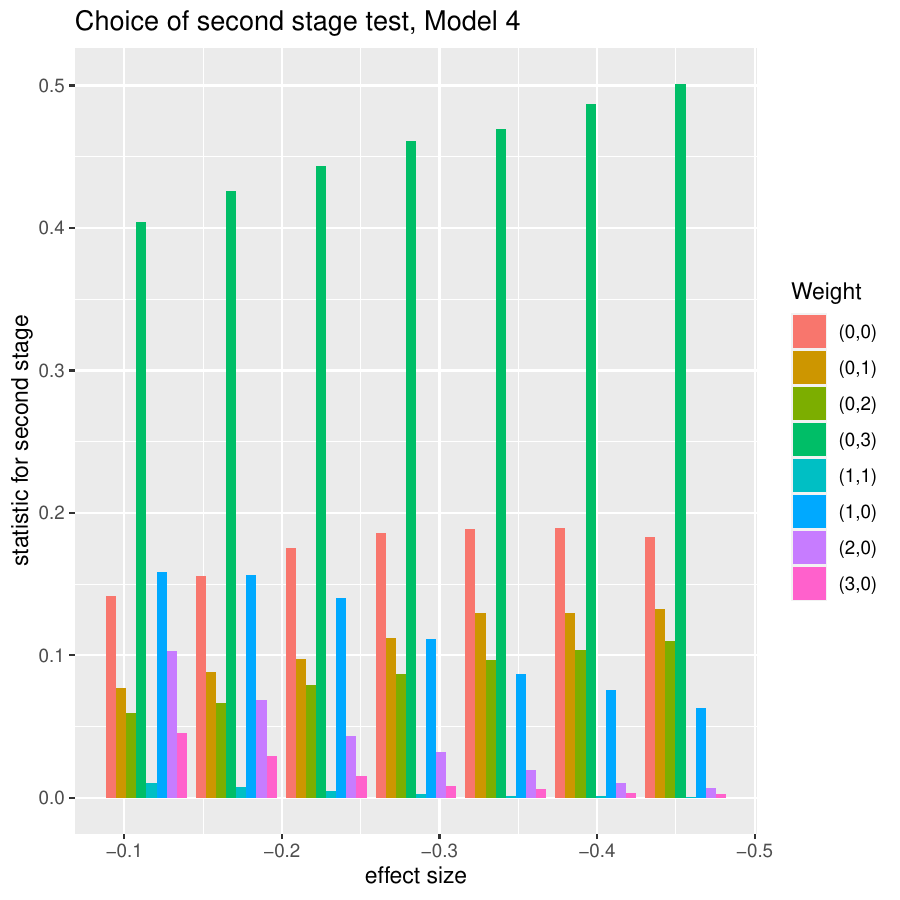}
	\end{subfigure}
    \begin{subfigure}[t]{0.3\textwidth}
		\includegraphics[width=\linewidth]{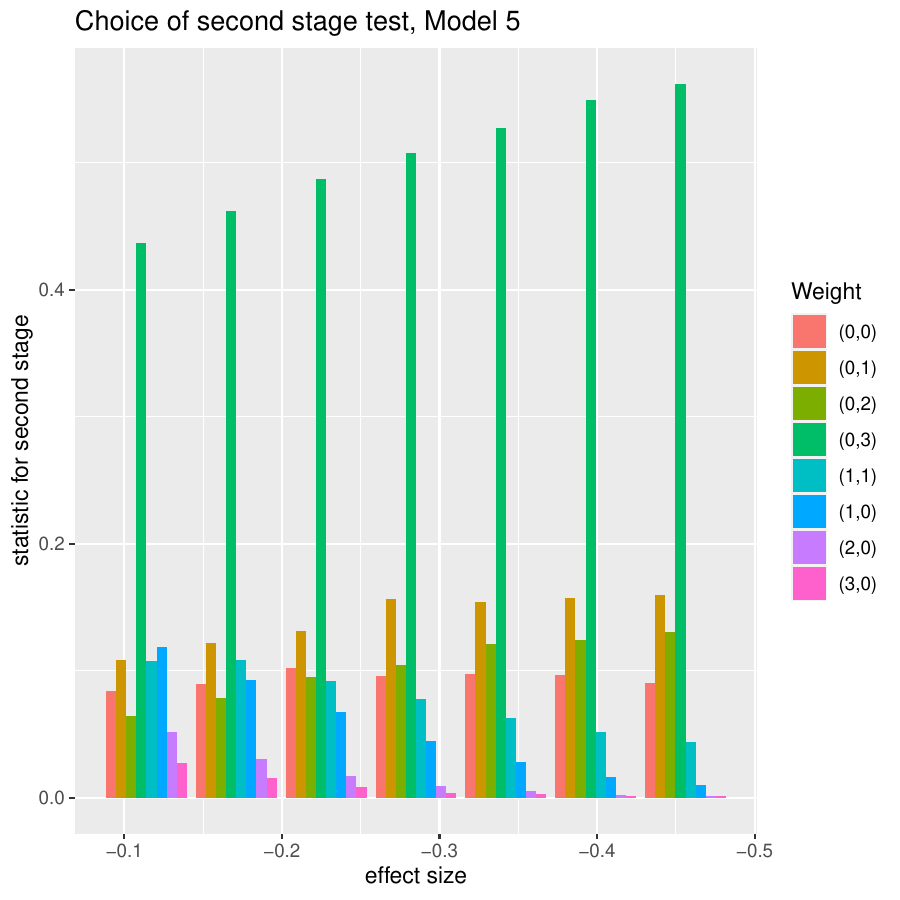}
	\end{subfigure}
    \begin{subfigure}[t]{0.3\textwidth}
		\includegraphics[width=\linewidth]{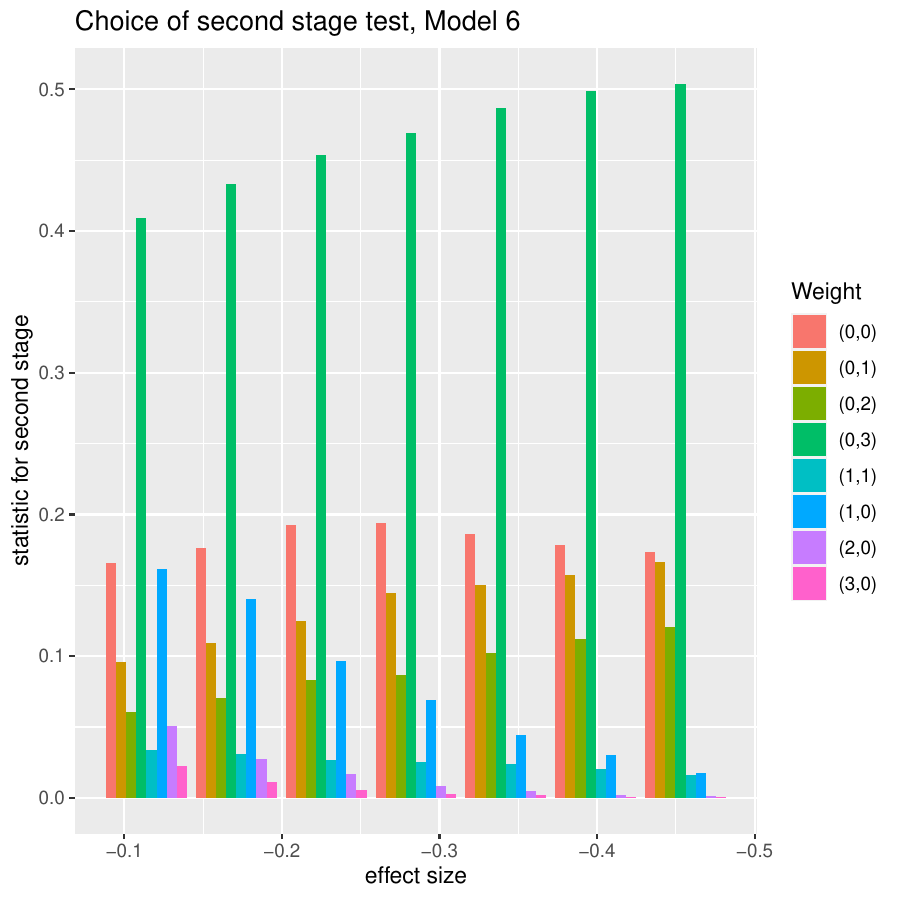}
	\end{subfigure}
    \begin{subfigure}{0.075\textwidth}
        $p=2$
        \vspace{2.5cm}
	\end{subfigure}
    \begin{subfigure}[t]{0.3\textwidth}
		\includegraphics[width=\linewidth]{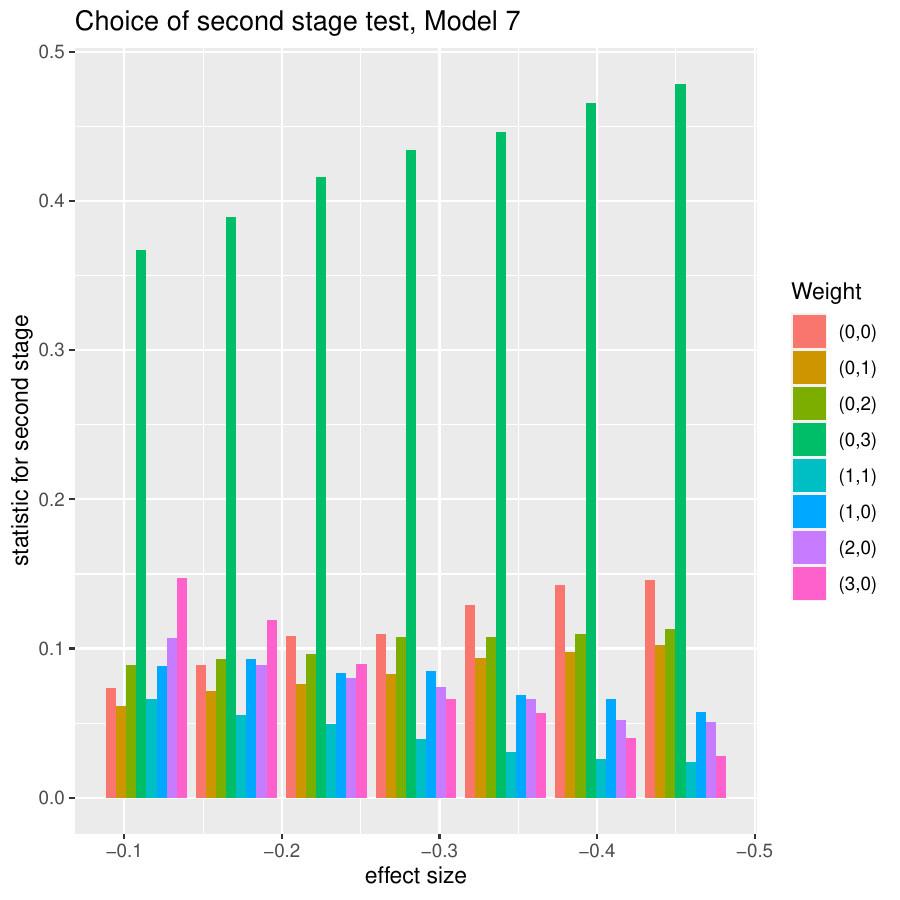}
	\end{subfigure}
    \begin{subfigure}[t]{0.3\textwidth}
		\includegraphics[width=\linewidth]{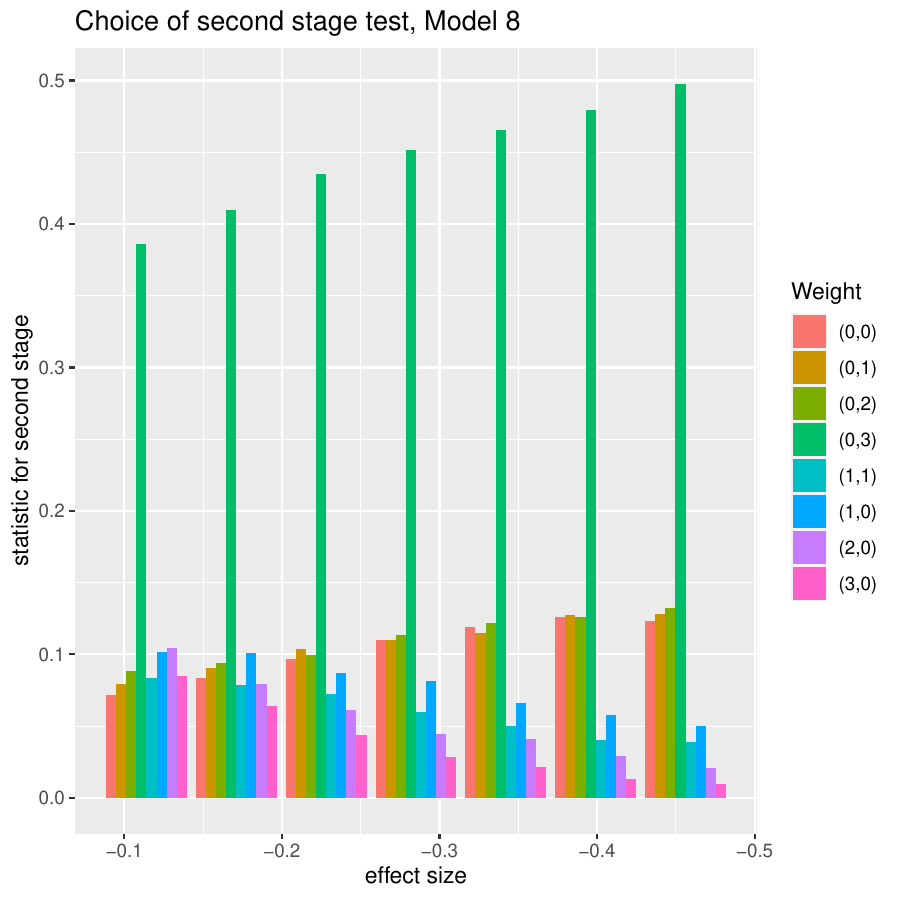}
	\end{subfigure}
    \begin{subfigure}[t]{0.3\textwidth}
		\includegraphics[width=\linewidth]{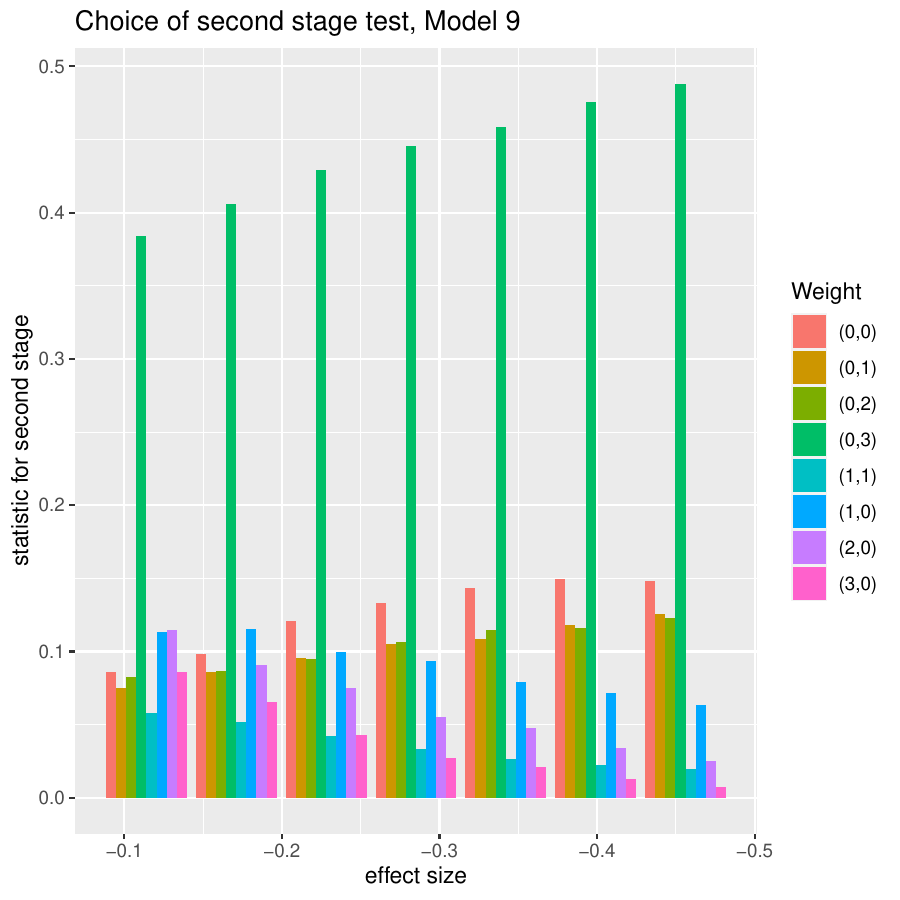}
	\end{subfigure}
	\caption{Choices of the log-rank test statistic with Fleming-Harrington weights for our nine different Royston-Parmar spline models. The rates refer to the total quantity of simulation runs in which the corresponding simulated trial proceeded to a second stage (i.e. no early termination). The plots are arranged in a grid where the columns refer to different scales and the rows to different numbers of interior knots $p$. These figures refer to the deviation type $(\rho^{\star}, \gamma^{\star})=(0,2)$, i.e. an early effects scenario.}
	\label{supp-fig:modelwise_choice_0_2}
\end{figure}

\clearpage

%\bibliography{Lib}
%\bibliographystyle{abbrv} 

\end{document}